\documentclass{article}
\usepackage{epsf}
\usepackage{epubtk}
\usepackage{longtable}
\usepackage{amssymb}
\usepackage{graphicx}

\newcommand{\Msun}{{\rm M_{\odot}}}

\newcommand{\pc}{{\rm pc}}
\newcommand{\trlx}{t_{\rm rlx}}
\newcommand{\trh}{t_{\rm rh}}
\newcommand{\tcr}{t_{\rm cr}}
\newcommand{\rc}{r_{c}}
\newcommand{\rh}{r_{h}}
\newcommand{\rt}{r_{t}}
\newcommand{\Myr}{{\rm Myr}}
\newcommand{\Gyr}{{\rm Gyr}}

\makeatletter
  \let\hangafter\@hangfrom
\makeatother

\newcounter{case}
\newenvironment{case}
  {\vspace{0.8 em}
   \refstepcounter{case}
   \hangafter{{Case~\thecase}:~}}
  {

  }
\renewcommand{\thecase}{\Alph{case}}

\newcommand{\mc}[1]{\multicolumn{1}{c}{#1}}
\newcommand{\qq}{\qquad}

\begin{document}

\title{Relativistic Binaries in Globular Clusters}

\author{\epubtkAuthorData{Matthew J.\ Benacquista}
                         {University of Texas at Brownsville\\
                         Center for Gravitational Wave Astronomy\\
                          80 Ft.\ Brown \\
                          Brownsville, Texas 78520}
                         {matthew.benacquista@utb.edu}
                         {http://phys.utb.edu/~benacquista/}
                        \and
                        \epubtkAuthorData{Jonathan M. B. Downing}
                         {Astronomisches Rechen-Institut\\
                         Zentrum f\"{u}r Astronomie der Universit\"a{}t
                         Heidelberg\\
                         12-14 M\"o{}nchhofstra\ss{}e\\
                         D-69120 Heidelberg\\
                         Germany}
                         {downin@ari.uni-heidelberg.de}{%
}
}
\date{}
\maketitle

%%%%%%%%%%%%%%%%%%%%%%%%%%%%%%%%%%%%%%%%%%%%%%%%%%%%%%%%%%%%%%%%%%%%%%%%%%%%%%%
%%%%%%%%%%%%%%%%%%%%%%%%%%%%%%%%%%%%%%%%%%%%%%%%%%%%%%%%%%%%%%%%%%%%%%%%%%%%%%%
%%%%%%%%%%%%%%%%%%%%%%%%%%%%%%%%%%%%%%%%%%%%%%%%%%%%%%%%%%%%%%%%%%%%%%%%%%%%%%%

\begin{abstract}
  Galactic globular clusters are old, dense star systems typically containing
  10\super{4}--10\super{7} stars. As an old population of stars, globular
  clusters contain many collapsed and degenerate objects. As a dense
  population of stars, globular clusters are the scene of many interesting
  close dynamical interactions between stars. These dynamical interactions can
  alter the evolution of individual stars and can produce tight binary systems
  containing one or two compact objects. In this review, we discuss
  theoretical models of globular cluster evolution and binary evolution,
  techniques for simulating this evolution that leads to relativistic
  binaries, and current and possible future observational evidence for this
  population. Our discussion of globular cluster evolution will focus on the
  processes that boost the production of hard binary systems and the
  subsequent interaction of these binaries that can alter the properties of
  both bodies and can lead to exotic objects. Direct {\it N}-body integrations
  and Fokker--Planck simulations of the evolution of globular clusters that
  incorporate tidal interactions and lead to predictions of relativistic
  binary populations are also discussed. We discuss the current observational
  evidence for cataclysmic variables, millisecond pulsars, and low-mass X-ray
  binaries as well as possible future detection of relativistic binaries with
  gravitational radiation.
\end{abstract}

\epubtkKeywords{accretion, accretion disks, astronomical observations,
astronomy, astrophysics, binary systems, black holes, dynamical systems,
gravitational wave sources, neutron stars, pulsars, radio astronomy,
stars, white dwarfs}

\newpage

%%%%%%%%%%%%%%%%%%%%%%%%%%%%%%%%%%%%%%%%%%%%%%%%%%%%%%%%%%%%%%%%%%%%%%%%%%%%%%%
%%%%%%%%%%%%%%%%%%%%%%%%%%%%%%%%%%%%%%%%%%%%%%%%%%%%%%%%%%%%%%%%%%%%%%%%%%%%%%%
%%%%%%%%%%%%%%%%%%%%%%%%%%%%%%%%%%%%%%%%%%%%%%%%%%%%%%%%%%%%%%%%%%%%%%%%%%%%%%%

\epubtkUpdate
    [Id=A,
     ApprovedBy=subjecteditor,
     AcceptDate={30 June 2008},
     PublishDate={30 June 2008}]{%
Added 16 new references and latest observations.
}

\newpage

%%%%%%%%%%%%%%%%%%%%%%%%%%%%%%%%%%%%%%%%%%%%%%%%%%%%%%%%%%%%%%%%%%%%%%%%%%%%%%%
%%%%%%%%%%%%%%%%%%%%%%%%%%%%%%%%%%%%%%%%%%%%%%%%%%%%%%%%%%%%%%%%%%%%%%%%%%%%%%%
%%%%%%%%%%%%%%%%%%%%%%%%%%%%%%%%%%%%%%%%%%%%%%%%%%%%%%%%%%%%%%%%%%%%%%%%%%%%%%%

\section{Introduction}
\label{section:introduction}
For the purposes of this review, relativistic binaries are systems containing
two stellar mass degenerate or collapsed objects that are in close orbits. In
the Galactic field, where these systems evolve in relative isolation, their
final properties are set solely by their initial conditions and are the result
of mass transfer, common envelope evolution, or other interactions that may
interrupt the course of single star evolution due to the presence of a nearby
neighbor. When considered as a fraction of the total stellar mass, the number
of relativistic binaries in Galactic globular clusters is overrepresented
compared to the Galactic field.  Thus, the dynamical interactions found in the
environment of dense stellar clusters provide additional channels for the
formation of these systems. Relativistic binaries reveal themselves
observationally as UV or X-ray sources and are potential sources of
gravitational radiation.

This review will concentrate on the Galactic globular cluster system for the
bulk of the text. We shall touch on extra-galactic globular cluster systems
briefly in the sections on observations (Section~\ref{sec:observations}) and
gravitational radiation prospects (Section~\ref{sec:grav_radiation}).

We begin in Section~\ref{sec:GCs} by looking at the physical structure and
general history of the galactic globular cluster system that leads to the
concentration of evolved stars, stellar remnants, and binary systems in the
cores of these clusters. Current observations of globular clusters that have
revealed numerous populations of relativistic binaries and their tracers are
presented in Section~\ref{sec:observations}. We also consider the prospects
for future observations in this rapidly changing area. Many relativistic
binaries are the product of stellar evolution in close binaries. In
Section~\ref{section:relativistic_binaries}, we will look at how mass transfer
between one star and a nearby companion can dramatically alter the evolution
of both stars. The enhanced production of relativistic binaries in globular
clusters results from dynamical processes that drive binaries toward tighter
orbits and that preferentially exchange more massive and degenerate objects
into binary systems.  Numerical simulations of globular cluster evolution,
which can be used to predict the rate at which relativistic binaries are
formed, are discussed in Section~\ref{section:dynamical_evolution}. These
models can be compared with the observable members of the population of
relativistic binaries in order to try and constrain the entire
population. Finally, we conclude with a brief discussion of the prospects for
observing these systems in gravitational radiation in
Section~\ref{sec:grav_radiation}.\epubtkUpdateA{Added explicit references
to sections.}

Readers interested in further studies of the structure and evolution of
globular clusters are invited to look at Binney and Tremaine~\cite{binney87},
Spitzer~\cite{spitzer87}, and Volumes I and II of Padmanabhan's {\it
Theoretical Astrophysics}~\cite{padmanabhan00, padmanabhan01} for a good
introduction to the physical processes involved. Review articles of Meylan and
Heggie~\cite{meylan97} and Meylan~\cite{meylan99} also provide a comprehensive
look at the internal dynamics of globular clusters. Although our focus is
mainly on the Galactic globular cluster system, the physics of globular
cluster systems associated with other galaxies is well covered in the review
article by Harris~\cite{harris91} as well as his lecture notes from the
Saas-Fee course on star clusters~\cite{harris01}. Carney has a thorough
introduction to evolution of stars in globular clusters~\cite{carney01}. An
observational perspective on the role of binaries in globular clusters is
presented in an excellent review by Bailyn~\cite{bailyn95}, while a good
introduction to the details of observing binary systems in general can be
found in {\it An Introduction to Close Binary
Stars}~\cite{hilditch01}. Although slightly out of date, the review of
binaries in globular clusters by Hut et al.~\cite{hut92a} is an excellent
introduction to the interaction between globular cluster dynamics and binary
evolution, as is a short article on globular cluster binaries by McMillan,
Pryor, and Phinney~\cite{mcmillan98}. Rappaport et al.~\cite{rappaport01} and
Rasio et al.~\cite{rasio01} have written reviews of numerical simulations of
binary populations in globular clusters. An excellent introduction to the
astrophysics and numerical techniques relevant to globular cluster dynamics
can be found in the book by Heggie and Hut~\cite{heggie03}.  Finally, a
shorter and more observationally focused review of compact objects in globular
clusters can be found in Maccarone and Knigge~\cite{MaccaroneKnigge07}.

\newpage

%%%%%%%%%%%%%%%%%%%%%%%%%%%%%%%%%%%%%%%%%%%%%%%%%%%%%%%%%%%%%%%%%%%%%%%%%%%%%%%
%%%%%%%%%%%%%%%%%%%%%%%%%%%%%%%%%%%%%%%%%%%%%%%%%%%%%%%%%%%%%%%%%%%%%%%%%%%%%%%
%%%%%%%%%%%%%%%%%%%%%%%%%%%%%%%%%%%%%%%%%%%%%%%%%%%%%%%%%%%%%%%%%%%%%%%%%%%%%%%

\section{Globular Clusters}
\label{sec:GCs}
Globular clusters are gravitationally bound associations of $10^{4}-10^{7}$
stars, distinct both from their smaller cousins, open clusters, and the
larger, dark matter dominated dwarf galaxies that populate the low-mass end of
the cosmological web of structure. Globular clusters are normally 
associated with a host galaxy and most galaxies, including the Milky Way, are
surrounded and penetrated by a globular cluster system.  A good estimate of
the number of globular clusters in the Milky Way is the frequently updated
catalogue by Harris~\cite{harris96}, which has 157 entries as of 2010.  Although
fairly complete, a few new clusters have been discovered in recent years at low
Galactic latitudes \cite{hurt00,kobulnicky05} and there may be more hidden
behind the  galactic disc and bulge.  The distribution of known globular
clusters in the Galaxy is given in Figure~\ref{fig:GCinGalaxy}.  Other
galaxies contain many more globular clusters (M87 alone may have over 10 000
\cite{Harris09}) and the luminosity of globular clusters normalized to the
host galaxy luminosity seems to be higher for elliptical than disc galaxies
\cite{BrodieStrader06}.

% ==========
\epubtkImage{GCcoordinates.png}{
  \begin{figure}[htbp]
    \def\epsfsize#1#2{0.6#1}
    \centerline{\epsfbox{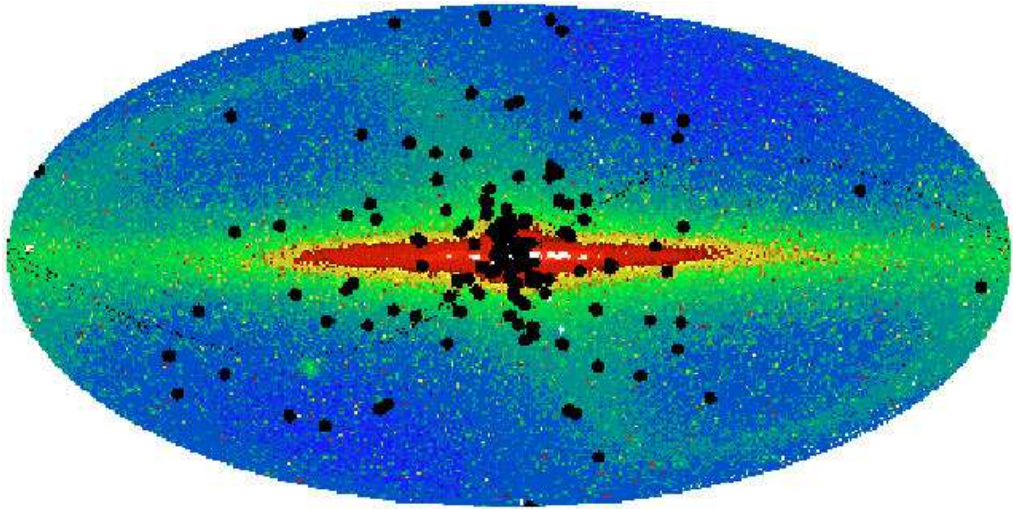}}
    \caption{Globular cluster distribution about the galaxy. Positions are
      from Harris~\cite{harris96} and are plotted as black circles on top of
      the COBE FIRAS 2.2 micron map of the Galaxy using a Mollweide
      projection. Figure taken from Brian Chaboyer's
      website~\cite{chaboyerweb}.}
    \label{fig:GCinGalaxy}
  \end{figure}
}
% ==========

Milky way globular clusters are old, having typical ages of 13 Gyr and an age
spread of less than 5 Gyr \cite{carretta00}.  This is on the order of the age of
the Galaxy itself, thus Galactic globular clusters are thought to be left over
from its formation.  By contrast other galaxies such as the small and large
Magellanic clouds (SMC and LMC) have intermediate age globular clusters ($< 3$
Gyr old, e.g.~\cite{MouldAaronson79,MiloneEtAl09}) and in some galaxy mergers,
such as the Antennae, massive star-forming regions that may become globular
clusters are observed \cite{eggers05}.  Taken together, this implies that
globular clusters of all ages are relatively common objects in the universe.

\subsection{Stellar Populations in Globular Clusters}
\label{sec:StellarPops}

Most of the detailed information on stellar populations in globular clusters
comes from those in the Milky Way since only they are close enough for stars
to be individually resolved.  The stars in individual Galactic globular
clusters all tend to have the same iron content \cite{Gratton04} so globular
clusters are thought to be internally chemically homogeneous.  The
colour-magnitude diagram (CMDs) for most Galactic globular clusters (e.g. M80,
Figure~\ref{fig:M80_CMD}) also indicate a single stellar population with a
distinct main-sequence, main-sequence turn-off, horizontal and giant branch.
The single main sequence turn-off in particular indicates a co-eval stellar
population.  This leads to a so-called ``simple stellar population'' model for
globular clusters where all stars have the same composition and age and differ
only by their masses, set by the initial mass function (IMF). This simple
picture has been challenged in recent years as observations have shown
systematic star-to-star light element variations in globular clusters
\cite{Gratton04}.  This has lead to the development of self-enrichment models
where a single globular cluster experiences several bursts of star formation,
each enriched by pollution from the previous generation
\cite{ConroySpergel10}.  How multiple populations will affect the CMD of a
globular cluster is shown in Figure~\ref{fig:NGC288_CMD}  The importance of
these scenarios for relativistic binaries has not yet been explored. Here we
outline the consequences of the co-eval model.

% =========
\epubtkImage{M80CMD.png}{
  \begin{figure}[htbp]
    \def\epsfsize#1#2{0.5#1}
    \centerline{\epsfbox{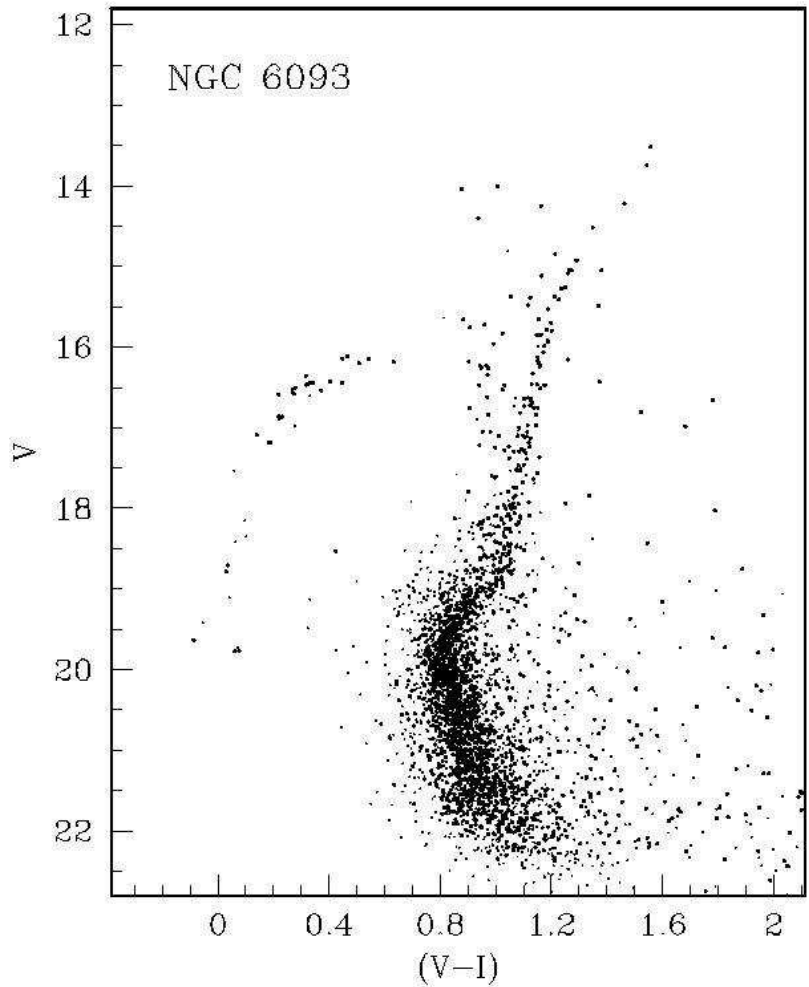}}
    \caption{Colour--magnitude diagram for M80. Figure taken from the catalogue
      of 52 globular clusters~\cite{rosenberg00}. The entire catalogue is
      available at the Padova Globular Cluster Group
      website~\cite{padovagroup}.}
    \label{fig:M80_CMD}
  \end{figure}
}
% ==========
\epubtkImage{NGC288_age_spread.png}{
 \begin{figure}[htbp]
  \def\epsfsize#1#2{0.9#1}
  \centerline{\epsfbox{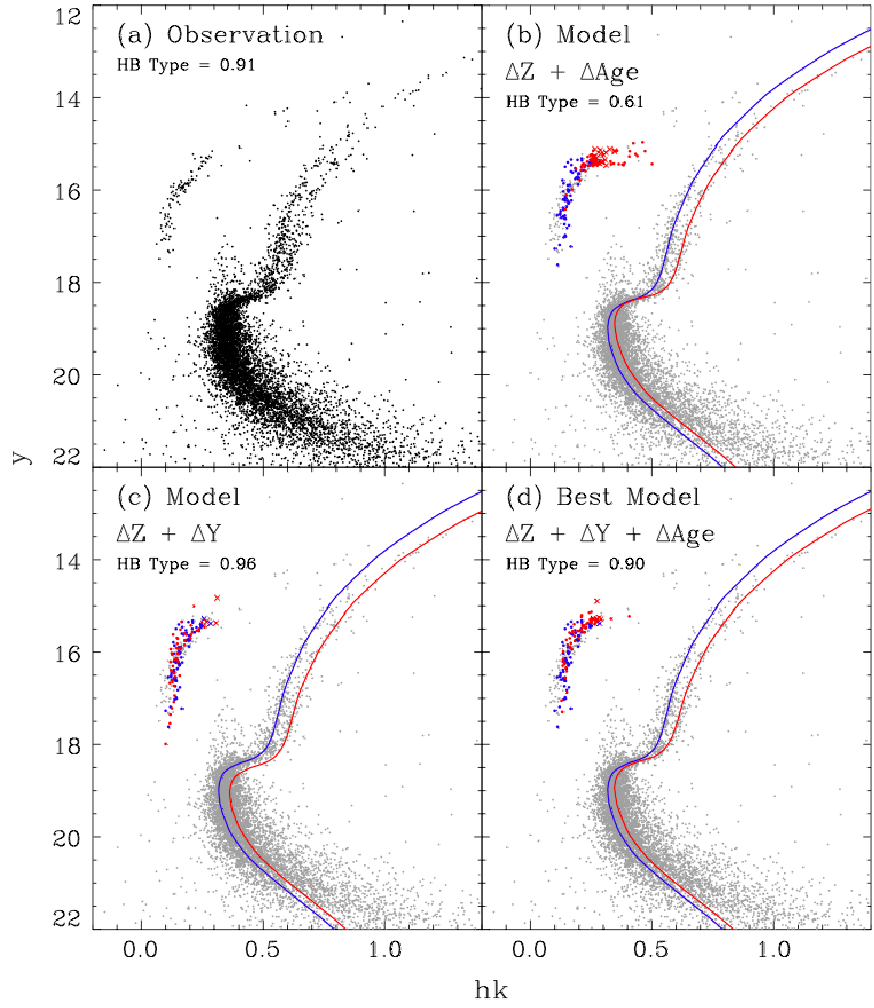}}
  \caption{CMD for NGC288 showing evidence of two populations, and  models
    incorporating different metallicities, helium fraction, and age. The best
    fit model has an age difference of $\Delta~{\rm Age} = 1.5~{\rm
      Gyr}$. Figure taken from Roh {\it et al.}~\cite{roh11}.}
  \label{fig:NGC288_CMD}
 \end{figure}
}
%==========

The IMF is thought to be universal \cite{bastian10} and is usually taken to be
a power-law of the form
\begin{equation}
  \frac{dN}{dM} \propto M^{-\alpha_{i}}
\end{equation}
where $\alpha_{i}$ can take different values for different mass ranges.  For
values above $\sim 1 \Msun$ $\alpha_{i}$ is usually assumed to have a single
value, the so-called Salpeter slope, of $\sim 2.35$ \cite{Salpeter55}.  There
is much more debate about the value of $\alpha_{i}$ in low-mass regime.  One
popular choice, the Kroupa IMF, is a broken power law with exponent values of
\begin{equation}
  \begin{array}{l l}
    \alpha_{0} = 0.3 \pm 0.7, & 0.01 \le M/\Msun < 0.08 \\
    \alpha_{1} = 1.3 \pm 0.5, & 0.08 \le M/\Msun < 0.50 \\
    \alpha_{2} = 2.3 \pm 0.3, & 0.50 \le M/\Msun < 1.00 \\
    \alpha_{3} = 2.3 \pm 0.7, & 1.00 \le M/\Msun
  \end{array}
\end{equation}
\cite{kroupa01d,KroupaWeidner03}.  Another possibility, introduced by
Chabrier~\cite{Chabrier03}, uses a log-normal distribution of masses below
$1 \Msun$.  In all cases it is clear that the IMF strongly favours low masses
so stars massive enough to form neutron stars (NSs) and black holes (BHs) will
be rare.

Since Galactic globular clusters are old, their stellar populations are highly
evolved.  Most of the stars above $\sim 0.8 \Msun$ have already moved off the
main-sequence.  Those just above $0.8 \Msun$ will be on the red giant branch
and are readily visible in most optical images of globular clusters, such as
the image of M80 shown in Figure~\ref{fig:M80}.  Those that are more massive
have evolved past the giant and horizontal branches and become low-luminosity
compact remnants.  The fact that the original high-mass population is no
longer visible produces an intrinsic uncertainty in our knowledge of the
high-mass end of the IMF of globular cluster stars.

% ==========
\epubtkImage{M80.png}{
  \begin{figure}[htbp]
    \def\epsfsize#1#2{1.0#2}
    \centerline{\epsfbox{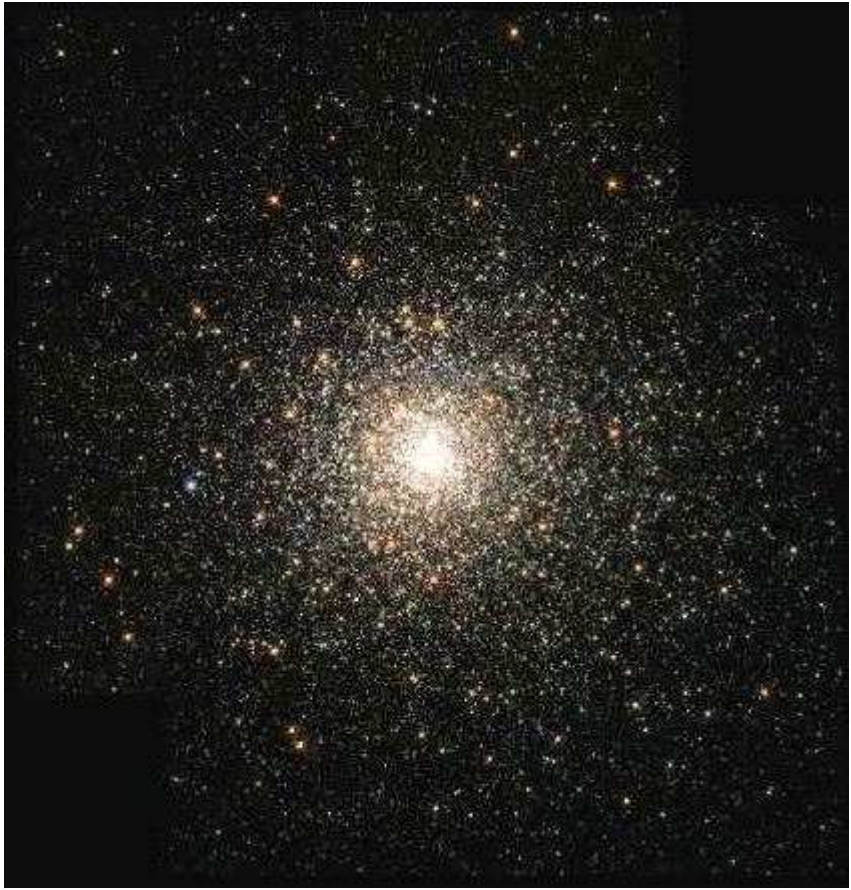}}
    \caption{Hubble Space Telescope photograph of the dense globular cluster
      M80 (NGC 6093).}
    \label{fig:M80}
  \end{figure}
}
% ==========

The picture of Galactic globular cluster stellar populations that emerges from
this analysis is of a simple, co-eval, chemically homogeneous set of luminous,
low-mass population II stars combined with a low-luminosity population of
high-mass stellar remnants.  It is interactions with members of this remnant
population that will be of particular interest for producing relativistic
binaries.

\subsection{The Structure of Globular Clusters}
\label{sec:GCstructure}

Globular clusters are roughly spherical N-body systems that can be described
by a core-halo structure.  The core is highly concentrated, reaching densities
of up to $10^{6} \Msun/\pc^{3}$, and strongly self-gravitating.  The
surrounding halo is of much lower density and is less strongly
self-gravitating.  The structure of a globular cluster can be classified using
three basic radii: the core radius ($\rc$), the half-mass radius ($\rh$), and
the tidal radius ($\rt$).  The core radius is often taken to be the radius at
which the space density drops to about one third, and thus the surface density
to one half, of its central value \cite{heggie03} (other definitions are
possible).  Observationally this corresponds to the radius at which the
surface brightness drops to half of its central value.  The half-mass radius
is simply the radius that contains half of the mass of the system.  The
corresponding observational value is the half-light radius, which contains
half the light of the system (the two radii do not necessarily agree).  The
tidal radius is the radius at which the gravitational field of the host galaxy
becomes more important than the self gravity of the star cluster.  For a given
cluster, cluster orbit, and galaxy model the tidal radius of the cluster can,
in principle, be clearly defined by comparing the effect of the galactic
versus globular cluster gravitational field on a test mass.  Observationally
tidal radii can be difficult to determine due to the low stellar density of
globular cluster halos and an imperfect knowledge of the gravitational
potential of the host galaxy.  Median values for $\rc$, $\rh$, and $\rt$ in
the Galaxy are $\sim 1 \pc$, $\sim 3 \pc$ and $\sim 35 \pc$ respectively
\cite{heggie03}.

There are also two important timescales that characterize globular cluster
evolution: the crossing time ($\tcr$) and the relaxation time ($\trlx$).  The
crossing time is simply the time required for a star traveling at a typical
velocity to cross some characteristic cluster radius.  Thus $\tcr \sim
R/v$ where, for example, $\rc$ or $\rh$ might be typical radii of interest and
$v$ could be the velocity dispersion ($\sim 10$ km s$^{-1}$).  $\tcr$ is also,
roughly speaking, the orbital timescale for the cluster.  For typical values
of $\rh$ and $v$, $\tcr$ for Galactic globular clusters is on the order of
$0.1 - 1 \Myr$ but is longer at the tidal radius and much shorter in the core.

The relaxation time describes how long it takes for orbits to be significantly
altered by stellar encounters.  In particular $\trlx$ is often defined as
the time necessary for the velocity of a star to change by an order of itself
\cite{binney87}.  This can be thought of as the time necessary for a cluster
to lose the memory of its initial conditions or, more exactly, the time
necessary for stellar encounters to transform an arbitrary velocity
distribution to a Maxwellian \cite{spitzer87}.  The relaxation time is related
to the number and strength of encounters and thus to the number density and
energy of a typical star in the cluster.  It can be shown that the mean
relaxation time in a globular cluster is given by
\cite{binney87,spitzer87}
\begin{equation}
  \label{eq:Trlx}
  \trlx \propto \frac{N}{\ln N} \tcr.
\end{equation}
Taking a typical value of $N = 10^{5}$ and $\tcr$ as before, typical values of
$\trlx$ are $0.1 - 1 \Gyr$.  Thus, with ages typically greater than 10 Gyr,
Galactic globular clusters are expected to be dynamically relaxed objects.  In
reality the value of $\trlx$ varies significantly within a globular cluster
due to the highly inhomogeneous density distribution of the core-halo
structure.  It is possible for the core of a globular cluster to be fully
relaxed while the halo remains unrelaxed after 13 Gyr.  By making various
approximations for $\tcr$ it is possible to relate $\trlx$ to local cluster
properties.  For instance the criterion of Meylan and Heggie
\cite{meylan97,Larson70}:
\begin{equation}
  \trlx = \frac{0.065 \langle v^{2} \rangle^{3/2}}{\rho \langle m \rangle \ln
    (\gamma N)}
\end{equation}
relates $\trlx$ to the local mass density ($\rho$), the mass-weighted mean
squared velocity ($\langle v^{2} \rangle$), and the average mass ($\langle m
\rangle$).  The criterion by Spitzer \cite{spitzer87}:
\begin{equation}
  \trlx = \frac{0.138N^{1/2}r^{3/2}}{\langle m \rangle^{1/2}G^{1/2} \ln
    (\gamma N)}
\end{equation}
relates $\trlx$ to a characteristic radius ($r$), the average mass, and the
number of stars in the system.  In practise $\rh$ is normally used for the
characteristic radius in the Spitzer criterion and $\trlx$ is renamed the
half-mass relaxation time, $\trh$.  The factor $\ln \gamma N$ that appears in
both definitions is called the Coulomb logarithm and describes the relative
effectiveness of small and large angle collisions.  The exact value of
$\gamma$ is a matter of some debate and ranges from $0.02-0.4$ depending on
the mass distribution of the system \cite{GierszHeggie94,giersz96}.  Both
criteria are used extensively in stellar dynamics.

On timescales shorter than a relaxation time individual encounters do not
govern the overall evolution of a stellar system and the granularity of the
gravitational potential can be ignored.  On these timescales the background
structure of the cluster can be modelled using a static distribution
function, $f$, that describes the probability of finding a star at a
particular location in a 6-D position-velocity phase space.  Formally $f$
depends on position, velocity, mass, and time, thus we have
$f(\vec{x},\vec{v},m,t)$.  For times less than $\trlx$, however, the evolution
of $f$ is described by the collisionless Boltzmann equation:
\begin{equation}
  \label{eq:CollBolt}
  \frac{\partial f}{\partial t} + \vec{v} \cdot \vec{\nabla}f -
  \vec{\nabla}\phi \cdot \frac{\partial f}{\partial \vec{v}} = 0
\end{equation}
and the explicit time dependence can be removed: $\partial f/\partial t = 0$.
The gravitational potential, $\phi$, is given by Poisson's equation:
\begin{equation}
  \label{eq:Poss}
  \vec{\nabla}^{2}\phi = 4\pi G\rho
\end{equation}
and can be calculated at any position by integrating the distribution function
over mass and velocity:
\begin{equation}
  \vec{\nabla}^{2}\phi = 4\pi G \int f(\vec{x},\vec{v},m) d^{3}\vec{v} dm.
\end{equation}
Solutions to Equation~\ref{eq:CollBolt} are often described in terms of the
relative energy per unit mass, $\mathcal{E}  \equiv \Psi - v^{2}/2$ where
$\Psi = -\phi + \phi_{0}$ is the relative potential and $\phi_{0}$ is defined
such that no star has an energy less that zero ($f > 0$ for $\mathcal{E} > 0$
and $f = 0$ for $\mathcal{E} < 0$).  A simple class of solutions to
Equation~\ref{eq:CollBolt} are Plummer models \cite{Plummer1911}:
\begin{equation}
  f(\mathcal{E}) = F\mathcal{E}^{7/2},
\end{equation}
the stellar-dynamical equivalent of an $n=5$ ploytrope \cite{heggie03}.
Another class of models that admit anisotropy and a distribution in angular
momentum, $L$, are known as King-Mitchie models
\cite{King66,MichieBodenheimer63}.  The basic distribution function is given
by:
\begin{equation}
  f(\mathcal{E},L) = \rho_{1}(2\pi\sigma^{2})^{-3/2}
  \exp(\frac{-L^{2}}{2r_{a}^{2}\sigma^{2}}) \left[
    e^{\mathcal{E}/\sigma^{2}} - 1 \right] , \mathcal{E} > 0,
\end{equation}
where $\sigma$ is the velocity dispersion, $r_{a}$ the anisotropy radius where
the velocity distribution changes from nearly isotropic to nearly radial, and
$\rho_{1}$ is a constant related to the density.  Although not as well
theoretically supported as the single-mass case, King-Mitchie models have been
extended to include a spectrum of stellar masses \cite{GunnGriffin79} and even
external gravitational field \cite{HeggieRamamani95}.  Multi-mass King models
in particular are often fit to observed globular cluster cluster surface
brightness profiles in order to determine their masses.  A good example of the
construction of a multi-mass King-Mitchie model is found in the appendix of
Miocchi~\cite{miocchi05}.

\subsection{The Dynamical Evolution of Globular Clusters}
\label{sec:GCDynamics}

Although static models can be used to describe the instantaneous structure of
a globular cluster there is no stable equilibrium for self-gravitating systems
\cite{binney87} and therefore their structure changes dramatically over
time.  Accessible descriptions of globular cluster evolution are given in Hut
et al.~\cite{hut92a}, Meylan and Heggie \cite{meylan97} and
Meylan~\cite{meylan99}  and have also been the subject of several texts
(e.g. Spitzer~\cite{spitzer87} and Heggie and Hut~\cite{heggie03}).  Here we
merely outline some of the more important aspects of globular cluster
evolution.

The initial conditions of globular clusters are not well constrained since 
they seem to form only very early in the history of galaxy formation or in
major mergers \cite{geyer02,BrodieStrader06} -- both situations quite
different from the environment of our Galaxy today. However it is possible to
make some general statements.  Like all stars, the stars in globular clusters
collapse out of molecular gas.  Due to their small age spread and chemical
homogeneity, it seems certain that all the stars in a single globular cluster
formed out of the same material over a short time period.  Thus it seems
likely that each globular cluster formed by the collapse of a single giant
molecular cloud \cite{BrodieStrader06}.  The details of this collapse are
unclear.  Sub-clumps in the cloud may have collapsed individually and then
merged later, affecting the course of local star formation.  The star
formation may have been sufficiently extended that early-forming stars could
pollute late-forming stars, explaining some observed light-element abundance
anomalies in otherwise homogeneous star clusters \cite{Gratton04}.  In
particular, the star formation efficiency (the fraction of gas converted into
stars)  is a major issue for globular cluster formation.  If anything less
than 100\% of the primordial gas forms stars then the resulting globular
cluster will be less massive than its parent cloud.  This remaining gas will
be expelled by a combination of radiation pressure from young stars and energy
injection from supernovae.  If the star formation efficiency is low, then the
amount of mass loss through gas expulsion can leave the cluster out of virial
equilibrium and may lead to its immediate dissolution
\cite{Goodwin97, GoodwinBastian06, Weidner07}, a process called ``infant
mortality''.  It is estimated that over 50\% of young clusters in the local
universe are destroyed in this manner \cite{GoodwinBastian06} and most
surviving clusters will lose a large fraction of their stars.  Even if the
cluster survives the gas expulsion, the rapid change in potential will cause
the energy of individual stars to change in a mass-independent manner
\cite{binney87}.  This process, called violent relaxation, means that the
positions, velocities, and masses of cluster stars will be initially
uncorrelated.

Equipartition of energy dictates that the most massive stars should have the
lowest kinetic energies \cite{binney87,spitzer87}.  Thus, as soon as the
residual gas is expelled from a young globular cluster, massive stars with
large kinetic energies will start to transfer this energy to low-mass stars
through stellar encounters.  As the massive stars lose kinetic energy they
will sink to the centre of the cluster while the low-mass stars gain kinetic
energy and move to the halo.  This process is known as mass-segregation and
proceeds on a timescale $t_{\rm   ms} \propto m_{i}/\langle m \rangle$
\cite{Spitzer69, Watters00, Khalisi07}.  Due to mass-loss from stellar
evolution, compact remnants rapidly become the most massive objects in
globular clusters as the star population ages.  Thus, they are strongly
affected by mass segregation and are particularly likely to be found in
cluster cores.  Mass segregation continues until energy equipartition has been
achieved.  There are, however, initial conditions for which it is impossible
to achieve energy equipartition \cite{Watters00, Khalisi07} and it is formally
impossible to halt mass segregation, leading to a singularity.  This
phenomenon was first noted by Spitzer \cite{Spitzer69} and is thus called the
``Spitzer instability''.  In reality, the massive objects in such systems form
a strongly interacting subsystem, dynamically decoupled from the rest of the
cluster.  Due to their high masses, black holes in star clusters are
particularly likely to experience the Spitzer instability and this has been
the starting point for several investigations of BH binaries in star clusters
(e.g. \cite{Oleary06}).

The longer-term evolution of star clusters is driven by two-body relaxation,
where the orbits of stars are perturbed by encounters with their neighbours.
The theory of two-body relaxation was first quantified by Chandrasekhar in
1942~\cite{Chandrasekhar42}.  Two-body relaxation becomes important on
timescales longer than the local relaxation time.  The evolution of a globular
cluster over these timescales can still be described (at least formally) by
the Boltzmann equation but with a collisional term added to the right-hand
side.  Equation~\ref{eq:CollBolt} then takes on the form:
\begin{equation}
  \label{eq:BoltGamma}
  \frac{\partial f}{\partial t} + \vec{v}\cdot\vec{\nabla}f -
  \vec{\nabla}\phi\cdot\frac{\partial f}{\partial \vec{v}} = \Gamma[f].
\end{equation}
Equation~\ref{eq:BoltGamma} is sometimes called the collisional Boltzmann
equation and the term $\Gamma[f]$ describes the effect of two-body (and in
principle higher-order) interactions on the distribution function.  Practically
speaking it is not possible to evaluate $\Gamma[f]$ analytically and various
numerical approximations must be employed.  Approaches include the
Fokker-Planck method, where $\Gamma[f]$ is approximated in the weak scattering 
limit by an expansion in powers of the phase-space parameters; the Monte Carlo
method, where $\Gamma[f]$ is approximated by a Monte Carlo selection of weak
encounters over a time shorter than the relaxation time; or direct N-body
integration, where rather than solving Equation~\ref{eq:BoltGamma} the orbits
of each star in the cluster are explicitly integrated.  Each method has its
strengths and weaknesses and will be discussed further in
Section~\ref{section:dynamical_evolution}.

Thermodynamically speaking, strongly self-gravitating systems have a negative
heat capacity.  This can be understood by relating the kinetic energy of the
system to a dynamical temperature:
\begin{equation}
  \frac{1}{2}mv_{\rm ave}^{2} = \frac{3}{2}k_{B}T
\end{equation}
\cite{binney87}.  This, in combination with the virial theorem, can be
used to define a total internal energy for the cluster:
\begin{equation}
  E = -\frac{3}{2}Nk_{B}T
\end{equation}
where $N$ is the number of bodies in the system.  Finally, this can be used to
calculate a heat capacity:
\begin{equation}
  \label{eq:NegHeat}
  C = \frac{dE}{dT} = -\frac{3}{2}Nk_{B}.
\end{equation}
Since all the constants on the right hand side of Equation~\ref{eq:NegHeat}
are positive, $C$ is always negative.  A negative heat capacity means that
heating a self-gravitating system actually causes it to lose energy.  For a
core-halo star cluster, the core is strongly self-gravitating while the halo
is not, so the halo acts as a heat bath for the core.  Any perturbation in
which the core becomes dynamically hotter than the halo causes energy to flow
into the halo.  The negative heat capacity means that the core becomes even
hotter, increasing the flow of energy to the halo in a runaway process.  This
causes the core to contract, formally to a singularity.  The runaway process
is called the gravothermal catastrophe and the consequent shrinking of the
core is called core collapse.  It affects all self-gravitating systems and was
first noted in the context of star clusters by Anotnov \cite{Antonov62}.  In
equal-mass systems core collapse will occur after $12-20 \trh$
\cite{spitzer87} but may be accelerated in systems with a spectrum of masses
due to mass segregation.  Core collapse not only appears in analytic,
theoretical models, but has also been repeated in a variety of numerical
simulations  such as the model shown in Figure~\ref{fig:CoreCollapse}
\cite{joshi00}.  Furthermore, the Harris catalogue lists several Galactic
globular clusters that from there surface brightness profiles are thought to
have experiences a core collapse event \cite{harris96}.

% ==========
\epubtkImage{GCevolution.png}{
  \begin{figure}[htbp]
    \def\epsfsize#1#2{0.6#1}
    \centerline{\epsfbox{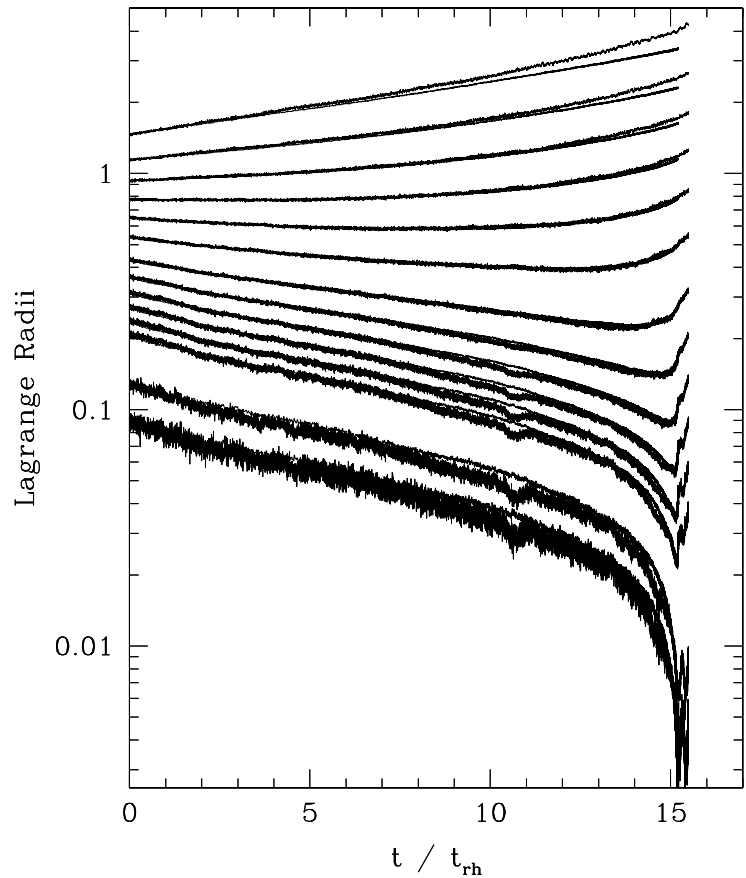}}
    \caption{Lagrange radii indicating the evolution of a Plummer
      model globular cluster for an $N$-body simulation and a Monte
      Carlo simulation. The radii correspond to radii containing 0.35,
      1, 3.5, 5, 7, 10, 14, 20, 30, 40, 50, 60, 70, and 80\% of the
      total mass. Figure taken from Joshi et al.~\cite{joshi00}.}
    \label{fig:CoreCollapse}
  \end{figure}
}
% ==========

Core collapse can be halted, at least temporarily, by an energy source in the
core.  For stars (also self-gravitating systems) this energy source is nuclear
burning.  In star clusters tightly bound binaries perform a similar role.
Stars in the core scatter off these binaries, gaining kinetic energy at the
expense of the orbital energy of the binary, and the collective effect causes
the core to return to equilibrium or even to re-expand.  In analogy to nuclear
burning in stars, this process in globular clusters is called ``binary
burning''.  The binaries taking part in binary burning may be either
primordial (binaries where the stars were born bound to each other) or
dynamically formed by a variety of interactions that will be discussed in
Subsection~\ref{subsection:globular_cluster_processes}.

Stars can escape from a star cluster if they gain a velocity greater than
the cluster's escape velocity, $\langle v_{e}^{2} \rangle = -4U/M$ where $U$
is the potential energy of the star cluster and $M$ its total mass.  Using the
virial theorem it is possible to show that $\langle v_{e}^{2} \rangle = 4
\langle v^{2} \rangle$ where $2\langle v^{2} \rangle$ is the RMS velocity in
the cluster \cite{binney87}.  There are two means through which a star can
reach the escape velocity.  The first is ejection where a single strong
interaction, such as occurs during binary burning, gives the star a sufficient
velocity impulse to exceed $v_{e}$.  This process is highly stochastic.  The
second is evaporation where a star reaches escape velocity due to a large
number of weak encounters during the relaxation process.  Relaxation tends to
maintain a local Maxwellian in the velocity distribution and, since a
Maxwellian distribution always has a fraction $\gamma = 7.38 \times 10^{-3}$
stars with $v > 2v_{\rm RMS}$, there are always stars in the cluster with a
velocity above the escape velocity.  Thus, it is the fate of all star
clusters to evaporate.  The evaporation time can be estimated as:
\begin{equation}
  \label{eq:Tevap}
  t_{\rm evap} \approx \frac{\trlx}{\gamma} = 136\trlx.
\end{equation}
This is much longer than a Hubble time so few globular clusters are directly
affected by evaporation.  Evaporation can, however, be accelerated by the
presence of a tidal field.  In this case stars whose orbits extend beyond
$\rt$ are stripped from the cluster and lost.  Tidal dynamics are more
complicated than a simple radial cutoff would imply and detailed prescriptions
taking into account orbital energy and angular momentum as well as a
time-varying field for star clusters in elliptical orbits are necessary to
capture all of the important  processes \cite{Takahashi98,Ernst09}.  It
seems likely that the ultimate fate of most globular clusters is destruction
due to tidal effects \cite{gnedin97,GielesBaumgardt08}. Since most compact
objects are likely to live deep in the core of globular clusters where
ejection will normally be due to violent interactions, the details of tidal
stripping are unlikely to be critical for the treatment of relativistic
binaries in star clusters.

\newpage

%%%%%%%%%%%%%%%%%%%%%%%%%%%%%%%%%%%%%%%%%%%%%%%%%%%%%%%%%%%%%%%%%%%%%%%%%%%%%%%
%%%%%%%%%%%%%%%%%%%%%%%%%%%%%%%%%%%%%%%%%%%%%%%%%%%%%%%%%%%%%%%%%%%%%%%%%%%%%%%
%%%%%%%%%%%%%%%%%%%%%%%%%%%%%%%%%%%%%%%%%%%%%%%%%%%%%%%%%%%%%%%%%%%%%%%%%%%%%%%

\section{Observations}
\label{sec:observations}
By their very nature, relativistic binaries are compact and faint, so that
observations of these systems are difficult unless they are in an interacting
phase. Furthermore, observations of binaries that have segregated to the
centers of clusters can be complicated by crowding issues. Nonetheless,
observations across a broad spectrum using a variety of ground- and
space-based telescopes have revealed populations of binaries and their tracers
in many Milky Way and extra-galactic globular clusters.

Although they generally are not binaries, blue straggler stars may be used as
tracers of the underlying binary population and the dynamical interaction rate
within individual globular clusters. Blue stragglers are found on the main
sequence above and to the left of the main sequence turn off in the CMD of a
globular cluster (see Figure~\ref{M3BSS}). As their name suggests, these stars
are too hot and too massive to still be on the main sequence if they are
coeval with the stellar population of the cluster. They are thought to be
rejuvenated through mass transfer, merger, or direct collision of
binaries~\cite{fregeau04, davies04, piotto04, knigge09, perets09}. As such,
they can be interpreted as tracers of binary interactions within globular
clusters. However, many of the populations that have been observed exhibit a
bimodal radial distribution~\cite{ferraro97,  zaggia97, ferraro04, sabbi04,
  warren06, lanzoni07, dalessandro08, beccari08, dalessandro09} for which the
inner population can be interpreted as tracing dynamical interactions and the
outer population as representative of the primordial binary population. Recent
observations of the blue straggler populations of 13 globular clusters
indicates a correlation between the specific frequency of blue stragglers and
the binary fraction in the globular cluster~\cite{sollima08}. This supports
observations which also seem to suggest that binary coalescences are the
dominant formation mechanism for blue stragglers in globular
clusters~\cite{leigh07}.\epubtkUpdateA{Added paragraph and references
  to Sollima~(2008) and Leigh~(2007).}

\epubtkImage{LeighBSS_CMD.png}{
  \begin{figure}[htbp]
    \def\epsfsize#1#2{1.1#1}
    \centerline{\epsfbox{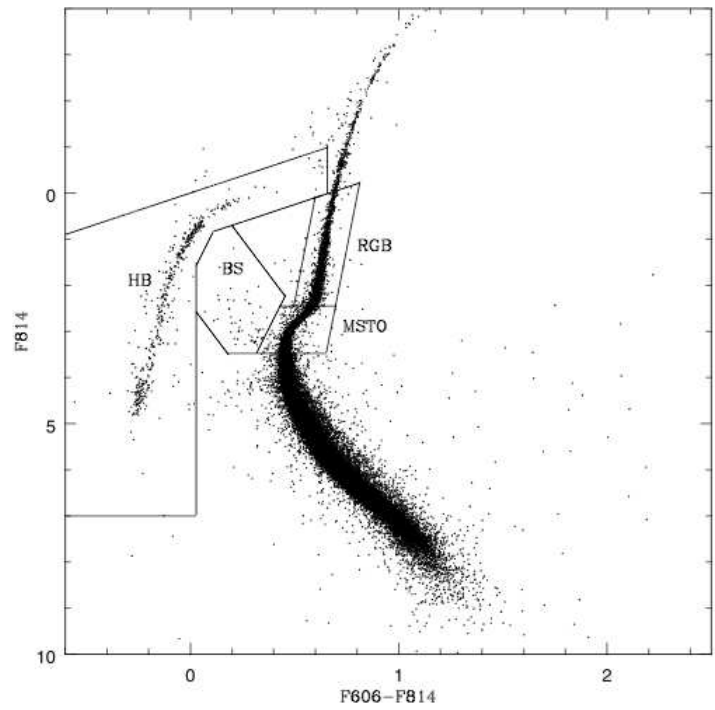}}
    \caption{CMD of NGC 6205 from the Hubble Space Telescope Advanced Detector
      for Surveys. Different populations are identified as horizontal branch
      (HB), red giant branch (RGB), main sequence turn off (MSTO), and blue
      stragglers (BS). Figure taken from Leigh, Sills, \&
      Knigge~\cite{leigh11c}.}
    \label{M3BSS}
  \end{figure}}

Blue stragglers are some of the most visible and populous evidence of the
dynamical interactions that can also give rise to relativistic
binaries. Current observations that have been revealing the blue straggler
population and their variable counterparts (SX Phe stars) are the ACS Survey
of Galactic Globular Clusters~\cite{sarajedini07} and the Cluster AgeS
Experiment (CASE)~\cite{caseweb}. For a good description of surveys that  use
far-ultraviolet in detecting these objects, see Knigge~\cite{knigge04}. For
somewhat older but still valuable reviews on the implications of blue
stragglers on the dynamics of globular clusters, see Hut~\cite{hut93} and
Bailyn~\cite{bailyn95}. 

The globular cluster population of white dwarfs can be used to determine the
ages of globular clusters~\cite{moehler04}, and so they have been the focus of
targeted searches despite the fact that they are arguably the faintest
electromagnetically detectable objects in globular clusters. These searches
have yielded large numbers of globular cluster white dwarfs. For example, a
recent search of $\omega$~Centauri has revealed over 2000 white
dwarfs~\cite{monelli05}, while Hansen et al.~\cite{hansen04} have detected 222
white dwarfs in M4. Deep ACS observations of NGC 6397~\cite{richer08} have
identified a substantial population of approximately 150 white
dwarfs~\cite{strickler07}.\epubtkUpdateA{Added sentence and references to
  Richer~(2008) and Strickler~(2007).} Dieball et al.~\cite{dieball10} have
found approximately 30 white dwarfs and about 60 probably cataclysmic
variables and white dwarf binaries in M80. In general, however, these searches
uncover single white dwarfs. Optical detection of white dwarfs in binary
systems tends to rely on properties of the accretion process related to the
binary type. Therefore, searches for cataclysmic variables (CVs, see
Section~\ref{subsection:cataclysmic_variables}) generally focus on
low-luminosity X-ray sources~\cite{johnston96, grindlay01a, verbunt01} and on
ultraviolet-excess stars~\cite{dieball05a, dieball07, grindlay99, knigge02a,
  margon81}, but these systems are usually a white dwarf accreting from a low
mass star.\epubtkUpdateA{Added reference to Dieball~(2007).} The class of
``non-flickerers'' which have been detected recently~\cite{cool98, taylor01}
have been explained as He white dwarfs in binaries containing dark CO white
dwarfs~\cite{edmonds99, grindlay01b, hansen03,strickler09}.

Pulsars, although easily seen in radio, are difficult to detect when they
occur in hard binaries, due to the Doppler shift of the pulse
intervals. Thanks to an improved technique known as an ``acceleration
search''~\cite{middleditch84}, which assumes a constant acceleration of the
pulsar during the observation period, more short orbital period binary pulsars
are being discovered~\cite{camilo00, camilo01, damico01a, damico01b,
  freire01b, fruchter00, ransom01}. For a good review and description of this
technique, see Lorimer~\cite{lorimer01}. The progenitors of the ultracompact
millisecond pulsars (MSPs) are thought to pass through a low-mass X-ray binary 
phase~\cite{deutsch00, grindlay01a, ivanova08, rappaport87, rasio00a} (LMXBs,
see Section~\ref{subsection:lmxbs}). These systems are very bright and when
they are in an active state, they can be seen anywhere in the Galactic
globular cluster system. There are, however, several additional LMXBs that are
currently quiescent~\cite{grindlay01a, heinke03c, verbunt04, guillot09,
  guillot11} (qLMXBs). As these systems turn on, they can be added to the list
of known LMXBs, which is currently at 15~\cite{heinke11}.  Additional evidence
of a binary spin-up phase for MSPs comes from measurements of their masses,
which indicate a substantial mass-transfer phase during the spin-up. Several
observed globular cluster MSPs in binary systems are seen to have masses above
the canonical mass of $1.4\,M_{\odot}$~\cite{freire08}.\epubtkUpdateA{Added last
  two sentences and references to Ivanova~(2008) and Freire~(2008).}

Although there are many theoretical predictions of the existence of black
holes in globular clusters (see, e.g., \cite{miller02, portegieszwart00b,
  miller04, dantona04}), there are very few observational hints of
them. Measurements of the kinematics of the cores of M15~\cite{gebhardt00,
  guhathakurta96}, NGC~6752~\cite{drukier03}, and $\omega$
Centauri~\cite{noyola08} provide some suggestions of a large, compact
mass.\epubtkUpdateA{Added $\omega$ Centauri and reference to Noyola~(2008).}
One proposed explanation would be a large black hole of $\sim 10^{3} \Msun$ in
the cluster core.  A black holes in this mass range are much larger than
stellar mass black holes ($10-100 \Msun$) but much smaller than the
supermassive black holes ($\gtrsim 10^{6}$) found in galactic centers.  As
such they are called intermediate mass black holes (IMBH) and may potentially
form through mergers of stars or stellar-mass black holes in globular
clusters.  However, these observations can also be explained without requiring
an IMBH~\cite{macnamara03,   pasquali04} and the existence of such objects is
questionable.  VLA observations of M80, M62, and M15 do not indicate any
significant evidence of radio emission, which can be used to place some limits
on the likelihood of an accreting black hole in these clusters. However,
uncertainties in the expected gas density makes it difficult to place any
upper limits on a black hole mass~\cite{bash08}.\epubtkUpdateA{Added two
  sentences and references to Bash~(2008) to reflect latest observations of
  black holes.} The unusual millisecond pulsar in the outskirts of NGC~6752
has also been argued to be the result of a dynamical interaction with a
possible binary intermediate mass black hole in the core~\cite{colpi02}. If
the velocity dispersion in globular clusters follows the same correlation to
black hole mass as in galactic bulges, then there may be black holes with
masses in the range $1 - 10^3 M_{\odot}$ in many globular
clusters~\cite{zheng01}. Recent  Hubble Space Telescope observations of the
stellar dynamics in the core of 47~Tuc have placed an upper bound of
$1500\,M_{\odot}$ for any intermediate mass black hole in this
cluster~\cite{mclaughlin06}.\epubtkUpdateA{Added sentence and reference to
  McLaughlin~(2006).} Stellar mass black hole binaries may also be visible as
low luminosity X-ray sources, but if they are formed in exchange interactions,
they will have very low duty cycles and hence are unlikely to be
seen~\cite{kalogera04}. For a good recent review on neutron stars and black
holes in globular clusters, see Rasio et
al.~\cite{rasio07}. Heinke~\cite{heinke11} has an excellent review of X-ray
observations of Galactic globular clusters, including CVs, LMXBs and qLMXBs,
MSPs, and other active binaries.

Recent observations and catalogs of known binaries are presented in
the following Subsections.

%%%%%%%%%%%%%%%%%%%%%%%%%%%%%%%%%%%%%%%%%%%%%%%%%%%%%%%%%%%%%%%%%%%%%%%%%%%%%%%
%%%%%%%%%%%%%%%%%%%%%%%%%%%%%%%%%%%%%%%%%%%%%%%%%%%%%%%%%%%%%%%%%%%%%%%%%%%%%%%

\subsection{Cataclysmic variables}
\label{subsection:cataclysmic_variables}

Cataclysmic variables (CVs) are white dwarfs accreting matter from a
companion that is usually a dwarf star or another white dwarf. They
have been detected in globular clusters through
identification of the white dwarf itself or through evidence of the
accretion process. White dwarfs managed to avoid detection until
observations with the Hubble Space Telescope revealed
photometric sequences in several globular clusters~\cite{cool96,
cool98, paresce95, renzini96, richer95, richer97,
taylor01, hansen04}. Spectral identification of white dwarfs in
globular clusters has begun both from the ground with the
VLT~\cite{moehler00, moehler04} and in space with the Hubble Space
Telescope~\cite{cool98, edmonds99, taylor01, monelli05}. With spectral
identification, it will be possible to identify those white dwarfs in
hard binaries through Doppler shifts in the $\mathrm{H}\beta$ line. This
approach has promise for detecting a large number of the expected
double white dwarf binaries in globular clusters. Photometry has also
begun to reveal orbital periods~\cite{neill02, edmonds03b, kaluzny03} of
CVs in globular clusters.

Accretion onto the white dwarf may eventually lead to a dwarf nova
outburst. Identifications of globular cluster CVs have been made
through such outbursts in the cores of M5~\cite{margon81},
47~Tuc~\cite{paresce94}, NGC~6624~\cite{shara96b}, M15~\cite{shara04},
and M22~\cite{anderson03, bond05}. With the exception
of V101 in M5~\cite{margon81}, original searches for dwarf novae
performed with ground based telescopes proved unsuccessful. This is
primarily due to the fact that crowding obscured potential dwarf novae
up to several core radii outside the center of the
cluster~\cite{shara94, shara95a}. Since binaries tend to settle into
the core, it is not surprising that none were found significantly
outside of the core. Subsequent searches using the improved resolution
of the Hubble Space Telescope eventually revealed a few dwarf novae
close to the cores of selected globular clusters~\cite{shara96a,
shara95b, shara96b, shara04, anderson03}. To date, there have been 14
found~\cite{pietrukowicz08,servillat11}, using the Hubble Space Telescope and
Las Campanas Observatory (CASE).

A more productive approach has been to look for direct evidence of the
accretion around the white dwarf. This can be in the form of excess UV
emission and strong $\mathrm{H}\alpha$ emission~\cite{ferraro01,
grindlay91, knigge02a, knigge02b, dieball05a} from the accretion
disk. This technique has resulted in the discovery of candidate CVs in
47~Tuc~\cite{ferraro01, knigge02a}, M92~\cite{ferraro00a},
NGC~2808~\cite{dieball05a}, NGC~6397~\cite{cool98, edmonds99, taylor01}, and
NGC 6712~\cite{ferraro00b}. The accretion disk can also be discerned by very
soft X-ray emissions. These low luminosity X-ray binaries are characterized by
a luminosity $L_\mathrm{X} < 10^{34.5} \mathrm{\ erg/s}$, which distinguishes
them from the low-mass X-ray binaries with $L_\mathrm{X} > 10^{36} \mathrm{\
  erg/s}$. Initial explanations of these objects focused on accreting white
dwarfs~\cite{bailyn91}, and a significant fraction of them are probably
CVs~\cite{verbunt04, webb05}. There have been 10 identified candidate CVs in
6752~\cite{pooley02a}, 15 in 6397~\cite{cohn10}, 19 in 6440~\cite{ pooley02b},
2 in $\omega$~Cen~\cite{gendre03}, 5 in Terzan~5~\cite{heinke03a}, 22 in
47~Tuc~\cite{edmonds03a}, 5 in M80~\cite{heinke03b}, 7 in M54~\cite{ramsay06},
2--5 in NGC~288~\cite{kong06}, 4 in M30~\cite{lugger07}, 4 in
NGC~2808~\cite{servillat08}, 1 in M71~\cite{huang10}, and 1 in
M4~\cite{bassa04}.\epubtkUpdateA{Added new observations in M54,   NGC~288,
  M30, and NGC~2808. New references to Ramsay~(2006), Kong~(2006),
  Lugger~(2007), and Servillat~(2008).} However, some of the more energetic
sources may be LMXBs in quiescence~\cite{verbunt04}, or even candidate QSO
sources~\cite{bassa04}.

The state of the field at this time is one of rapid change as Chandra
results come in and optical counterparts are found for the new X-ray
sources. A living catalog of CVs was created by Downes et al.~\cite{downes01},
but it has been allowed to lapse and is now archival~\cite{downes05,downes06}.
It may still be the best source for confirmed CVs in globular clusters up to
2006.

%%%%%%%%%%%%%%%%%%%%%%%%%%%%%%%%%%%%%%%%%%%%%%%%%%%%%%%%%%%%%%%%%%%%%%%%%%%%%%%
%%%%%%%%%%%%%%%%%%%%%%%%%%%%%%%%%%%%%%%%%%%%%%%%%%%%%%%%%%%%%%%%%%%%%%%%%%%%%%%

\subsection{Low-mass X-ray binaries}
\label{subsection:lmxbs}

The X-ray luminosities of low-mass X-ray binaries are in the range
$L_\mathrm{X} \sim 10^{36} \mbox{\,--\,} 10^{38} \mathrm{\ erg/s}$.
The upper limit is close to the Eddington limit for accretion onto a
neutron star, so these systems must contain an accreting neutron star
or black hole. All of the LMXBs in globular clusters contain an
accreting neutron star as they also exhibit X-ray bursts, indicating
thermonuclear flashes on the surface of the neutron
star~\cite{johnston96}. Compared with $\sim$~100 such systems in the
galaxy, there are 15 LMXBs known in globular clusters. The globular
cluster system contains roughly 0.1\% of the mass of the galaxy and
roughly 10\% of the LMXBs. Thus, LMXBs are substantially
over-represented in globular clusters.

Because these systems are so bright in X-rays, the globular cluster population
is completely known -- we expect  new LMXBs to be discovered in the globular
cluster system only as quiescent (qLMXBs) and transient systems become
active. The 15 sources are in 12 separate clusters. Five have orbital periods
greater than a few hours, six ultracompact systems have measured orbital
periods less than 1 hour, and four have undetermined orbital periods. A member
of the ultracompact group, 4U 1820-30 (X1820-303) in the globular cluster
NGC 6624, has an orbital period of 11 minutes~\cite{stella87}. This is one of
the shortest known orbital periods of any binary and most certainly indicates
a degenerate companion. The orbital period, X-ray luminosity, and host
globular clusters for these systems are given in Table~\ref{LMXB_Properties}.

The improved resolution of Chandra allows for the possibility of
identifying optical counterparts to LMXBs. If an optical counterpart
can be found, a number of additional properties and constraints for
these objects can be determined through observations in other
wavelengths. In particular, the orbital parameters and the nature of
the secondary can be determined. So far, optical counterparts have
been found for X0512--401 in NGC 1851~\cite{homer01b}, X1745--203 in
NGC 6440~\cite{verbunt00}, X1746--370 in NGC 6441~\cite{deutsch98},
X1830--303 in NGC 6624~\cite{king93}, X1832--330 in NGC
6652~\cite{deutsch00, heinke01}, X1850--087 in NGC
6712~\cite{cudworth88, bailyn88, nieto90}, X1745-248 in Terzan
5~\cite{heinke03a}, and both LMXBs in NGC
7078~\cite{auriere84, nwhite01}. Continued X-ray observations will also
further elucidate the nature of these systems~\cite{mukai00}.

\begin{table}[htbp]
  \caption{Low-mass X-ray binaries in globular clusters: Host
    clusters and LMXB properties.}
  \label{LMXB_Properties}
  \vskip 4mm

  \begin{minipage}{\textwidth}
    \renewcommand{\arraystretch}{1.2}
    \centering
    \renewcommand{\footnoterule}{}
    \begin{tabular}{l|lrrr}
      \hline \hline
      LMXB Name &
      \mc{Cluster} &
      \mc{$ L_\mathrm{X} $} &
      \mc{$ P_\mathrm{orb} $} &
      \mc{Ref.} \\
      & &
      \mc{($ \times 10^{36} \mathrm{\ erg/s} $)} &
      \mc{(hr)} & \\
      \hline
      X0512--401  & NGC 1851 &   1.9  \qq{} & 0.28 &
      \cite{deutsch00, sidoli01, zurek09} \\
      X1724--307\epubtkFootnote{Sidoli et al.~\cite{sidoli01} give
        X1724--304.} &
      Terzan 2 &  4.3  \qq{} & --- &
      \cite{deutsch00, sidoli01} \\
      X1730--335  & Liller 1 &   2.2  \qq{} &     --- &
      \cite{deutsch00, sidoli01} \\
      X1732--304  & Terzan 1 &   0.5  \qq{} &     --- &
      \cite{deutsch00, sidoli01} \\
      X1745--203  & NGC 6440 &   0.9  \qq{} &     8.7 &
      \cite{deutsch00, sidoli01,altamirano08} \\
      CX-2 & NGC 6440 & 1.5-0.4 \qq{} & 0.96 & \cite{heinke10,altamirano10} \\
      X1745--248  & Terzan 5 &    --- \qq{} &     --- &
      \cite{deutsch00} \\
      J17480-2446 & Terzan 5 & 37 \qq{} & 21.25 & \cite{bordas10,miller11}\\
      X1746--370  & NGC 6441 &   7.6  \qq{} &   5.70  &
      \cite{deutsch00, podsiadlowski02, sidoli01} \\
      X1747--313  & Terzan 6 &   3.4  \qq{} &  12.36  &
      \cite{deutsch00, podsiadlowski02, sidoli01} \\
      X1820--303  & NGC 6624 &  40.6  \qq{} &   0.19  &
      \cite{deutsch00, podsiadlowski02, sidoli01} \\
      X1832--330  & NGC 6652 &   2.2  \qq{} &   0.73  &
      \cite{deutsch00, podsiadlowski02} \\
      X1850--087  & NGC 6712 &   0.8  \qq{} &   0.33  &
      \cite{deutsch00, podsiadlowski02, sidoli01} \\
      X2127+119-1 & NGC 7078 &   3.5  \qq{} &  17.10  &
      \cite{deutsch00, podsiadlowski02, sidoli01} \\
      X2127+119-2 & NGC 7078 &      1.4 \qq{} &   0.38 &
      \cite{deutsch00, podsiadlowski02, nwhite01,dieball05b} \\
      \hline \hline
    \end{tabular}
    ~\\ [-1 em]
  \end{minipage}
  \renewcommand{\arraystretch}{1.0}
\end{table}

The 15 bright LMXBs are thought to be active members of a larger population of
lower luminosity quiescent low mass X-ray binaries
(qLMXBs)~\cite{wijnands05}. Early searches performed with ROSAT data (which
had a detection limit of $10^{31} \mathrm{\ erg/s}$) revealed roughly 30
sources in 19 globular clusters~\cite{johnston96}. A more recent census of the
ROSAT low luminosity X-ray sources, published by Verbunt~\cite{verbunt01},
lists 26 such sources that are probably related to globular clusters. Recent
observations with the improved angular resolution of Chandra have begun to
uncover numerous low luminosity X-ray candidates for CVs~\cite{grindlay01a,
  grindlay01b, heinke01, homer01a, heinke03a, heinke03b, edmonds03a,
  edmonds03b, gendre03, pooley02a, pooley02b}. For a reasonably complete
discussion of recent observations of qLMXBs in globular clusters, see Verbunt
and Lewin~\cite{verbunt04} or Webb and Barret~\cite{webb05} and references
therein. Table~\ref{qLMXBs} lists the 27 qLMXBs known to date.

\begin{table}[htbp]
  \caption{Quiescent Low-mass X-ray binaries in globular clusters: Host
    clusters, ID numbers, and references.}
  \label{qLMXBs}
  \vskip 4mm

  \begin{minipage}{\textwidth}
    \renewcommand{\arraystretch}{1.2}
    \centering
    \begin{tabular}{l|l|r}
      \hline \hline
      qLMXB Name &
      Cluster &
      Ref. \\
      \hline
      171411-293159  & NGC 6304 &   \cite{guillot09} \\
      171421-292917  & NGC 6304 &   \cite{guillot09} \\
      171433-292747  & NGC 6304 &   \cite{guillot09} \\
      No. 3 & $\omega$ Cen & \cite{guillot09,heinke03b}\\
      Ga & M13 & \cite{guillot09,heinke03b}\\
      X5 & 47 Tuc & \cite{guillot09,heinke03b}\\
      X7 & 47 Tuc & \cite{guillot09,heinke03b}\\
      No.24 & M28 & \cite{guillot09,heinke03b}\\
      A-1 & M30 & \cite{guillot09,heinke03b}\\
      U24 & NGC 6397 & \cite{guillot09,heinke03b}\\
      CX2 & M80 & \cite{guillot09,heinke03b}\\
      CX6 & M80 & \cite{guillot09,heinke03b}\\
      C2 & NGC 2808 & \cite{guillot09}\\
      16 & NGC 3201 & \cite{guillot09}\\
      CX1 & NGC 6440 & \cite{heinke03b}\\
      CX2 & NGC 6440 & \cite{heinke03b}\\
      CX3 & NGC 6440 & \cite{heinke03b}\\
      CX5 & NGC 6440 & \cite{heinke03b}\\
      CX7 & NGC 6440 & \cite{heinke03b}\\
      CX10 & NGC 6440 & \cite{heinke03b}\\
      CX12 & NGC 6440 & \cite{heinke03b}\\
      CX13 & NGC 6440 & \cite{heinke03b}\\
      W2 & Terzan 5 & \cite{heinke03b}\\
      W3  & Terzan 5 & \cite{heinke03b}\\
      W4 & Terzan 5 & \cite{heinke03b}\\
      W8 & Terzan 5 & \cite{heinke03b}\\
      J180916-255623 & NGC 6553 & \cite{guillot11}\\
      \hline \hline
    \end{tabular}
    ~\\ [-1 em]
  \end{minipage}
  \renewcommand{\arraystretch}{1.0}
\end{table}

%%%%%%%%%%%%%%%%%%%%%%%%%%%%%%%%%%%%%%%%%%%%%%%%%%%%%%%%%%%%%%%%%%%%%%%%%%%%%%%
%%%%%%%%%%%%%%%%%%%%%%%%%%%%%%%%%%%%%%%%%%%%%%%%%%%%%%%%%%%%%%%%%%%%%%%%%%%%%%%

\subsection{Millisecond pulsars}
\label{subsection:millisecond_pulsars}

The population of known millisecond pulsars (MSPs) is one of the fastest
growing populations of relativistic binaries in globular clusters. Several
ongoing searches continue to reveal millisecond pulsars in a number of
globular clusters. Previous searches have used deep multi-frequency imaging to
estimate the population of pulsars in globular clusters~\cite{fruchter00}. In
this approach, the expected number of pulsars beaming toward the earth,
$N_\mathrm{puls}$, is determined by the total radio luminosity observed when
the radio beam width is comparable in diameter to the core of the cluster. If
the minimum pulsar luminosity is $L_\mathrm{min}$ and the total luminosity
observed is $L_\mathrm{tot}$, then, with simple assumptions on the neutron
star luminosity function,
\begin{equation}
  N_\mathrm{puls} = \frac{L_\mathrm{tot}}{L_\mathrm{min}
  \ln{\left(L_\mathrm{tot}/L_\mathrm{min}\right)}}.
  \label{luminosity_number}
\end{equation}
In their observations of 7 globular clusters, Fruchter and Goss have recovered
previously known pulsars in NGC~6440, NGC~6539, NGC~6624, and
47~Tuc~\cite{fruchter00}. Their estimates based on
Equation~(\ref{luminosity_number}) give evidence of a population of between 60
and 200 previously unknown pulsars in Terzan~5, and about 15 each in Liller~1
and NGC~6544~\cite{fruchter00}. Additional Fermi LAT work uses $\gamma$-ray
emission to estimate total numbers of $\simeq 2600-4700$ MSPs in the Galactic
globular cluster system, which corresponds to estimates from encounter
rates~\cite{abdo10}.

Current searches include the following: Arecibo, which is searching over 22
globular clusters~\cite{hessels04}; Green Bank Telescope (GBT), which is
working both alone and in conjunction with Arecibo~\cite{jacoby02, hessels04};
the Giant Metrewave Radio Telescope (GMRT), which is searching over about 10
globular clusters~\cite{freire04}; and Parkes, which is searching over 60
globular clusters~\cite{damico03}. Although these searches have been quite
successful, they are still subject to certain selection effects such as
distance, dispersion measure, and acceleration in compact
binaries~\cite{camilo05}. For an excellent review of the properties of all
pulsars in globular clusters, see the review by Camilo and
Rasio~\cite{camilo05} and references therein. The properties of known pulsars
in binary systems with orbital period less than one day are listed in
Table~\ref{MSP_Properties}, which has been extracted from the online catalog
maintained by Freire~\cite{freireweb}.\epubtkUpdateA{Changed reference of
  table from Camilo and Rasio~(2006) to Freire.}

With the ongoing searches, it can be reasonably expected that the
number of millisecond pulsars in binary systems in globular clusters
will continue to grow in the coming years.

\begin{table}[htbp]
  \caption[Short orbital period binary millisecond pulsars in globular
  clusters. Host clusters and orbital properties.]{Short orbital period binary
    millisecond pulsars in globular clusters. Host clusters and orbital
    properties. See \url{http://www.naic.edu/~pfreire/GCpsr.html} for
    references to each pulsar.\epubtkUpdateA{Added new entries to reflect
      current observations. Deleted references column and directed reader to
      online catalog for updated references.}}
  \label{MSP_Properties}
  \vskip 4mm

  \renewcommand{\arraystretch}{1.2}
  \centering
  \begin{tabular}{l|rllrl}
    \hline \hline
    Pulsar &
    \mc{$ P_\mathrm{spin} $} &
    \mc{Cluster} &
    \mc{$ P_\mathrm{orb} $} &
    \mc{$ e $} &
    \mc{$ M_2 $}  \\
    &
    \mc{(ms)} &
    &
    \mc{(days)} &
    &
    \mc{($ M_{\odot} $)} \\
    \hline
    J0024--7204I &  3.485  & 47~Tuc &  0.229  & $<$~0.0004\phantom{00}  &
     0.015   \\
    J0023--7203J &  2.101  & 47~Tuc &  0.121  & $<$~0.00004\phantom{0}  &
     0.024   \\
    J0024--7204O &  2.643  & 47~Tuc &  0.136  & $<$~0.00016\phantom{0} 
    &  0.025 \\
    J0024--7204P &  3.643  & 47~Tuc &  0.147  & --- \qq{} &  0.02  \\
    J0024--7204R &  3.480  & 47~Tuc &  0.066  & --- \qq{} &  0.030  \\
    J0024--7203U &  4.343  & 47~Tuc &  0.429  &  0.000015  & 
    0.14  \\
    J0024--7204V &  4.810  & 47~Tuc &  0.227  & --- \qq{} &  0.34(?)  \\
    J0024--7204W &  2.352  & 47~Tuc &  0.133  & --- \qq{} &  0.14  \\
    J0024--7204Y &  2.196  & 47~Tuc &  0.522  & --- \qq{} &  0.16  \\
    J1518+0204C &  2.484  & M5 &  0.087  & --- \qq{} &  0.038  \\
    J1641+3627D &  3.118  & M13 &  0.591  & --- \qq{} &  0.18  \\
    J1641+3627E &  2.487  & M13 &  0.118  & --- \qq{} &  0.02  \\
    J1701--3006B &  3.594  & M62 &  0.145  & $<$~0.00007\phantom{0}  & 
    0.14  \\
    J1701--3006C &  3.806  & M62 &  0.215  & $<$~0.00006\phantom{0} & 
    0.08  \\
    J1701--3006E &  3.234  & M62 &  0.16  & --- \qq{} &  0.035  \\
    J1701--3006F &  2.295  & M62 &  0.20  & --- \qq{} &  0.02  \\
    B1718--19 &  1004.03\phantom{0}  & NGC 6342 &  0.258  & $<$~
    0.005\phantom{000}
     &  0.13  \\
    J1748--2446A &  11.563  & Terzan 5 &  0.076  & --- \qq{} &  0.10  \\
    J1748--2446M & 3.569  & Terzan 5 &  0.443  & --- \qq{} & 0.16  \\
    J1748--2446N &  8.667  & Terzan 5 &  0.386  &  0.000045  &  0.56 \\
    J1748--2446O &  1.677  & Terzan 5 &  0.259  & --- \qq{} & 0.04 \\
    J1748--2446P &  1.729  & Terzan 5 &  0.363  & --- \qq{} & 0.44  \\
    J1748--2446V &  2.073  & Terzan 5 &  0.504  & --- \qq{} & 0.14 \\
    J1748--2446ae & 3.659 & Terzan 5 & 0.171 & --- \qq{} & 0.019 \\
    J1748-2446ai & 21.228 & Terzan 5 & 0.851 & 0.44 & 0.57\\
    J1748--2021D & 13.496 & NGC6440 & 0.286 & --- \qq{} & 0.14 \\
    J1807--2459A &  3.059  & NGC 6544 &  0.071  & --- \qq{} & 0.010  \\
    J1824--2452G & 5.909 & M28 & 0.105 & --- \qq{} & 0.011 \\
    J1824--2452H & 4.629 & M28 & 0.435 & --- \qq{} & 0.2 \\
    J1824-2452I & 3.932 & M28 & 0.459 & --- \qq{} & 0.2\\
    J1824--2452J & 4.039 & M28 & 0.097 & --- \qq{} & 0.015 \\
    J1824-2452L & 4.100 & M28 & 0.226 & --- \qq{} & 0.022\\
    J1836-2354A & 3.354 & M22 & 0.203 & --- \qq{} & 0.020\\
    J1905+0154A & 3.193 & NGC6749 & 0.813 & --- \qq{} & 0.09 \\
    J1911--5958A &  3.266  & NGC 6752 &  0.837  & $<$~0.00001\phantom{0} 
    &  0.22 \\
    J1911+0102A &  3.619  & NGC 6760 &  0.141  & $<$~0.00013\phantom{0} 
    &  0.02  \\
    J1953+1846A & 4.888 & M71 & 0.177 &  --- \qq{} & 0.032 \\
    B2127+11C &  30.529  & M 15 &  0.335  &  0.681\phantom{000}  & 
    1.13 \\
    J2140--2310A &  11.019  & M30 &  0.174  &  $<$~0.00012\phantom{0} & 
    0.11  \\
    \hline \hline
  \end{tabular}
  \renewcommand{\arraystretch}{1.0}
\end{table}

%%%%%%%%%%%%%%%%%%%%%%%%%%%%%%%%%%%%%%%%%%%%%%%%%%%%%%%%%%%%%%%%%%%%%%%%%%%%%%%
%%%%%%%%%%%%%%%%%%%%%%%%%%%%%%%%%%%%%%%%%%%%%%%%%%%%%%%%%%%%%%%%%%%%%%%%%%%%%%%

\subsection{Black holes}
\label{subsection:black_holes}

There have been no confirmed observations of black hole binaries in Galactic
globular clusters. All of the Galactic globular cluster high luminosity LMXBs
exhibit the X-ray variability that is indicative of nuclear burning on the
surface of a neutron star. It is possible that some of the recently discovered
low luminosity LMXBs may house black holes instead of neutron
stars~\cite{verbunt04}, it is more likely that they are simply unusual neutron
star LMXBs in quiescence~\cite{wijnands05}. Finally, there is very
circumstantial evidence for the possible existence of an intermediate mass
black hole (IMBH) binary in NGC 6752 based upon an analysis of the MSP binary
PSR A~\cite{colpi03, colpi02}.  Further evidence for black holes in globular
clusters will probably have to wait for the next generation of gravitational
wave detectors.

\subsection{Extragalactic Globular Clusters}
\label{subsection:extragalactic_observations}

Recently the X-ray observatories XMM-Newton, Chandra, and Swift have been
combined with HST observations to search for extragalactic globular cluster
binaries. These have yielded a number of systems found in M31, M104, NGC 1399,
NGC 3379, NGC 4278, NGC 4472, and NGC 4697.

Three super soft X-ray sources have been found in M31 globular
clusters~\cite{henze09,pietsch10,henze10}. These are assumed to be from
classical novae, and two have been associated with optical
novae~\cite{henze09,henze10}. Stiele et al.~\cite{stiele11} have found 36
LMXBs and 17 candidate LMXBs that are co-located with M31 globular clusters,
while Peacock et al.~\cite{peacock10} claim 41 LMXBs in 11\% of the M31
globular clusters. Barnard et al.~\cite{barnard11} have found 4 black hole
x-ray binaries associated with M31 globulars. Finally, there is the suspected
intermediate mass black hole in G1~\cite{angelini01, distefano02}.

Searches for bright X-ray sources in other nearby galaxies have yielded LMXB's
thought to contain black holes based on their luminosity. There are 2 in NGC
1399~\cite{shih10,irwin10,maccarone11b}, and Paolillo et al.~\cite{paolillo11}
estimate that 65\% of globular clusters in NGC 1399 host LMXBs. Brassington et
al.~\cite{brassington10} have found 3 LMXBs in globular clusters associated
with NGC 3379, of which 1 is a presumed black hole. Fabbiano et
al.~\cite{fabbiano10} have found 4 LMXBs in globular clusters associated with
NGC 4278, all of which are bright enough to be black holes. Maccarone and
collaborators~\cite{maccarone07,maccarone11a} have found 2 LMXBs containing
black holes in NGC 4472 globular clusters. Kim et al.~\cite{kim09} claim 75
globular clusters LMXBs in NGC 3379, 4278, and 4697. Finally, Li et
al.~\cite{li10} have found 41 X-ray sources that are presumably LMXBs in
globular clusters in M104.

\newpage

%%%%%%%%%%%%%%%%%%%%%%%%%%%%%%%%%%%%%%%%%%%%%%%%%%%%%%%%%%%%%%%%%%%%%%%%%%%%%%%
%%%%%%%%%%%%%%%%%%%%%%%%%%%%%%%%%%%%%%%%%%%%%%%%%%%%%%%%%%%%%%%%%%%%%%%%%%%%%%%
%%%%%%%%%%%%%%%%%%%%%%%%%%%%%%%%%%%%%%%%%%%%%%%%%%%%%%%%%%%%%%%%%%%%%%%%%%%%%%%

\section{Relativistic Binaries}
\label{section:relativistic_binaries}
Relativistic binaries are binary systems containing two degenerate
or collapsed objects and an orbital period such that they will be
brought into contact within a Hubble time. (Note that this definition
also includes binaries which are already in contact.) Frequently, 
in the process of becoming a relativistic binary, a binary will exist
with a single degenerate or collapse object  and a normal star. 
These systems are tracers of relativistic binaries. Outside of dense
stellar clusters, most relativistic binary systems arise from
primordial binary systems whose evolution drives them to tight,
ultracompact orbits. Dynamical processes in globular clusters can
drive wide binary systems toward short orbital periods and can also
insert degenerate or collapsed stars into relativistic orbits with
other stars. Before addressing specific evolutionary scenarios, we
will present the generic features of binary evolution that lead to the
formation of relativistic binaries.

%%%%%%%%%%%%%%%%%%%%%%%%%%%%%%%%%%%%%%%%%%%%%%%%%%%%%%%%%%%%%%%%%%%%%%%%%%%%%%%
%%%%%%%%%%%%%%%%%%%%%%%%%%%%%%%%%%%%%%%%%%%%%%%%%%%%%%%%%%%%%%%%%%%%%%%%%%%%%%%

\subsection{Binary evolution}
\label{subsection:binary_evolution}

The evolution of a binary system of two main-sequence stars can significantly
affect the evolution of both component stars if the orbital separation is
sufficiently small. If the orbital period is less than about 10 days, tidal
interactions will have circularized the orbit during the pre- and early
main-sequence phase~\cite{goldman91, zahn89a, zahn89b}. Both stars start in
the main sequence with the mass of the primary $M_\mathrm{p}$ and the mass of
the secondary $M_\mathrm{s}$, defined such that $M_\mathrm{p} \geq
M_\mathrm{s}$. The binary system is described by the orbital separation $r$,
and the mass ratio of the components $q \equiv M_\mathrm{s}/M_\mathrm{p}$. The
gravitational potential of the binary system is described by the Roche model
where each star dominates the gravitational potential inside regions called
Roche lobes. The two Roche lobes meet at the inner Lagrange point along the
line joining the two stars. Figure~\ref{roche_lobe} shows equipotential
surfaces in the orbital plane for a binary with $q = 0.4$. If either star
fills its Roche lobe, matter will stream from the Roche lobe filling star
through the inner Lagrange point to the other star in a process known as Roche
lobe overflow (RLOF). This mass transfer affects both the evolution of the
components of the binary as well as the binary properties such as orbital
period and eccentricity.

Roche lobe overflow can be triggered by the evolution of the binary properties
or by evolution of the component stars. On the one hand, the orbital
separation of the binary can change so that the Roche lobe can shrink to
within the surface of one of the stars. On the other hand, stellar evolution
may eventually cause one of the stars to expand and fill its Roche lobe. When
both stars in the binary are main-sequence stars, the latter process is more
common. Since the more massive star will evolve first, it will be the first to
expand and fill its Roche lobe. At this stage, the mass exchange can be
conservative (no mass is lost from the binary) or non-conservative (mass is
lost). Depending on the details of the mass exchange and the evolutionary
stage of the mass-losing star there are several outcomes that will lead to the
formation of a relativistic binary. The primary star can lose its envelope,
revealing its degenerate core as either a helium, carbon-oxygen, or
oxygen-neon white dwarf; it can explode as a supernova, leaving behind a
neutron star or a black hole; or it can simply lose mass to the secondary so
that they change roles. Barring disruption of the binary, its evolution will
then continue. In most outcomes, the secondary is now the more massive of the
two stars and it may evolve off the main sequence to fill its Roche lobe. The
secondary can then initiate mass transfer or mass loss with the result that
the secondary can also become a white dwarf, neutron star, or black hole.

The relativistic binaries that result from this process fall into a number of
observable categories. A WD--MS or WD--WD binary may eventually become a
cataclysmic variable once the white dwarf begins to accrete material from its
companion. If the companion is a main-sequence star, RLOF can be triggered by
the evolution of the companion. If the companion is another white dwarf, then
RLOF is triggered by the gradual shrinking of the orbit through the emission
of gravitational radiation. Some WD--WD cataclysmic variables are also known
as AM CVn stars. If the total mass of the WD--WD binary is above the
Chandrasekhar mass, the system may be a double degenerate progenitor to a Type
Ia supernova.

The orbit of a NS--MS or NS--WD binary will shrink due to the emission of
gravitational radiation. At the onset of RLOF, the binary will become  either
a low-mass X-ray binary (if the donor star is a white dwarf or main sequence
star with $M \leq 2\,M_{\odot}$), or a high-mass X-ray binary (if the donor is
a more massive main-sequence star). These objects may further evolve to become
millisecond pulsars if the neutron star is spun up during the X-ray binary
phase~\cite{davies98, rasio00a}. A NS--NS binary will remain virtually
invisible unless one of the neutron stars is observable as a pulsar. A BH--MS
or BH--WD binary may also become a low- or high-mass X-ray binary. If the
neutron star is observable as a pulsar, a BH--NS binary will appear as a
binary pulsar. BH--BH binaries will be invisible unless they accrete matter
from the interstellar medium. A comprehensive table of close binary types that
can be observed in electromagnetic radiation can be found in
Hilditch~\cite{hilditch01}.

\epubtkImage{rochelobe.png}{
  \begin{figure}[htbp]
    \def\epsfsize#1#2{0.7#1}
    \centerline{\epsfbox{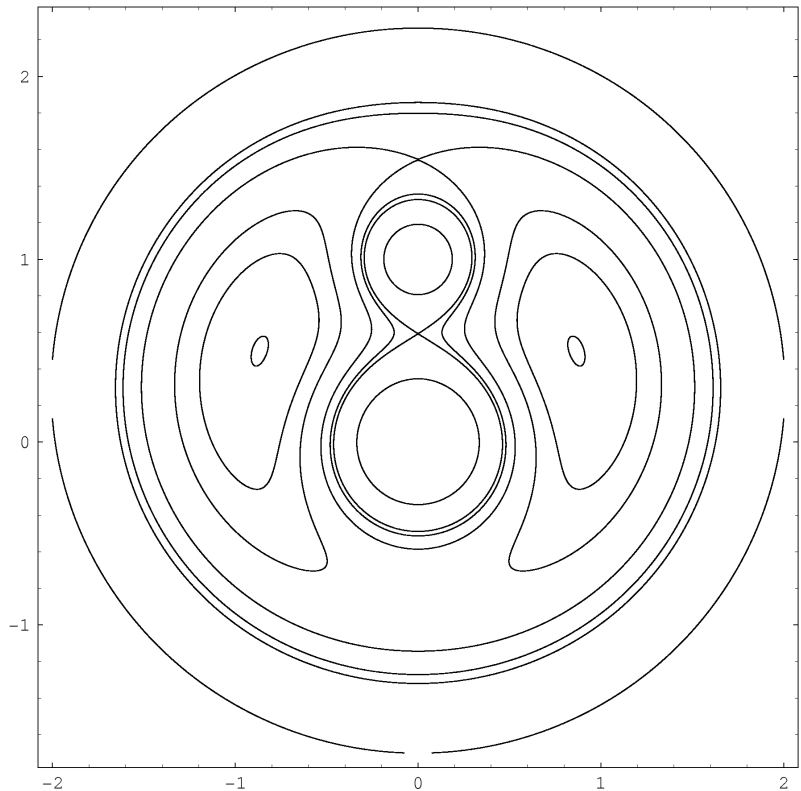}}
    \caption{Cross section of equipotential surfaces in the
      orbital plane of a binary with $q = 0.4$. The values of the
      potential surfaces are 5.0, 3.9075, 3.8, 3.559, 3.2,
      3.0, and 2.8. The units have been normalized to the orbital
      separation, so $a = 1$.}
    \label{roche_lobe}
  \end{figure}}

The type of binary that emerges depends upon the orbital separation and the
masses of the component stars. During the evolution of a $10\,M_{\odot}$ star,
the radius will slowly increase by a factor of about two as the star
progresses from zero age main sequence to terminal age main sequence. The
radius will then increase by about another factor of 50 as the star
transitions to the red giant phase, and an additional factor of 10 during the
transition to the red supergiant phase. These last two increases in size occur
very quickly compared with the slow increase during the main-sequence
evolution. Depending upon the orbital separation, the onset of RLOF can occur
any time during the evolution of the star. Mass transfer can be divided into
three cases related to the timing of the onset of RLOF:

\begin{case}%
  If the orbital separation is small enough (usually a few days), the
  star can fill its Roche lobe during its slow expansion through the
  main-sequence phase while still burning hydrogen in its core.
  \label{case_a}
\end{case}
\begin{case}%
  If the orbital period is less than about 100 days, but longer than a
  few days, the star will fill its Roche lobe during the rapid
  expansion to a red giant with a helium core. If the helium core
  ignites during this phase and the transfer is interrupted, the
  mass transfer is case BB.
  \label{case_b}
\end{case}
\begin{case}%
  If the orbital period is above 100 days, the star can evolve to the
  red supergiant phase before it fills its Roche lobe. In this case,
  the star may have a CO or ONe core. \\
  \label{case_c}
\end{case}

\noindent
The typical evolution of the radius for a low metallicity star is shown in
Figure~\ref{radial_evolution}. Case~\ref{case_a} mass transfer occurs during
the slow growth, Case~\ref{case_b} during the first rapid expansion, and
Case~\ref{case_c} during the final expansion phase. The nature of the remnant
depends upon the state of the primary during the onset of RLOF and the orbital
properties of the resultant binary depend upon the details of the mass
transfer.

\epubtkImage{radialevolution.png}{
  \begin{figure}[htbp]
    \def\epsfsize#1#2{0.5#1}
    \centerline{\epsfbox{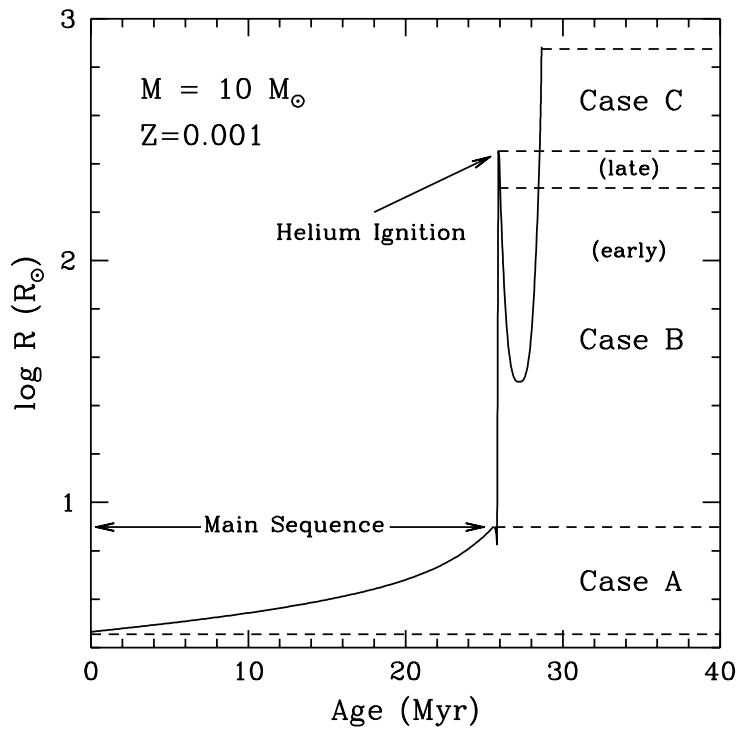}}
    \caption{Evolution of the radius for a $10\,M_{\odot}$ star with a
      metallicity of {\it Z}~=~0.001. Figure taken from Pfahl et
      al.~\cite{pfahl02a}.} 
    \label{radial_evolution}
  \end{figure}}

%%%%%%%%%%%%%%%%%%%%%%%%%%%%%%%%%%%%%%%%%%%%%%%%%%%%%%%%%%%%%%%%%%%%%%%%%%%%%%%
%%%%%%%%%%%%%%%%%%%%%%%%%%%%%%%%%%%%%%%%%%%%%%%%%%%%%%%%%%%%%%%%%%%%%%%%%%%%%%%

\subsection{Mass transfer}
\label{subsection:mass_transfer}

Although there are still many unanswered theoretical questions about the
nature of the mass transfer phase, the basic properties of the evolution of a
binary due to mass transfer can easily be described. The rate at which a star
can adjust to changes in its mass is governed by three time scales. The
dynamical time scale results from the adiabatic response of the star to
restore hydrostatic equilibrium, and can be approximated by the free fall time
across the radius of the star,
\begin{equation}
  t_\mathrm{dyn} \simeq \left(\frac{2 R^3}{G M}\right)^{1/2}
  \!\!\!\! \sim
  40\left[ \left( \frac{R}{R_{\odot}} \right)^3
  \frac{M_{\odot}}{M}\right]^{1/2} \!\!\!\!\!\!\!\!\! \mathrm{\ min},
  \label{dynamical_time}
\end{equation}
where $M$ and $R$ are the mass and radius of the star. The thermal equilibrium
of the star is restored over a longer period given by the thermal time scale
\begin{equation}
  t_\mathrm{th} \simeq \frac{G M^2}{R L} \sim 3 \times 10^7 
  \left(\frac{M}{M_{\odot}}\right)^2
  \frac{R_{\odot}}{R}\frac{L_{\odot}}{L} \mathrm{\ yr},
  \label{thermal_time}
\end{equation}
where $L$ is the luminosity of the star. Finally, the main-sequence lifetime
of the star itself provides a third time scale, which is also known as the
nuclear time scale:
\begin{equation}
  t_\mathrm{nuc} \sim 7 \times 10^9
  \frac{M}{M_{\odot}}\frac{L_{\odot}}{L} \mathrm{\ yr}.
  \label{nuclear_time}
\end{equation}

The rate of mass transfer/loss from the Roche lobe filling star is governed by
how the star's radius changes in response to changes in its mass. Hjellming
and Webbink~\cite{hjellming87} describe these changes and the response of the
Roche lobe to mass changes in the binary using the radius-mass exponents,
$\zeta \equiv d \ln{R}/d \ln{M}$, for each of the three processes described in
Equations~(\ref{dynamical_time}, \ref{thermal_time}, \ref{nuclear_time}) and
defining
\begin{equation}
  \zeta_\mathrm{L} = (1 + q)\frac{d \ln{R_\mathrm{L}}}{d \ln{q}}
  \label{roche_lobe_exponent}
\end{equation}
for the Roche lobe radius-mass exponent. If $\zeta_\mathrm{L} >
\zeta_\mathrm{dyn}$, the star cannot adjust to the Roche lobe, then the mass
transfer takes place on a dynamical time scale and is limited only by the rate
at which material can stream through the inner Lagrange point. If
$\zeta_\mathrm{dyn} > \zeta_\mathrm{L} > \zeta_\mathrm{th}$, then the mass
transfer rate is governed by the slow expansion of the star as it relaxes
toward thermal equilibrium, and it occurs on a thermal time scale. If both
$\zeta_\mathrm{dyn}$ and $\zeta_\mathrm{th}$ are greater than
$\zeta_\mathrm{L}$, then the mass loss is driven either by stellar evolution
processes or by the gradual shrinkage of the orbit due to the emission of
gravitational radiation. The time scale for both of these processes is
comparable to the nuclear time scale. A good analysis of mass transfer in
cataclysmic variables can be found in King et al.~\cite{king96}.

Conservative mass transfer occurs when there is no mass loss from the system,
and therefore all mass lost from one star is accreted by the other
star. During conservative mass transfer, the orbital elements of the binary
can change. Consider a system with total mass $M = M_1 + M_2$ and semi-major
axis $a$. The total orbital angular momentum
\begin{equation}
  J = \left[\frac{GM_1^2M_2^2 a}{M}\right]^{1/2}
  \label{orbital_J}
\end{equation}
is a constant, and we can write $a \propto (M_1 M_2)^{-2}$. Using Kepler's
third law and denoting the initial values by a subscript $i$, we find:
\begin{equation}
  \frac{P}{P_i} = \left[\frac{M_{1i} M_{2i}}{M_1 M_2}\right]^3\!\!\!.
  \label{period_change}
\end{equation}
Differentiating Equation~(\ref{period_change}) and noting that conservative
mass transfer requires $\dot{M}_1 = - \dot{M}_2$ gives:
\begin{equation}
  \frac{\dot{P}}{P} = \frac{3\dot{M}_1\left(M_1-M_2\right)}{M_1M_2}.
  \label{Pdot}
\end{equation}
Note that if the more massive star loses mass, then the orbital period
decreases and the orbit shrinks. If the less massive star is the donor, then
the orbit expands. Usually, the initial phase of RLOF takes place as the more
massive star evolves. As a consequence, the orbit of the binary will shrink,
driving the binary to a more compact orbit.

In non-conservative mass transfer, both mass and angular momentum can be
removed from the system. There are two basic non-conservative processes that
are important for the formation of relativistic binaries -- the
common-envelope process and the supernova explosion of one component of the
binary. The result of the first process is often a short-period, circularized
binary containing a white dwarf. Although the most common outcome of the
second process is the disruption of the binary, occasionally this process will
result in an eccentric binary containing a neutron star or a black hole.

Common envelope scenarios result when one component of the binary expands so
rapidly that the mass transfer is unstable and the companion becomes engulfed
by the donor star. This can happen if the mass transfer rate is so great that
it exceeds the Eddington mass accretion rate of the accretor, or when the
donor expands past the outer Lagrange point~\cite{paczynski76,webbink76}. The
companion can then mechanically eject the envelope of the donor star. There
are two approaches to determining the outcome of the process of ejection of
the envelope.

The most commonly used approach is the ``$\alpha$-prescription'' of
Webbink~\cite{webbink84}, in which the energy required to eject the envelope
comes from the orbital energy of the binary and thus the orbit shrinks. The
efficiency of this process determines the final orbital period after the
common envelope phase. This is described by the efficiency parameter
\begin{equation}
  \alpha_\mathrm{CE} = \frac{\Delta E_\mathrm{bind}}{\Delta E_\mathrm{orb}},
  \label{CE_efficiency}
\end{equation}
where $\Delta E_\mathrm{bind}$ is the binding energy of the mass stripped from
the envelope and $\Delta E_\mathrm{orb}$ is the change in the orbital energy
of the binary. The result of the process is the exposed degenerate core of the
donor star in a tight, circular orbit with the companion. This process can
result in a double degenerate binary if the process is repeated twice or if
the companion has already evolved to a white dwarf through some other
process.  If enough orbital energy is lost it can also lead to a merger of the
binary components.  A brief description of the process is outlined by
Webbink~\cite{webbink84}, and a discussion of the factors involved in
determining $\alpha_\mathrm{CE}$ is presented in Sandquist et
al.~\cite{sandquist00}.

The other approach that is used is the ``$\gamma$-prescription'' of Nelemans
et al.~\cite{nelemans00}. In this approach, other energy sources such as tidal
heating or luminosity of the donor may assist in the unbinding of the
envelope. The material lost by the envelope is assumed to carry away angular
momentum and reduce the total angular momentum of the system such that:
\begin{equation}
J_f = J_i\left(1-\gamma\frac{\Delta M}{M_{\rm tot}}\right)
\end{equation}
where $J_i$ and $J_f$ are the initial and final angular momenta of the binary,
$M_{\rm tot}$ is the total mass of the binary prior to mass loss, and $\Delta
M$ is the mass lost in the ejection of the envelope.

The effect on a binary of mass loss due to a supernova can be quite
drastic. Following Padmanabhan~\cite{padmanabhan01}, this process is outlined
using the example of a binary in a circular orbit with a semi-major axis
$a$. Let $v$ be the velocity of one component of the binary relative to the
other component. The initial energy of the binary is given by
\begin{equation}
  E = \frac{1}{2} \left(\frac{M_1 M_2}{M_1 + M_2}\right) v^2 -
  \frac{G M_1 M_2}{a} = - \frac{G M_1 M_2}{2a}.
  \label{initial_energy}
\end{equation}
Following the supernova explosion of $M_1$, the expanding mass shell will
quickly cross the orbit of $M_2$, decreasing the gravitational force acting on
the secondary. The new energy of the binary is then
\begin{equation}
  E^{\prime} = \frac{1}{2} \left( \frac{M_\mathrm{NS} M_2}{M_\mathrm{NS} +
      M_2} \right) v^2 - \frac{G M_\mathrm{NS} M_2}{a},
  \label{final_energy_1}
\end{equation}
where $M_\mathrm{NS}$ is the mass of the remnant neutron star. We have assumed
here that the passage of the mass shell by the secondary has negligible effect
on its velocity (a safe assumption, see Pfahl et al.~\cite{pfahl02a} for a
discussion), and that the primary has received no kick from the supernova (not
necessarily a safe assumption, but see Davies and Hansen~\cite{davies98} or
Pfahl et al.~\cite{pfahl02b} for an application to globular cluster
binaries). Since we have assumed that the instantaneous velocities of both
components have not been affected, we can replace them by $v^2 = G\left(M_1 +
  M_2\right)/a$, and so
\begin{equation}
  E^{\prime} = \frac{G M_\mathrm{NS} M_2}{2a}\left(\frac{M_1+M_2}
  {M_\mathrm{NS} + M_2} -2\right).
  \label{final_energy_2}
\end{equation}
Note that the final energy will be positive and the binary will be disrupted
if $M_\mathrm{NS} < (1/2)(M_1 + M_2)$. This condition occurs when the mass
ejected from the system is greater than half of the initial total mass,
\begin{equation}
  \Delta M > \frac{1}{2}\left(M_1 + M_2\right),
  \label{deltaM}
\end{equation}
where $\Delta M = M_1 - M_\mathrm{NS}$. If the binary is not disrupted, the
new orbit becomes eccentric and expands to a new semi-major axis given by
\begin{equation}
  a^{\prime} = a\left(\frac{M_1 + M_2 - \Delta M}
  {M_1 + M_2 - 2\Delta M}\right),
  \label{final_a}
\end{equation}
and orbital period
\begin{equation}
  P^{\prime} = P\left(\frac{a^{\prime}}{a}\right)^{3/2}
  \left(\frac{2a^{\prime}-a}{a^{\prime}}\right)^{1/2}\!\!\!\!\!\!\!.
  \label{final_P}
\end{equation}
Note that we have not included any mention of the expected velocity kick that
newly born neutron star or black hole will receive due to asymmetries in the
supernova explosion.  These kicks can be quite substantial, up to several
hundred kilometers per second and, at least for observed pulsars, seem to be
drawn from a Maxwellian distribution with a peak at 265 km s$^{-1}$
\cite{HobbsEtAl05}.  In most cases, the kick  will further increase the
likelihood that the binary will become unbound, but occasionally the kick
velocity will be favorably oriented and the binary will remain intact.  If the
kick is higher than the escape velocity of the cluster, it will also remove
the remnant from the system.  This mechanism may be very important in
depleting the numbers of neutron stars and black holes in globular clusters.

We have seen that conservative mass transfer can result in a tighter binary if
the more massive star is the donor. Non-conservative mass transfer can also
drive the components of a binary together during a common envelope phase when
mass and angular momentum are lost from the system. Direct mass loss through a
supernova explosion can also alter the properties of a binary, but this
process generally drives the system toward larger orbital separation and can
disrupt the binary entirely. With this exception, the important result of all
of these processes is the generation of tight binaries with at least one
degenerate object.

The processes discussed so far apply to the generation of relativistic
binaries anywhere. They occur whenever the orbital separation of a progenitor
binary is sufficiently small to allow for mass transfer or common envelope
evolution. Population distributions for relativistic binaries are derived from
an initial mass function, a distribution in mass ratios, and a distribution in
binary separations. These initial distributions are then fed into models for
binary evolution such as StarTrack~\cite{belczynski02} or
SeBa~\cite{portegieszwart96, nelemans01a} in order to determine rates of
production of relativistic binaries. The evolution of the binary is often
determined by the application of some simple operational formulae such as
those described by Tout et al.~\cite{tout97} or Hurley et
al.~\cite{hurley00}. For example, Hils, Bender, and Webbink~\cite{hils90}
estimated a population of close white dwarf binaries in the disk of the galaxy
using a Salpeter mass function, a mass ratio distribution strongly peaked at
1, and a separation distribution that was flat in $\ln(a)$. Other estimates
of relativistic binaries differ mostly by using different
distributions~\cite{belczynski01, iben97, nelemans01a, nelemans01b}.

%%%%%%%%%%%%%%%%%%%%%%%%%%%%%%%%%%%%%%%%%%%%%%%%%%%%%%%%%%%%%%%%%%%%%%%%%%%%%%%
%%%%%%%%%%%%%%%%%%%%%%%%%%%%%%%%%%%%%%%%%%%%%%%%%%%%%%%%%%%%%%%%%%%%%%%%%%%%%%%

\subsection{Globular cluster processes}
\label{subsection:globular_cluster_processes}

In the galactic field stellar densities are low enough that stars and binaries
rarely encounter each other.  In this environment binaries evolve in isolation
with their properties and fate determined by solely by their initial conditions
and the processes described in Sections~\ref{subsection:binary_evolution}
and~\ref{subsection:mass_transfer}.  In star cluster stellar densities are
much higher and encounters between stars, stars and binaries, and binaries and
binaries become increasingly common.  Such encounters can affect a binary's
parameters and dramatically alter the evolution that would be expected if the
same binary were isolated.  Some encounter outcomes that are of particular
interest for relativistic binaries include:
\begin{itemize}
\item Reduction or increase of the period during distant encounters.
\item Exchange interactions where binary membership is changed in close
  encounters.
\item Binary formation during strong few-body interactions.
\item Encounter-induced binary mergers.
\end{itemize}
Different types of interactions can produced each of these outcomes and we
will describe these briefly in the terms normally used by stellar dynamicists
- the number of bodies and type of objects involved in the interaction.  In
all cases these interactions must be distinguished from the very weak
interactions that drive two-body relaxation of clusters as described in
Section~\ref{sec:GCDynamics}.  The interactions affecting binary parameters
result from close encounters, the outcomes of which cannot be described
statistically using the language of relaxation theory.  Note also that the
term binding energy is normally spoken of as though it were a positive
quantity.  Thus binaries with the largest (most negative) binding energy are
the most tightly bound.

\subsubsection{Single-Single Interactions}

As the name suggests, single-single interactions are encounters between two
single stars.  If the periastron of the encounter is sufficiently small, the
stars may excite tidal oscillations in each other at the cost of some of their
relative kinetic energies.  If sufficient kinetic energy is dissipated, the
stars can become bound and form a new binary.

The exact nature of the oscillations excited in the stars are not important,
only that they dissipate sufficient kinetic energy.  Furthermore, to obtain a
bound orbit it is only necessary to dissipate sufficient kinetic energy that
the total energy \emph{at apocenter} is reduced to less than zero.  The basic
condition for tidal capture is:
\begin{equation}
  \Delta E_{\rm T1} + \Delta E_{\rm T2} \ge \frac{1}{2}\mu v_{a}^{2}
\end{equation}
where $\Delta E_{\rm T1,T2}$ is the energy associated with the tidal
oscillation in each star, $\mu$ is the reduced mass, and $v_{a}$ is the
relative velocity at apocenter (or infinity) \cite{fabian75}.  Only a fraction
$\sim (v_{a}/v_{p})^{2}$, where $v_{p}$ is the velocity at pericenter, of the
energy must be converted to tidal oscillations in order for a capture to
occur.  For example, an encounter with a pericenter velocity of $\sim 100 {\rm
  km s^{-1}}$ in a cluster with a velocity dispersion of $\sim 10 {\rm km
  s^{-1}}$ only needs to dissipate $\sim 1\%$ of the apocenter kinetic energy 
in order for a capture to occur.  This mechanism favours the creation of
binaries in encounters with large ratios between apo and pericenter velocity.
Thus it tends to produce highly eccentric binaries with very small pericenter
separations.  Such binaries would then be expected to circularize over time
due to further tidal effects.

This process was once thought to be the dominant path for dynamically creating
binaries in star clusters \cite{fabian75,binney87} since two-body interactions
are more likely than higher-order encounters \cite{spitzer87}.  It has been
realized, however, that this is not the case since stars must approach very
closely, to within a few stellar radii, in order for tidal oscillations to
dissipate enough energy for a capture to occur \cite{fabian75, LeeOstriker86,
  PressTeukolsky77}.  In these situations it is actually more likely,
depending upon the stellar equation of state, for the stars to merge rather
that form a binary \cite{RayEtAl87, McMillanEtAl90, bailyn91, hut92a,
  rasio00a}.  Furthermore, due to the difficulty in exciting tides in
degenerate matter, tidal capture is unlikely to form binaries from two compact
objects.

\subsubsection{Three-body Interactions}
\label{sec:3bdyInt}

When three single-stars interact it is possible for one of the stars to gain
kinetic energy and, if the relative kinetic energy of the remaining two stars
is sufficiently reduced, they can form a bound pair.  In general the least
massive body will gain the highest velocity, since it is the easiest to
accelerate, and the two more massive bodies will remain (this can also be seen
as a consequence of energy equipartition).  Heggie \cite{heggie75} showed that
if $E_{b}$ is the binding energy of the new binary, the rate of three-body
binary formation is $\propto E_{b}^{-7/2}$.  Thus  three-body interactions
tend to favour the creation of binaries with low binding energies.  The
binding energy can be increased by later encounters (see
Section~\ref{sec:BSint}) and there is about a 10\% probability that a binary
formed by three-body interactions will gain sufficient binding energy to
survive permanently \cite{GoodmanHut93, spitzer87}.

Three-body binary formation interactions are less common than tidal
interactions since the probability of a close encounter between three stars is
smaller than the probability of an encounter between two.  Thus stellar
densities need to be higher for three-body binary formation to be efficient
than for tidal interactions \cite{GoodmanHut93, Hut85, spitzer87}.  However,
because it is not necessary to excite tides, pericenter passages need not be
as close in order to form binaries during these interactions and mergers are a
less common outcome of three-body interactions.  It is worth noting that, in
equal-mass systems, neither tidal nor three-body interactions are likely to
produce many binaries over the lifetime of a star cluster \cite{Hut85,
  GoodmanHut93, binney87, spitzer87}.  This is not necessarily the case for
clusters with a mass function due to the very high central densities and
strong encounters that can be achieved by mass segregation.  Nevertheless,
primordial binaries and the interactions involving them are critical for
producing large numbers of relativistic binaries in globular clusters.

\subsubsection{Binary - Single Interactions}
\label{sec:BSint}

Binary-single interactions, although formally still three-body interactions,
differ because two of the stars are already bound.  In such an encounter
several outcomes are possible depending on the relative kinetic energy of the
star and binary and on the binding energy of the binary itself.

If the kinetic energy of the star is less than the binding energy of the
binary, energy equipartition requires that the star be accelerated and the
binding energy of the binary be increased (the period will become shorter).
If, however, the kinetic energy of the star is greater than the binding energy
of the binary, the star will donate kinetic energy to the binary and the
binding energy will decrease (the period will become longer).  For equal-mass
systems this introduces a simple yet important classification for binaries.
If $|E_{b}| < k_{B}T$ (Section~\ref{sec:GCDynamics}) then the binary is said
to be ``soft'' and, on average, will tend to have its period increased by
encounters (softening).  If $|E_{b}| > k_{B}T$ than the binary is said to be
``hard'' and will tend to have its period decreased by encounters (hardening).
A hard binary gains on average $\langle \Delta E_{b} \rangle \sim 0.2-0.4
E_{b}$ per interaction assuming a Maxwellian velocity distribution
\cite{heggie75, hut92b} (although this is only fully valid in the case of very
hard binaries \cite{Hut85}).  Since the encounter rate is proportional to the
semi-major axis of the binary ($\propto 1/E_{b}$), the energy of a hard binary
increases by $\Delta E_{b,rlx} \sim 0.6k_{b}T/\trlx$ per relaxation time
\cite{binney87}.  This leads to the ``Heggie-Hills Law'' \cite{heggie75,
  Hills75} that states  ``hard binaries get harder while soft binaries get
softer''.  The end result of binary hardening can be a merger while the end
result of binary softening can be the disruption of the binary.

In a multi-mass system the division between hard and soft binaries is not so
clear since the relative energies depend upon the specific masses of all the
bodies involved.  For an individual encounter between a binary with member
masses $m_{1}$ and $m_{2}$ and a single star of mass $m_{3}$ travelling at
velocity $v$, it is possible to define a critical velocity \cite{HutBahcall83}:
\begin{equation}
  v_{c}^{2} = \frac{m_{1}+m_{2}+m_{3}}{m_{3}(m_{1}+m_{2})}
\end{equation}
such that for $v_{c} < v$ the binary cannot capture the single star and the
star can disrupt the binary (softening) and for $v_{c} > v$ then the binary
can capture the single star and cannot be disrupted in the process
(hardening).  Thus the distinction between hard and soft binaries still exists
in multi-mass systems.  It has been shown by Hills that it is also possible to
use the ratio of orbital speeds to define whether the binary will gain or
lose energy \cite{hills90}.

The reference to capture alludes to another important process that can occur
in binary-single interactions: exchange.  In an exchange interaction one of
the original binary members is replaced by the single star so the binary
membership changes.  As with other encounters, equipartition of energy favours
the ejection of the lowest-mass object and exchange encounters are a way of
introducing massive objects into binaries.  It has been shown that in the
limit of $m_{3} \gg m_{1}$ or $m_{2}$, the probability for a massive object to
be exchanged into the binary is $\sim 1$ \cite{HillsFullerton80}.  Thus it is
possible to create a compact binary from a binary where neither member was
originally compact entirely by dynamical means.  This may be particularly
effective for BHs since they are the most massive objects in evolved star
clusters \cite{sigurdsson93}.

\subsubsection{Binary - Binary Interactions}
\label{sec:BBint}

There are many possible outcomes for binary-binary interactions, particularly
in the multi-mass case, that depend rather sensitively upon the initial
conditions of the encounter.  Therefore a quantitative theory of these
interactions is lacking.  For sufficiently distant encounters both binaries
will ``see'' each other as single centres of mass and a hardening or softening
interaction will occur in both binaries, depending on their binding energies
and relative kinetic energy.  There are also some general results for close
encounters where all stars have equal mass.  If one of the binaries is much
harder than the other, it can be exchanged into the other binary as a single
star.  This produces a hierarchical three-body system, the future of which
will be decided by further interactions \cite{spitzer87}.  One or both
binaries can also be disrupted during the encounter.  Numerical experiments
show that at least one of the binaries is disrupted in $\sim$88\% of cases
\cite{Mikkola83, Mikkola84a, Mikkola84b, gao91}.  It is also possible for one
or, more rarely, both binaries to exchange members with each other.  Thus
binary-binary interactions can provide all of the effects of binary-single
interactions but can result in extensive binary destruction as well.

Higher-order interactions (those involving more than four bodies) are also
possible but will be quite rare and are even less amenable to quantitative
description than are binary-binary encounters.

%%%%%%%%%%%%%%%%%%%%%%%%%%%%%%%%%%%%%%%%%%%%%%%%%%%%%%%%%%%%%%%%%%%%%%%%%%%%%%%
%%%%%%%%%%%%%%%%%%%%%%%%%%%%%%%%%%%%%%%%%%%%%%%%%%%%%%%%%%%%%%%%%%%%%%%%%%%%%%%
%%%%%%%%%%%%%%%%%%%%%%%%%%%%%%%%%%%%%%%%%%%%%%%%%%%%%%%%%%%%%%%%%%%%%%%%%%%%%%%

\newpage

%%%%%%%%%%%%%%%%%%%%%%%%%%%%%%%%%%%%%%%%%%%%%%%%%%%%%%%%%%%%%%%%%%%%%%%%%%%%%%%
%%%%%%%%%%%%%%%%%%%%%%%%%%%%%%%%%%%%%%%%%%%%%%%%%%%%%%%%%%%%%%%%%%%%%%%%%%%%%%%
%%%%%%%%%%%%%%%%%%%%%%%%%%%%%%%%%%%%%%%%%%%%%%%%%%%%%%%%%%%%%%%%%%%%%%%%%%%%%%%

\section{Dynamical Evolution}
\label{section:dynamical_evolution}
The evolution of the compact binary population in a globular cluster depends
on the interplay between the global evolution of the cluster as described in
Section~\ref{sec:GCDynamics}, the few-body interactions described in
Section~\ref{subsection:globular_cluster_processes}, and the binary stellar
evolution described in Section~\ref{subsection:binary_evolution}.  Simulations
that hope to describe the compact binary population of a globular cluster must
be able to couple these processes and describe their mutual interaction.

The evolution of globular clusters is primarily governed by point-mass
gravitational attraction between individual stars and thus their overall
structure can be described in terms of classical Newtonian $N$-body dynamics.
Other processes, however, also play a role.  Stellar evolution obviously
affects the number and properties of compact objects and compact binaries so
describing it accurately is important for producing the correct initial
compact population.  Stellar evolution proceeds on the same timescales as
relaxation and core-collapse in globular clusters and can result in
significant mass-loss from individual stars.  This mass-loss can change the
mass of the cluster enough to affect its dynamical evolution.  Consequently
the global dynamical evolution of a globular cluster is coupled to nuclear
burning in individual stars and this must be taken into account.  Furthermore,
stellar evolution affects the orbital parameters of the binary population and,
as we have seen in Section~\ref{sec:GCDynamics}, binaries can be an energy
source of for the cluster and can slow down or halt core collapse.  Thus the
details of both binary evolution and few-body interactions can affect the
global dynamics of a globular cluster.  Finally, the loop is closed by the
fact that the global dynamics a globular cluster can affect  the properties of
the binary population by bringing stars and binaries close enough that they
can interact and perturb each other.  We are interested in the binary
population that is the end result of this complex interplay of different
processes operating on very different spacial and temporal scales.  Simulating
star cluster dynamics is a problem that has a long history and has promoted
contact between various branches of astrophysics.  The MODEST (MOdeling DEnse
STellar systems) collaboration, a collection of various working groups from
different backgrounds, maintains a website that provides some up-to date
information about efforts to model the dynamical and stellar evolution of star
clusters \cite{modestweb}.

%%%%%%%%%%%%%%%%%%%%%%%%%%%%%%%%%%%%%%%%%%%%%%%%%%%%%%%%%%%%%%%%%%%%%%%%%%%%%%%
%%%%%%%%%%%%%%%%%%%%%%%%%%%%%%%%%%%%%%%%%%%%%%%%%%%%%%%%%%%%%%%%%%%%%%%%%%%%%%%

\subsection{Star Cluster Simulation Methods}
\label{sec:SimMethods}
Broadly speaking, there are three approaches that can be used to simulate the
dynamical evolution of star clusters.  The first, direct $N$-body integration,
treats the dynamical evolution of a star cluster by numerically solving
Newton's equations.  The second, the distribution function method, represents
the the star cluster using a 6$N$ dimensional distribution function and then
describes its evolution using a collisional form of the Boltzmann equation.
Both of these methods have their strengths and weaknesses.  Several authors
have compared these techniques (e.g. \cite{heggie98, KimEtAl08, HeggieEtAl06,
  TrentiEtAl07}) and in general find that they agree well.  The summary of the
MODEST-2 meeting \cite{sills03} gives a fairly recent review of efforts to
implement these techniques and use them in simulations.  The third method, the
encounter rate technique, does not self-consistently simulate the evolution of
star clusters but can be uses a static cluster model combined with individual
scattering experiments to estimate the effect of dense stellar systems on a
specific population.  In the next three sections we outline these techniques.
We then conclude the chapter by summarizing some of the most recent results
about compact binary populations in star clusters derived using these methods.

\subsubsection{Direct $N$-body Integration}
\label{sec:directNbody}

Direct $N$-body simulations are the most conceptually straightforward method
of simulating star clusters.  Each star is explicitly represented by a
massive (non-test) particle and the gravitational interactions are calculated
by numerically integrating the $3N$ coupled differential equations of motion
in classical Newtonian gravity:
\begin{equation}
  \label{eq:Newton}
  \ddot{\vec{r}}_{i} = - \sum_{i \ne j} G \frac{ \left( m_{j} \vec{r}_{i} -
      \vec{r}_{j} \right) }{ | \vec{r}_{j} - \vec{r}_{i} |^{3}}
\end{equation}
Here $m_{j}$ is the mass of particle $j$ and $\vec{r}_{i,j}$ are the positions
of particles $i$ and $j$ respectively.  A great deal of general information on
direct $N$-body methods can be found in the review article by
Spurzem~\cite{spurzem99} and Aarseth's book on $N$-body simulations
\cite{aarseth03}.

There are currently two major families of codes used for direct $N$-body
simulations: the \texttt{Kira}/STARLAB/AMUSE environment and the NBODY$X$
series of codes.  \texttt{Kira} integrates the $N$-body equations of motion
using a 4$^{th}$-order Hermite predictor-corrector scheme
\cite{MakinoAarseth92}.  \texttt{Kira} is available as part of the star
cluster evolution code STARLAB \cite{starlabweb}, a code that also includes a
stellar and binary evolution package, \texttt{SeBa}, based on the analytic
tracks of Hurley et al.~\cite{hurley00, HurleyEtAl02}, and various other
routines for generating initial conditions and dealing with tidal processes.
\texttt{Kira} is also available in the more recent star cluster simulation
environment AMUSE (Astrophysical Multipurpose Software Environment)
\cite{amuseweb}.  AMUSE is designed to be highly modular and allows the user a
choice between various integrators, stellar evolution routines, and
prescriptions for dealing with stellar collisions.  The NBODY$X$ series of
codes are a long-standing feature of the stellar dynamics community and have
been in constant development by Aarseth since the late 1960s \cite{NBODYweb}.
These codes employ the same basic 4$^{th}$-order Hermite predictor-corrector
scheme as \texttt{Kira} and include (in some versions) stellar evolution based
on the same Hurley et al. analytic tracks.  The NBODY$X$ codes, however, treat
binary evolution rather differently and are highly optimized for computational
efficiency.  Aarseth has produced two excellent reviews of the general
properties and development of the NBODY$X$ codes~\cite{aarseth99a, aarseth99b}
and includes further details in his book on $N$-body
simulations~\cite{aarseth03}.  A fairly complete summary of common $N$-body
codes and related applications can also be found on the NEMO
website~\cite{nemoweb}.

The primary advantage of direct $N$-body simulations is the small number of
simplifying assumptions made in treating the dynamical interactions.  All
gravitational interactions are explicitly integrated so the direct $N$-body
method can, in principle, resolve all of the microphysics involved in the
dynamical evolution of globular clusters.  In this sense direct $N$-body
simulations represent star clusters ``as they really are''.  Furthermore,
since all trajectories involved in any interaction are known, it is easy to
study each interaction in detail.  Finally, because direct $N$-body
simulations produce a star-by-star models of globular clusters, it is
relatively straightforward to couple them with additional stellar physics such
as stellar evolution, stellar equations of state, stellar collision models and
so on.  Therefore, within the limits of numerical error, the results of direct
$N$-body simulations are very reliable and they represent the gold standard in
globular cluster modelling.

Direct $N$-body simulations are, however, very computationally expensive.
Over a relaxation time, the computational cost of a direct $N$-body simulation
intrinsically scales as $N^{3-4}$ \cite{heggie03}.  Two factors of $N$ come
from calculating the pairwise interactions, another factor comes from the
increasing number of interactions per crossing time as $N$ increases, and a
final factor of $N/\log{N}$ come from the increasing length of $\trlx$
compared to the crossing timescale (see Equation~\ref{eq:Trlx}).  Another
problem is the very large range of scales present in globular clusters
\cite{hut93,heggie03}.  Distances, for instance, can range from a few tens of
kilometers (neutron star binaries) to several tens of parsecs (the size of the
cluster itself).  The timescales can have an even larger range, from several
milliseconds (millisecond pulsar binaries) to several Gigayears (the
relaxation times of very large globular clusters).  In the most extreme cases
it may be necessary to know positions to 1 part in $10^{14}$ while a cluster
may need to be integrated forward on timescales $10^{20}$ times shorter than
its lifetime.  Furthermore, $N$-body systems are known to be chaotic
\cite{Miller64,Miller66,goodman93}.  This means that the integration schemes
must be very accurate (4$^{th}$-order or better) and, consequently, they too are
very computationally expensive.  Obviously for large $N$ these simulations
rapidly become numerically intractable!

There are several methods that can be used to mitigate these problems.  The
Hermite scheme, for instance, requires only the first and second derivatives of
acceleration, even though it is a 4$^{th}$-order method.  Since dynamical
evolution proceeds more quickly in regions of high stellar density, the stars
in cluster halos do not need to have their phase space parameters updated as
frequently as those in cluster cores.  Both STARLAB and NBODY$X$ include
individual adaptive timesteps to ensure that during each force calculation,
only the necessary particles have their parameters updated.  Some of the
NBODY$X$ codes also include a neighbour scheme so that not all pairwise forces
need to be explicitly calculated at each timestep.  Binaries tight enough that
their internal evolution is largely unaffected by external forces from the
cluster can also be removed from the global dynamical evolution and replaced
with a single centre-of-mass particle.  The evolution of the binary itself can
then be followed using an efficient special-purpose code.  STARLAB calculates
the evolution of binaries using a semi-analytic Keplarian two-body code that
admits perturbations from the cluster environment.  The NBODY$X$ codes use a
regularization technique to achieve the same effect with the additional
advantage of removing the singularity created by very close binaries.  Both of
these methods avoid the global timestep of the star cluster being reduced to
that of its tightest binary.

Attempts have been made to create parallel direct $N$-body codes with
NBODY6++~\cite{spurzem99} being the most successful example.  A major problem
with this type of parallization is that each particle in an $N$-body
calculation formally requires information about every other particle for each
force calculation.  This means that parallel $N$-body codes are communication
dominated and do not scale well with the number of processors.  Another
approach has been to use hardware acceleration.  The GRAPE (GRavity PipE),
invented by Makino~\cite{makino98}, is a special-purpose computer designed to
rapidly calculate the $1/r^{2}$ force necessary for direct $N$-body
simulations.  The current generation, the GRAPE-6, was reported to have a peak
speed of 64 T-flops in 2002 \cite{grapeweb}.  A single GRAPE 6 board, the
GRAPE-6A, is also available as PCI card.  The next generation GRAPE, the
GRAPE-DR, is predicted to have a peak speed of $\sim 2$
P-flops~\cite{Makino07}.  It is not yet clear when (or if) the GRAPE-DR will
become commercially available.  Another very promising possibility is hardware
acceleration using GPUs (Graphical Processing Units)~\cite{SpurzemEtAl11}.
GPUs provide a similar service to the GRAPE while having the advantages of
being available off-the-shelf and commercially supported.  AMUSE, NBODY6 and
NBODY6++ have been modified to make use of the GPU.

Despite developments in both hardware and software, current direct $N$-body
simulations are practically relegated to $N \lesssim 10^{5}$, about an order of
magnitude lower than the number of stars in a typical globular cluster.
Furthermore, such simulations can take several months to run, making large
parameter studies impossible.  For this reason more approximate methods
employing distribution functions are still widely used in stellar dynamics.

\subsubsection{Distribution Function Methods}
\label{sec:FokkerPlanck}

Some of the computational problems associated with  direct $N$-body systems
can be avoided by describing globular clusters in terms of distribution
functions rather than as ensembles of particles.  The distribution function
for stars of mass $m$ at time $t$, $f_{m}(\vec{x},\vec{v},t)$, is
$6N$-dimensional in position-velocity phase space.  The number of stars in the
phase space volume $d^{3}x \, d^{3}v$ is given by $dN = f_{m}d^{3}x \,
d^{3}v$.  This description requires that either the phase space volume $d^{3}x
\, d^{3}v$ is small enough to be infinitesimal yet large enough to be
statistically meaningful or, more commonly, that $f_{m}$ be interpreted as the
probability of finding a star of mass $m$ in the volume $d^{3}x \, d^{3}v$ at
time $t$.  The time evolution of the cluster is modelled by calculating the
time evolution of $f_{m}$.  The effect of gravity is divided into two
components.  The first is a smoothed potential, $\phi$, that describes the 
effect of the overall gravitational field of the cluster.  $\phi(\vec{x})$ is
found by solving the Poisson equation, which can be written in terms of the
distribution function as:
\begin{equation}
  \label{eq:FPpoisson}
  \nabla^{2}\phi = 4\pi G \sum_{i} \left[ m_{i} \int
    f_{m_{i}}(\vec{x},\vec{v},t) d^{3}\vec{v} \right]
\end{equation}
The second component is a collision term, $\Gamma[f]$, that can be though of
as the effect of random close encounters between individual stars.  In
this picture $\phi$ represents the smooth background gravitational field of
the cluster while $\Gamma[f]$ represents the granularity induced by the
mass concentration in individual stars.  The time evolution of $f_{m}$ is
calculated using a modified form of the Boltzmann Equation: 
\begin{equation}
  \label{eq:FokkerPlanck}
  \frac{\partial f_{m}}{\partial t} + \vec{v} \cdot \nabla f_{m} -
  \nabla\phi \cdot \frac{\partial f_{m}}{\partial \vec{v}} = \Gamma[f].
\end{equation}
In stellar dynamics Equation~\ref{eq:FokkerPlanck} is usually called the
Fokker-Planck equation.

Equation~\ref{eq:FokkerPlanck} can be directly numerically solved if a
specific expression for $\Gamma[f]$ can be found.  One common choice is to
consider a cluster in the weak scattering limit where distant two-body
interactions drive dynamical evolution.  This is essentially equivalent to
assuming the cluster evolves only by two-body relaxation as described in
Section~\ref{sec:GCDynamics}.  In this case $\Gamma[f]$ can be described
by an expansion in the phase space coordinates ($\vec{x},\vec{v} \in \vec{w})$
and takes on the form \cite{binney87}:
\begin{equation}
  \label{eq:FPcoll}
  \Gamma[f] = - \sum_{i=1}^{6} \frac{\partial}{\partial w_{i}} \left[
    f(\vec{w}) \langle \Delta w_{i} \rangle \right] + \frac{1}{2}
  \sum_{i,j=1}^{6} \frac{\partial^{2}}{\partial w_{i} \partial w_{j}} \left[
    f(\vec{w}) \langle \Delta w_{i} \Delta w_{j} \rangle \right]
\end{equation}
where $\langle \Delta X \rangle$ is the expectation value of the phase space
variable over some time $\Delta t$.  In practise it is not efficient to solve
Equations~\ref{eq:FokkerPlanck} and~\ref{eq:FPcoll} in Cartesian coordinates,
however a careful choice of $\Delta t$ can alleviate this problem.  The key
observation is that there are two relevant timescales for evolution in the
Fokker-Planck approximation: the crossing time, $\tcr$, that governs changes
in position, and the relaxation time, $\trlx$, that governs changes in
energy.  In the case where $\tcr \ll \trlx$ changes in position are
essentially periodic and at any given time the orbit of a star can be written
in terms of its energy, $E$, and the magnitude of its angular momentum vector,
$J$.  By making a careful choice of $\Delta t$ and suitable assumptions about
the symmetry of the potential and the velocity dispersion
\cite{cohn79,spitzer87} it is possible to describe the evolution of the
phase space structure of the cluster solely in terms of the orbital
parameters and Equations~\ref{eq:FokkerPlanck} and~\ref{eq:FPcoll} can be
written in terms of $E$ and $J$ alone.  In this representation
Equation~\ref{eq:FokkerPlanck} is called the orbit-averaged Fokker-Planck
equation.

Numerically solving the orbit-averaged Fokker-Planck equation is much less
computationally intensive than direct $N$-body simulations and thus can be used
to model full-sized globular clusters.  Since these simulations are much
faster (taking hours or days rather than months) they are also useful for
performing parameter space studies.  It is also possible, in principle, to
improve the accuracy of the Fokker-Planck method by including higher-order
terms in Equation~\ref{eq:FPcoll} \cite{SchneiderEtAl11}.  The Fokker-Planck
method can be generalized to include velocity anisotropies, tidal stripping,
non-spherical potentials and cluster rotation
\cite{EinselSpurzem99,kim02,KimEtAl08}.

The Fokker-Planck method also has some serious limitations.  Each stellar mass
in the cluster requires its own distribution function in
Equation~\ref{eq:FokkerPlanck}.  For more than a few distribution functions
the Fokker-Planck method becomes numerically intractable so the ability of the
method to resolve a continuous spectrum of masses limited.  Since the
Fokker-Planck method uses a distribution function, it does not automatically
provide a star-by-star representation of a globular cluster.  This means that
it can be difficult to compare the results of a Fokker-Planck simulation with
observations.  It is also difficult to include stellar evolution in
Fokker-Planck simulations because there are no individual stars to evolve.
For the same reason Fokker-Planck methods cannot easily model strong few-body
interactions \cite{gao91}, interactions that are essential for describing the
evolution of the compact binary fraction in star clusters.  Therefore
Fokker-Planck methods are of limited  use in studying binary populations in
star clusters.

Another approach to solving Equation~\ref{eq:FokkerPlanck} that avoids some
of the problems  of the Fokker-Planck approach is the Monte Carlo method.
This method was first developed by H\'e{}non~\cite{henon71a, henon71b,
  henon75} and significantly improved by
Stod\'o{}\l{}kiewicz~\cite{Stod82,Stod86}.  It does not use the distribution
function explicitly but rather represents the cluster as a collection of
particles just as would a direct $N$-body simulation.  Each particle is
characterized by a mass, a distance from the cluster centre, and a tangential
and radial velocity.  The underlying treatment of relaxation is the same as in
the Fokker-Planck approach but is calculated on a particle-by-particle basis.
At the beginning of a Monte Carlo timestep the code calculates the global
potential based on the current mass and radius of each particle.  This
potential is then used to generate a plane-rosette orbit for each particle,
defined by $E$ and $J$ (the orbit-averaged approximation).  Next each particle
has its orbit perturbed by a weak encounter, the parameters of which are
calculated between it an a neighbouring particle.  The key to the method is
that assuming weak scattering, this perturbation is statistically
representative of all weak encounters the particle will experience over a time
$\Delta t$.  Thus the results of the perturbation can be multiplied by an
appropriately chosen factor to represent the total perturbation the particle
receives over $\Delta t$.  Once the new orbits have been found, new positions
can be randomly chosen from a time-weighted average along the orbital path, a
new potential calculated, and the process can start again.

The Monte Carlo method is mathematically identical to the Fokker-Planck method
but, since each star is individually represented, it is much easier to couple
with additional physics.  Each star in the simulation can have a different
mass with no loss of efficiency so the method can resolve a continuous mass
spectrum.  Stellar evolution can be included in exactly the same way as a
direct $N$-body code.  It is also relatively easy to include strong few-body
interactions.  For each star or binary at each timestep a cross-section for a
strong interaction with another star or binary can be calculated and the
encounter resolved using either analytic prescriptions or explicit few-body
integration.  Thus the Monte Carlo method has many of the advantages of the
direct $N$-body method while being much faster.  Since there are a fixed
number of operations per particle, the Monte Carlo method scales almost
directly with $N$ rather than the $N^{3-4}$ dependence  of a direct $N$-body
code.  Thus a simulation of $N \sim 10^{6}$ can be completed in a matter of
days, rather than the months  required by the direct $N$-body method.  Monte
Carlo codes have been shown to re-produce observations \cite{HeggieGiersz08,
  GierszHeggie09, GierszEtAl08} and the results of direct $N$-body codes
\cite{Giersz06, Fregeau08, ChatterjeeEtAl10} quite well, although some
discrepancies remain \cite{TrentiEtAl07}.

There are several Monte Carlo codes currently available, in particular that
of Freitag~\cite{freitag01}, that of the Northwestern Group \cite{joshi00,
  joshi01, fregeau03, ChatterjeeEtAl10}, and that of Giersz~\cite{giersz98,
  Giersz01, Giersz06, GierszEtAl08}.  The Freitag code is focused on the
simulation of star clusters with a central massive black hole but the
Northwestern and Giersz codes are both focused on general cluster dynamics and
are roughly comparable in capabilities.

The Monte Carlo method does have some shortcomings.  It is limited by the weak
scattering approximation and is not always reliable in the core region where
interactions can be very strong.  Furthermore, the method can only be as good
as the assumptions that go into the treatment of few-body interactions.  In
order to solve the gravitational potential at each timestep efficiently, it is
necessary to assume that the cluster is spherically symmetric, also a
significant limitation.  Nonetheless, Monte Carlo methods have proved to be
quite effective for efficiently modelling star clusters.

Another option for solving equation~\ref{eq:FokkerPlanck} is to use an analogy
between between a globular cluster and a self-gravitating gaseous sphere
\cite{louis91, giersz94}.  These models are solved by taking moments of the
Boltzmann equation and, like the Fokker-Planck method, can have their accuracy
increased by including higher-order moments \cite{SchneiderEtAl11}.  These are
still continuum models and have difficulty dealing with strong few-body
interactions and stellar evolution.  A solution to this is the so-called
``hybrid'' code where single stars are treated by the gaseous moment equations
while strong few-body encounters are treated using a Monte Carlo method
\cite{giersz00, giersz03}.  So far this method have not been used to examine
compact binary populations in star clusters.

\subsubsection{Encounter Rate Techniques}
\label{sec:EncRate}

Encounter rate techniques are another method that has proven very useful for
understanding the effect of dense stellar environments on binary populations.
These methods do not deal with the global evolution of globular clusters but
rather deal with the interactions between an environment and a particular
stellar species.  Normally a stellar density distribution is established for
a globular cluster (or its core) and then the number of interactions per unit
time is calculated for a specific object using cross-sections.  The density
distribution, interactions and outcomes can be calculated either analytically
or numerically, depending on the complexity of the system in question.  These
methods are limited but are very useful for estimating which processes will be
the most significant for specific stellar populations.  There is no ``general
method'' for encounter rate techniques but we will discuss some specific
implementations in section~\ref{sec:CurrentResults}.

%%%%%%%%%%%%%%%%%%%%%%%%%%%%%%%%%%%%%%%%%%%%%%%%%%%%%%%%%%%%%%%%%%%%%%%%%%%%%%%
%%%%%%%%%%%%%%%%%%%%%%%%%%%%%%%%%%%%%%%%%%%%%%%%%%%%%%%%%%%%%%%%%%%%%%%%%%%%%%%

\subsection{Results from Models of Globular Clusters}
\label{sec:CurrentResults}

Over the past two decades there has been extensive work modeling the dynamical
evolution of globular clusters.  Published studies have employed encounter
rate techniques in static cluster backgrounds, Fokker-Planck and Monte Carlo
models and direct N-body simulations (e.g. \cite{benacquista99,
benacquista01c, miller02, pfahl02a, popov02, rappaport01, rasio00a,
schneider01, shara02, sigurdsson95, takahashi00, davies95a, davies95b,
davies97, davies98, Lee87, giersz03, Giersz06, GierszEtAl08, joshi00, joshi01,
fregeau03, ivanova05a, FregeauEtAl09a, FregeauEtAl09b, ChatterjeeEtAl10,
PortegiesZwartEtAl97, portegieszwart01, portegieszwart04a, hurley03,
HurleyEtAl05}).  These models can, if properly interpreted, provide a wealth
of information about stellar populations in globular clusters.  These models,
however, have limitations that must be understood.  The dynamical
approximations employed coupled with an imperfect understanding of stellar and
binary evolution make it impossible to compare the results of these
simulations to observations.  The problem is exacerbated by the fact that many
authors are more concerned with the development of numerical methods than on
producing astrophysical results.  Furthermore, few of these studies focus on
(and some do not even model) the compact binary population.  Those that do
often focus on the effect that the compact population will have on the cluster
dynamics rather than the effect the cluster dynamics will have on the compact
binary population.  Nevertheless many published simulations provide useful
insights into the behaviour of compact binary populations in dense stellar
systems.  Here we report on these results for binaries where the primary is a
WD, NS or BH, trying to give both a historical overview as well as details of
the latest results.  When reading past papers, attention must be paid to
terminology.  A neutron star binary, for instance, can imply a binary composed
of two neutron stars or simply a binary where a neutron star is paired with
any type of secondary depending on the authors preference.  Here we attempt to
clearly indicate the nature of the secondary in order to resolve any
ambiguities.

\subsubsection{Binaries with White Dwarf Primaries}
\label{sec:WDbinaries}
Two kinds of binaries where the white dwarf is the most massive degenerate
member have attracted particular attention from modelers: cataclysmic
variables (CVs) and double white dwarf binaries (DWDs).  As described in
Chapter~\ref{subsection:cataclysmic_variables}, CVs consist of white dwarfs
accreting matter from a companion, normally either a dwarf star or another
white dwarf.  CVs with a dwarf star companion are not, strictly speaking,
relativistic binaries as defined in the introduction.  Unlike many
double-degenerate binaries, however, CVs can be observed in the
electromagnetic spectrum and they are very useful for understanding the
behaviour of compact objects in globular clusters.  CVs with a helium white
dwarf companion are know as AM CVn systems and are a type of DWD.  DWDs, which
as the name suggests are binaries of two white dwarfs, are relativistic
binaries but not all are AM CVns.  The merger of DWDs can, however, be
observed as a SNe Ia and thus they too have attracted some attention from
globular cluster modelers.

White dwarfs are the lowest-mass compact stellar remnants and are not
necessarily significantly more massive than the main sequence turn-off of
present-day globular clusters.  Because of this they may not experience as
significant mass segregation as higher-mass stellar remnants and may not be
so clearly preferred for exchange interactions.  As such, modeling is crucial
to understanding their behaviour in globular clusters.

Lee (1987) \cite{Lee87} is an example of an early numerical study of the compact
binary population in globular clusters, focusing on the effect of the binaries
on the cluster dynamics.  It is based on Fokker-Planck models using two
different stellar mass components to represent main sequence stars and compact
objects.  It demonstrated that compact objects have a strong effect on the
cluster core, implying they experience many interactions and highlighting the
importance of the cluster environment for compact binaries.

One of the early, specific studies of CVs in globular clusters is found in Di
Stefano and Rappaport (1994) \cite{DiStefanoRappaport94} who performed a large
series of experiments using an encounter-rate technique.  Specifically, they
performed numerical two-body scattering experiments in static cluster
backgrounds coupled with binary stellar evolution prescriptions in order to
investigate the tidal capture and subsequent evolution of CVs.  The cluster
parameters were chosen to match the current states of 47~Tuc and $\omega$~Cen.
Since these clusters are frequently compared in studies of compact binaries,
some of their properties are given in Table~\ref{tab:47TucWCen} (more can be
found in the Harris Catalogue of globular clusters \cite{harris96}).  The
primary differences between these clusters are that $\omega$~Cen is
significantly more massive that 47~Tuc while 47~Tuc is much more concentrated
and dynamically older than $\omega$~Cen.

% ==========
\begin{table}
  \centering
  \begin{tabular}[6]{c | r r r r r}
    \hline
    \hline
    Cluster & $M_{tot}$ [M$_{\odot}$] & $\rho_{0}$ [M$_{\odot}$/pc] & $r_{c}$
    [pc] & $r_{t}$ [pc] & $t_{rh}$ [Gyr] \\
    \hline
    $\omega$ Cen & $3.34 \times 10^{6}$ & $4.0 \times 10^{3}$ & 3.4 & 74.0 &
    16.0 \\
    47 Tuc & $1.06 \times 10^{6}$ & $5.1 \times 10^{4}$ & 0.59 & 69.0 & 8.7 \\
    \hline
    \hline
  \end{tabular}
  \caption{Structural parameters for 47~Tuc and $\omega$~Cen taken from Davies
    and Benz \cite{davies95a}.  $M_{tot}$ is the total mass, $\rho_{0}$ the
    central density, $r_{c}$ the core radius, $r_{t}$ the tidal radius and
    $t_{rh}$ the current estimated half-mass relaxation
    time.\label{tab:47TucWCen}}
\end{table}
% ==========

Di Stefano and Rappaport report $150-200$ CVs in 47 Tuc, $\sim 45$ of which
have an accretion luminosity of $> 10^{33}$ erg s$^{-1}$, and are thus
visible, at the current epoch.  For $\omega$ Cen they predict $100-150$ and
$\sim 20$ respectively.  They claim a factor of $\lesssim 10$ enhancement in
the number of CVs over the field population.  The authors point out both that
this number of CVs was barely within the upper limits on the number of CVs in
clusters known observationally at the time and that the number could be even
higher when binary-single and binary-binary interactions were taken into
account.  Observations of CVs in globular clusters were in their infancy at
the time. However the over-production of CVs would remain a problem in later
works.

The next studies to touch on CVs in globular clusters were a series of papers
by Davies and collaborators in 1995-1997 \cite{davies95a, davies95b, davies97}
concerned with exotic binaries produced by dynamical interactions.  These were
also encounter rate models but included binary-single interactions as well as
tidal capture.  Davies and collaborators \cite{davies95a, davies95b} reported
many single-single interactions between white dwarfs and main sequence stars
but found that the majority of strong encounters will lead to a white dwarf
surrounded by the envelope of a disrupted star rather than a CV.  Overall
Davies predict more CVs in 47 Tuc (smaller, concentrated, dynamically older)
than than in $\omega$ Cen (larger, less concentrated, dynamically younger) but
more merged systems than CVs in both.  Davies cautions, however, that merging
may be over-estimated during tidal capture.  They also predict a reasonable
number of DWD mergers in both clusters, most of which are above a
Chandrasekhar mass and will thus lead to SNe Ia but they spend little time on
this population.

In 1997 Davies \cite{davies97} produced another encounter rate study
concentrating on the effect a cluster environment would have on a population
of binaries that would become CVs due to binary stellar evolution in
isolation.  This study observed that disruption of binaries in a dense stellar
environment is important and can reduce the number of CVs compared to the
field.  Davies found that most CV progenitors in cluster cores with number
densities $> 10^{3}$ are ionized during dynamical interactions.  Thus, if 
dynamical interactions form no new CVs, the cores of galactic globular
clusters will be depleted in CVs.  CV progenitors in the low-density halos of
globular clusters may survive to become CVs if the relaxation time of the
cluster is long enough that they do not mass-segregate to the core.  Davies
proposed that comparing the CV population in the cores and halos of globular
clusters could be used as a probe of their dynamical properties.  A cluster
with a core depleted in CVs compared to the halo will not have experienced
much dynamical binary formation while a cluster with a large number of CVs in
the core must have experienced dynamical binary formation at some point in its
history.

The first detailed study of DWDs in dense stellar environments became available
with the direct N-Body simulations of Shara and Hurley (2002)~\cite{shara02}.
They were specifically interested in the galactic rate of SNe Ia and if
globular clusters could significantly enhance it.  They used the NBODY4 code
to perform four 22 000 particle simulations with 10\% primordial binaries, two
different metallicities and detailed stellar evolution prescriptions.  These
simulations are in the open rather than globular cluster regime and were
terminated after only 4.5 Gyr when they had lost more than 75\% of their
initial mass.  They were run on 16 chip GRAPE-6 boards and took approximately
five days each to complete.

Shara and Hurley found a factor of 2-3 enhancement in the number of DWDs in a
clustered environment compared to a similar field population.  Exchange
interactions seem to be particularly important since of the 135 DWDs produced
between all four simulations, 93 are formed by exchanges.  The enhancement in
SNe Ia progenitors, DWDs with a total mass above a Chandrasekhar mass and a
gravitational wave inspiral timescale shorter than a Hubble time, is even
greater.  They find $\sim$4 SNe Ia progenitors per 2000 binaries, a factor of
$\sim$15 more than in the field.  Both exchange interactions and hardening
during fly-bys are responsible for the enhancement.  Any DWD that went through
a common envelope phase were assumed to be circularized and they found that
later encounters did not induce significant eccentricity in the binaries.
Thus they concluded that most DWDs in globular clusters will be circular.
They find that neither normal CVs nor AM CVns are significantly enhanced over
the field population however those in the clusters are often the result of
exchange interactions.  Thus while star clusters will not necessarily enhance
the number of AM CVns, their properties may be different from the AM CVns in
the field.  This also implies that most CVs in star clusters are the product
of dynamics and that pure binary stellar evolution channels for CV and AM CVn
creation are shut down in clustered environments.

Shara and Hurley (2006)~\cite{SharaHurley06} continued this work using a
single 100 000 simulation with 5000 primordial binaries.  This simulation took
six months to run to an age of 20 Gyr using a 32-chip GRAPE-6 board,
demonstrating the strong $N$-dependence of computation time in direct N-body
simulations.  The focus of this simulation was on normal CVs, 15 of which
were produced during the course of the simulation and three of which would be
visible at the current epoch.  This illustrates the limitation of direct
N-body methods.  After six months of computing time the authors had only 15
objects to analyze.  Further simulations, either for verification purposes or
to explore parameter space, are simply too computationally expensive.  By
contrast the encounter rate techniques were able to produce many simulations
over a large parameter space and produce a much larger number of binaries for
better statistics.  Nonetheless the much more accurate N-body simulations are
very important for verifying the approximate results.  Shara and Hurley found
that dynamical effects pushed several of the progenitors to the active CV
state in a shorter time that would have occurred in the field.  Indeed one of
these binaries would have merged without becoming a CV if not for an orbital
perturbation.  Furthermore the simulation produced two binaries formed by
exchange interactions that clearly have no analogs in the field.  The exchange
CVs are, on average, more massive than CVs in the field.  They confirm that
many CV progenitors are destroyed by dynamical interactions and that a field
population of stars with similar properties would produce 17 CVs over 20 Gyr, 
nine of which would be visible at the current epoch.  Thus Shara and Hurley
predict $\sim 2-3$ times \emph{less} CVs in globular clusters than in the
field.

This simulation, although primarily focused on CVs, includes some discussion
of the DWD population and generally confirms the picture presented in Shara
and Hurley 2002.  In particular short-period ($P \le 1$ day, comparable to the
largest period of SNe Ia progenitors in Shara and Hurley 2002) DWDs are
enhanced by a factor of 5-6 over the field population.  Although this is
somewhat less than the factor of $\sim$ 15 previously reported it is still
much larger than any possible enhancement in CVs.  Shara and Hurley propose
several reasons for more efficient DWD production in star clusters.  In a DWD
both components must be sufficiently massive to become white dwarfs within a
Hubble time while only one does in a standard CV.  Thus CVs are a much more
common outcome of pure binary stellar evolution than are DWDs.  DWDs are also,
on average, more massive than CVs and thus tend to mass-segregate further
towards the core and undergo more interactions. Finally, since each component
of a DWD is likely to be more massive than the main sequence star in a CV and
since exchanges tend to place massive objects in binaries, a DWD is a more
likely outcome of an exchange interaction between a main  sequence star and
two white dwarfs than is a CV.

One of the most complete projects on the compact binary population in the
recent literature is found in a series of papers by Ivanova and collaborators
\cite{rasio00a, ivanova05a, ivanova05b, IvanovaEtAl06, ivanova08,
  IvanovaEtAl10}.  These papers simulate the binary -- and particularly the
compact binary -- population of star clusters using a combination of Monte
Carlo simulations and well-developed encounter rate techniques coupled with
stellar and binary evolution algorithms tailored towards compact objects
\cite{hurley00, HurleyEtAl02, BelczynskiEtAl08}.  Ivanova et al. (2005a)
\cite{ivanova05a} consider the effect of cluster dynamics on the initial
binary fraction of globular clusters and show that dynamical disruption will
significantly reduce the binary fraction of for reasonable initial conditions.
Indeed they show that a globular cluster where 100\% of stars are initially
binaries will normally have a core binary fraction of only $\sim$10\% after
several Gyrs.  This implies that globular clusters may have had high binary
fractions when young and thus binary-single and binary-binary interactions may
be more important than previously thought.  They also demonstrate both the
binary burning that supports the cluster against core collapse and a build-up
of short-period white dwarf binaries in the core as shown in
Figure~\ref{ivanova47tuc}.

% ==========
\epubtkImage{ivanova47tuc.png}{
  \begin{figure}[htbp]
    \def\epsfsize#1#2{0.5#1}
    \centerline{\epsfbox{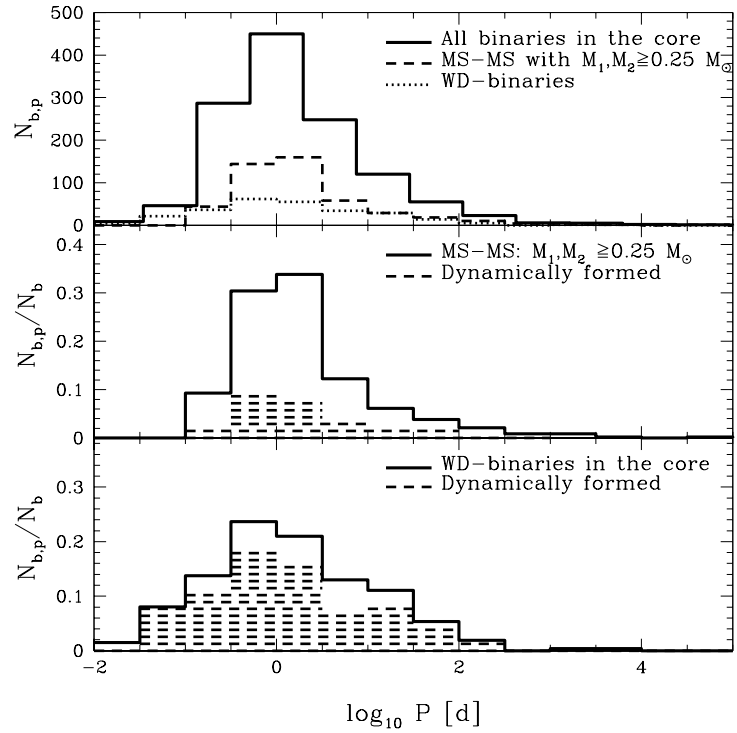}}
    \caption{Binary period distributions from the Monte Carlo
      simulation of binary fraction evolution in 47~Tuc. The bottom
      panel indicates the period distribution for binaries containing
      at least one white dwarf. $N_\mathrm{b}$ is the total number of
      binaries and $N_\mathrm{b,p}$ is the number of binaries per
      bin. Figure taken from Ivanova et al.~\cite{ivanova05a}.}
    \label{ivanova47tuc}
  \end{figure}}
% =========

The models were improved by Ivanova et al. (2006) \cite{IvanovaEtAl06} in order
to investigate the white dwarf binary population in globular clusters.  A more
sophisticated, time-evolving two-zone model was used as a cluster background.
Scattering encounters were chosen using a Monte Carlo method and numerical
few-body integration for single-single, binary-single and binary-binary
interactions was used to determine their outcome.  They confirm that CVs are
not highly over-produced in globular clusters relative to the field population
but find that CV production is moderately efficient in cluster cores.  In the
cores they predict an enhancement over the field population of a factor of
2-3.  As with Shara and Hurley, the globular cluster CVs have different
properties than field CVs.  In particular the globular cluster CVs are
slightly brighter and have a larger X-ray-to-optical flux.  Ivanova et
al. attribute this to slightly higher-mass white dwarfs in the cluster CVs, as
observed by Shara and Hurley.  The more massive white dwarfs lead to the
globular cluster CVs extracting more energy at the same mass transfer rate as
field CVs and having slightly higher magnetic fields. This would account for
the observed X-ray excess.

Ivanova et al. can shed more light on the dynamical processes that form CVs
than previous encounter rate studies since they include most of the likely
formation mechanisms in the same simulation.  Their results show that some
60-70\% of CVs in the core  form because of dynamical interactions while most
of the CVs in the halo originate from primordial binaries.  Overall they find
that only $\sim$25\% of CVs formed in the clusters would have become CVs in
the field.  They conclude  that tidal capture of main sequence stars is
largely unimportant for CVs but that collisions between red giants and main
sequence stars -- a situation where the envelope of the red giant can stripped
away leaving its core as white dwarf bound to the main sequence star -- can
provide up to 15\% the total number of CVs in the clusters.  Exchange
interactions are the most efficient channel, forming $\sim$ 32\% of all CVs
while collisions during binary-single or binary-binary encounters produce
another $\sim$ 13\%.  The remaining CVs are produced by primordial binaries
but $\sim$ 20\% of these experienced hardening encounters in fly-by
interactions and would not have become CVs in the field within a Hubble time.
Finally, Ivanova et al. claim that 60 \% of CVs did not form solely by common
envelope evolution, the most common channel for field CV production and thus
highlight the importance of cluster dynamics for the CV population.
Ultimately Ivanova et al. predict 1 detectable CV per 1000-2000 M$_{\odot}$ of
mass in Galactic globular clusters.  They specifically predict 35-40 CVs in 47
Tuc, of the same order as the 22 observed there at the time.  The discrepancy
could be due to stochasticity, some inherent deficiencies in the modeling or
because not all CVs in 47 Tuc have been observed. Observations of globular
clusters in the 2000s, particularly in the UV \cite{HeinkeEtAl05, dieball07,
  MaccaroneKnigge07} have indicated a larger number of CVs than previously
thought.  The combination of fewer CVs in current models and a larger number
of detected CVs has gone a long way towards solving the ``CV problem'' in
globular clusters.

Unlike Shara and Hurley, Ivanova et al. find little evidence for a strong
enhancement in the number of DWDs.  This could be ```real'' in the sense that
their simulations model larger systems than Shara and Hurely's work and thus
could be taken as evidence that while DWDs are enhanced in open clusters, they
are not in globular clusters.  It could also be a result of the different
numerical methods.  The two-zone model of Ivanova et al., for example, may not
be able to represent mass segregation as accurately as the direct N-body
models of Shara and Hurley and thus the effect of interactions in the dense
core could be underestimated.  It is also possible that the encounter
prescriptions could be missing some details that are important for DWD
creation.  More controlled comparisons of the two methods are necessary in
order to definitively state which prediction is more accurate.  Ivanova et
al. do, however, find a modest enhancement in the number of SNe Ia progenitors
over the field population.  They predict that this will not be important for
understanding the SNe Ia from Milky Way-type galaxies since only a very small
fraction of their masses is in globular clusters.  Elliptical galaxies,
however, tend to have a larger fraction of their mass in globular clusters and
in these galaxies the contribution of globular clusters to SNe Ia rate could
be important.  Ivanova et al. point out that DWD mergers and accretion induced
collapse of white dwarfs in binaries also produce neutron stars and may do so
without imparting a velocity kick.  If natal kicks are large enough to remove
most core-collapse neutron stars, merged white dwarfs could form a significant
fraction of the current neutron star population in globular clusters.

% ==========
\epubtkImage{willems07.png}{
  \begin{figure}[htbp]
    \def\epsfsize#1#2{0.8#1}
    \centerline{\epsfbox{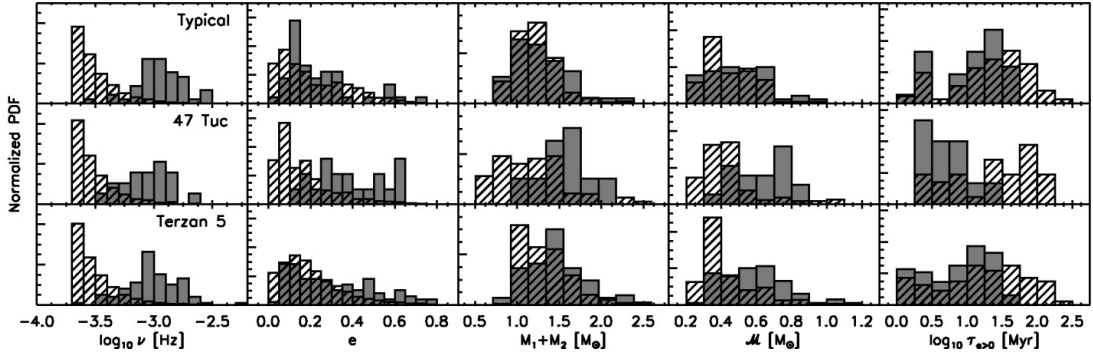}}
    \caption{Statistical properties of eccentric DWDs ($e>0.01$) in the LISA
      band ($P<5000$s) for three different cluster models.  Hatched histograms
      are for the entire eccentric DWD population while shaded histograms are
      fore the DWDs with at least two harmonics individually resolvable by
      LISA with an S/N $>$ 8 (5 year mission, 10 kpc distance).  The PDFs are
      normalized to unity.  From left to right: frequency, eccentricity, total
      mass, mass ratio and lifetime before merger.  Plot taken from Willems et
      al. 2007~\cite{willems07}.}
    \label{fig:willems07}
  \end{figure}}
% ==========

Willems et al. 2007 \cite{willems07} performed an extended analysis of the
simulations of Ivanova et al. 2006 in order to determine the number of
eccentric DWDs that may be in the low frequency gravitational wave band
accessible to space-based gravitational wave detectors, such as the Laser
Interferometer Space Antenna (LISA).  Taking the DWD eccentricities from the
10$^{\rm th}$ to 13$^{\rm th}$ Gyr of the Ivanova et al. simulations, Willems
et al. generate the probability distribution for eccentric DWDs in Galactic
globular clusters shown in Figure~\ref{fig:willems07}.  Unlike Shara and
Hurley, these results indicate that there will be a small but non-negligible
population of eccentric DWDs in the Galactic globular cluster system (see
Table~\ref{tab:willems07}).  They will have no analogs in the field since
field DWDs are expected to be circularized by common envelope  evolution.
Thus a detection of an eccentric DWD by LISA is both probable and a clear
indication of a dynamical origin.

% ==========
\begin{table}
  \centering
  \begin{tabular}[5]{l r r r r}
    \hline
    \hline
    GC Model & $\langle N \rangle$ & $P_{1}$ & $P_{2}$ & $P_{3}$ \\
    \hline
    Typical & 0.39 & 0.32 & 0.06 & 0.01 \\
    47 Tuc & 1.90 & 0.84 & 0.55 & 0.29 \\
    Terzan 5 & 1.60 & 0.80 & 0.48 & 0.22 \\
    \hline
    \hline
  \end{tabular}
  \caption{$\langle N \rangle$ is the mean number of eccentric binaries per
    cluster and $P_{N}$ is the Poisson probability that at least $N$ eccentric
    DWDs are present at any given time.  Data taken from Willems et
    al. 2007~\cite{willems07}.}
  \label{tab:willems07}
\end{table}
% ==========

The consensus that emerges from the simulations is that dynamical destruction
of potential CVs and dynamical formation of new CVs are equally important in
globular clusters.  Therefore the number of CVs per unit mass is unlikely to
be much higher in globular clusters than it is in the field.  Globular cluster
CVs may, however, be slightly more massive than field CVs and thus have
slightly higher X-ray fluxes.  Overall current simulations seem to be able to
re-produce the observed number of CVs in clusters reasonably well but still
tend to slightly over-predict the population.  Whether this is due to a
failing in the simulations or observational limitations is not yet clear.

It seems likely that the DWD population in globular clusters is larger per
unit stellar mass than that in the field however it is not clear if the
enhancement is by factors of ten or only factors of a few.  The disagreement
between various simulations could be due to differing numerical methods or
because scaling the DWD population from open to globular clusters is not
straightforward.  Systematic comparison of the different simulation methods
will be necessary in order to resolve this unambiguously.

\subsubsection{Binaries with Neutron Star Primaries}
\label{sec:NSbinaries}
Neutron star binaries were the first binaries with degenerate companions to be
specifically investigated in globular clusters.  LMXBs, neutron stars
accreting from a companion as described in Chapter~\ref{subsection:lmxbs}, have
attracted particular interest since results from 1975 \cite{Katz75,Clark75}
gave indications that they are over-abundant in globular clusters.
These papers immediately suggested this was a result of dynamical encounters
and by the end of the 1970s a debate over which type of encounter, whether
tidal capture by MS stars \cite{fabian75}, collisions with RG stars
\cite{Sutantyo75}, exchange interactions \cite{Hills76} or three-body
interactions \cite{Shull79}, was already underway.  Like CVs, LMXBs are not
necessarily relativistic by our definition but are electromagnetically active
and thus useful for calibration purposes.  The ultracompact systems discussed
in Chapter~\ref{subsection:lmxbs}, often called UCXBs, are almost certainly
double-degenerate and may consist of a neutron star accreting from a white
dwarf.

Neutron stars are, on average, more massive than the main sequence turn-off of
old globular clusters and also more massive than white dwarfs.  As such they
should both mass-segregate to cluster cores and be preferentially exchanged
into binaries during interactions.  Thus globular clusters may produce a large
number of double neutron star binaries (DNSs), the mergers of which will be
excellent sources of gravitational radiation for ground-based detectors and
may also be a source of short gamma-ray bursts~\cite{GrindlayEtAl06}.  Neutron
star production in globular clusters is, however, by no means certain.
Neutron stars are rare and thus primordial DNSs are not common.  Furthermore
large natal kicks can both break up primordial DNSs and remove neutron stars
from globular clusters.  Also, neutron stars are only a factor of two to three
times more massive than the average cluster mass, leaving the extent of their
mass segregation and their probability of exchange unclear.

Neutron stars are low-luminosity and, if not members of an X-ray binary, are
normally detected as MSPs.  Their presence was long-expected in globular
clusters due to the large population of LMXBs that are thought to be MSP
progenitors but  were not confirmed until 1987 \cite{LyneEtAl87}.  Since all
MSPs are thought to need a spin-up phase during accretion in a binary system,
even single MSPs can be used as a probe of binary evolution.  As discussed in
Chapter~\ref{subsection:millisecond_pulsars}, these are now known to be quite
common in globular clusters.  Of more interest for gravitational wave
detection are the many binaries containing MSPs and particularly the double
MSP binaries (DMSPs) that have been detected in some globular clusters.

Originally tidal capture or collisions were the favoured method for producing
X-ray binaries.  When the LMXB enhancement was first discovered, globular
clusters were thought to have very few primordial binaries and thus there
would be little opportunity for exchange interactions.  Since three-body
interactions tend to produce soft binaries they were largely ignored.  During
the 1980s studies employing static cluster models and encounter rate estimates
based either on analytic estimates (e.g. \cite{HutVerbunt83, Verbunt87,
  Bailyn88b, VerbuntMeylan88}) or few-body scattering experiments
(e.g. \cite{Krolik84,KrolikEtAl84}) were able to roughly re-produce the X-ray
binary population in globular clusters using only tidal capture and
collisions. They also indicated that hardening encounters might be important
in bringing dynamically formed binaries into the mass-transferring, and thus
X-ray, regime.  Verbunt and Meylan \cite{VerbuntMeylan88} showed that mass
segregation has a significant effect on the production of CVs and X-ray
binaries, an early indication that globular cluster structure needs to be
treated in some detail rather than as a homogeneous static background.

As stellar modeling improved, the central role of tidal capture began to be
called into question.  Several authors demonstrated that close encounters
between MS or RG stars lead to mergers rather than bound pairs
\cite{RayEtAl87,McMillanEtAl90,BailynEtAl90} and thus will not produce X-ray
binaries.  It was also realized that, since dynamical interactions in globular
cluster cores can destroy binaries \cite{ivanova05a}, a low binary fraction in
old Milky Way globular clusters did not necessarily imply a low primordial
binary fraction.  Thus, particularly in young clusters, interactions between
stars and binaries are likely to be important.

One of the earliest systematic attempts to simulate double-degenerate binaries
in globular clusters was the Rappaport et al. (1989) \cite{RappaportEtAl89}
study of the effect of a dense stellar environment on a wide binary containing
an MSP with a low-mass white dwarf companion.  The study utilized an encounter
rate technique where a large number of numerical binary-single scattering
experiments were conducted in a static cluster background with parameters
chosen to represent $\omega$~Cen and 47~Tuc.  They found that binary-single
interactions were capable of introducing significant eccentricity into some of
the binaries, enough to effect their gravitational wave inspiral timescales.
It also appeared that disruption interactions were unlikely, at least for
binaries with $P \le 300$ days so hard double-degenerate binaries should be
able to survive in globular cluster cores.  The results were marginally
consistent with the available observations but could not be fully reconciled
with predicted eccentricities as a function of age as estimated by spin-down
observations.

Sigurdsson and Phinney \cite{SigurdssonPhinney93,sigurdsson95} conducted
similar scattering experiments to those performed by Rappaport et
al. \cite{RappaportEtAl89} but with tighter binaries and a much extended mass
spectrum.  They discovered that exchanges were even more important than
previously thought, and indeed could be the dominant type of interaction
between massive stars and moderately hard binaries.  A secondary effect of
exchange is that when a light star is swapped for a massive one, the binary
will have a larger semi-major axis for the same binding energy, which
increases the cross section of the binary to undergo further interactions.
They also found that three-body interactions with very hard binaries tended to
produce a merger between two of the members and accretion onto the neutron
star during such an event could plausibly create a MSP.  Indeed Sigurdsson and
Phinney argued that such interactions are sufficient to explain the entire MSP
population of globular clusters.  They also showed that a small number of low
mass DMSPs can be formed in this way.

Sigurdsson and Hernquist \cite{SigurdssonHernquist92} considered the role of
binary-binary interactions in forming DMSPs.  They proposed a mechanism where
two neutron star-main sequence binaries interacted and both main sequence
stars  were disrupted.  The result was an DNS surrounded by an envelope that
could be accreted to spin up both members to produce a DMSP.  The study
employed both point-mass scattering experiments and an SPH representation of
the main sequence stars, when appropriate.  The model appeared plausible but
has not factored into later studies and would be a rather rare, if interesting,
outcome of binary-binary interactions.

Davies and collaborators \cite{davies95a, davies95b} specifically investigated
the population of LMXBs in globular cluster in the same simulations described
in Chapter~\ref{sec:WDbinaries}.  Davies and Benz (1995) \cite{davies95a}
considered the evolution of 1000 binaries injected into various cluster cores
modelled as King models with a concentration parameter $W_{0}$ between 3 and
12 -- a range that covers the concentrations of galactic globular clusters.
In this study they found that the number of LMXBs produced depends on the IMF,
assumptions about the how easily the envelope can be ejected during common
envelope evolution, assumptions about stellar evolution and the concentration
of the cluster.  Any of these parameters can affect some types of binaries by
factors of several.

Davies (1995) \cite{davies95b} concentrated on the blue straggler and compact
binary population of 47~Tuc and $\omega$~Cen.  Particular interest was shown
in comparing the efficiency of single-single and binary-single encounters in
producing various kinds of binaries.  Here the effect of different cluster
models became clear.  For the model of $\omega$~Cen (massive, low
concentration) blue stragglers are produced significantly more efficiently by
binary-single than by single-single interactions.  By contrast in 47~Tuc
(lower mass, higher concentration) binary-single and single-single
interactions are of comparable significance in blue straggler formation.  In
both systems binaries containing neutron stars seem to be more likely to be
formed by single-single interactions with an main sequence star rather than 
binary-single interactions.  Like the CVs, however, binaries so-formed, tend
to lead to disruption of the main sequence star and a ``smothering'' of the
neutron star with the main sequence star's envelope.  This may lead to the
formation of a Thorne-Zytkow object.  The neutron star will eventually accrete
the matter which could lead to black hole production (if enough matter is
accreted) or an MSP but will not lead to an X-ray binary.  A smothered system
can also be the result of a binary-single interactions but is a less common.
For the most optimistic set of parameters Davies predicts a 4:1 ratio of
smothered to non-smothered neutron star binaries in $\omega$~Cen and a 5:1
ratio in 47 Tuc.

Davies predicted that $\omega$~Cen would form more X-ray binaries than in 47
Tuc since the IMF in $\omega$~Cen model was skewed towards higher masses and
thus produced a larger number of neutron stars.  By contrast more CVs were
predicted in 47~Tuc.  Both clusters were predicted to produce more CVs than
LMXBs which was difficult to reconcile with observations at the time.  The
number of predicted LMXBs was consistent with contemporary observations of
47~Tuc and $\omega$~Cen (the possibility of one or two in each at the current
epoch).

A new model for neutron star-white dwarf binaries was introduced by Rasio et
al. (2000) \cite{rasio00a} who proposed that short-period MSP-white dwarf
binaries can be formed by exchanges with primordial binaries followed by a
common envelope phase.  In this model a $\sim$1-3 M$_{\odot}$ main sequence
star in a binary acquires an neutron star companion through an exchange
interaction and forms a new binary at a separation of 0.1-1.0 AU.  The main
sequence star undergoes Roche lobe overflow and has its envelope stripped by
the neutron star.  Spin-up during the accretion phase leaves an MSP in a close
orbit with a white dwarf.  Rasio et al. use Monte Carlo scattering simulations
to predict the MSP-white dwarf binary population produced by this scenario and
compare it to the population of short-period MSP-white dwarf binaries observed
in 47 Tuc.  The results, given in Figure~\ref{WD-NS_in_47Tuc}, show very good
agreement with the observations and confirm the importance of interactions for
the compact binary population in globular clusters.

% =========
\epubtkImage{WDNSin47Tuc.png}{
  \begin{figure}[htbp]
    \def\epsfsize#1#2{0.5#1}
    \centerline{\epsfbox{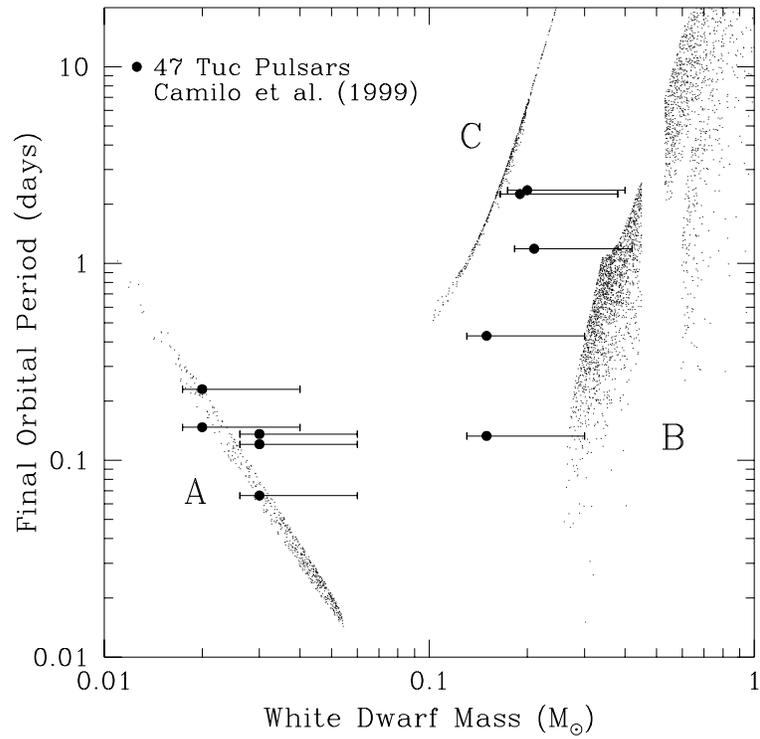}}
    \caption{Results of the Monte Carlo simulation of neutron star--white
      dwarf binary generation and evolution in 47~Tuc. Each small dot
      represents a binary system. The circles and error bars are the 10 binary
      pulsars in 47~Tuc with well measured orbits. Systems in A have evolved
      through mass transfer from the white dwarf to the neutron star. Systems
      in B have not yet evolved through gravitational radiation to begin RLOF
      from the white dwarf to the neutron star. Systems in C will not undergo
      a common envelope phase. Figure taken from Rasio et
      al.~\cite{rasio00a}.}
    \label{WD-NS_in_47Tuc}
  \end{figure}}
% ==========

The tail end of the systems in group B of Figure~\ref{WD-NS_in_47Tuc}
represents the neutron star--white dwarf binaries that are in very short
period orbits and are undergoing a slow inspiral due to gravitational
radiation. These few binaries can be used to make an order of magnitude
estimate on the population of such objects in the galactic globular cluster
system. If we consider that there are two binaries with orbital period less
than 2000~s out of $\sim 10^6\,M_{\odot}$ in 47~Tuc, and assume that this rate
is consistent throughout the globular cluster system as a whole, we find a
total of $\sim$~60 such binaries.  This is a crude estimate and may be
optimistic considering the results of Ivanova et al. \cite{IvanovaEtAl06,
  ivanova08} but is an example of how simulations can be extended to produce
gravitational wave detection estimates.

Ivanova et al. (2005b) \cite{ivanova05b} Re-introduced the plausibility of a
collision between a red giant and a neutron star forming an eccentric neutron
star-white dwarf UCXB.  The result is based on hydrodynamical simulations
\cite{LombardiEtAl06} that show, while a common envelope phase is likely after
such a collision, much of the envelope is ejected to infinity and an eccentric
UCXB is a common outcome of the interaction.  Ivanova et al. use the results
from 40 SPH simulations and simple, analytic cluster modeling to estimate that
it is possible for all UCXBs in globular clusters to be formed by this
mechanism.

Ivanova et al. (2008) \cite{ivanova08} extended the results of the CV study
described in Chapter~\ref{sec:WDbinaries} to cover neutron star binaries as
well.  The dynamical situation for neutron star binaries slightly different
than for CVs.  About half of the neutron star binaries in the core formed as
the result of an exchange interaction with only a few percent more formed by
either tidal capture or collision.  The rest of the binaries are primordial.
Very few neutron star binaries will become X-ray binaries and Ivanova et
al. estimate there is only a $\sim$3\% chance that any given neutron star will
ever be in an LMXB.  Furthermore, most primordial binaries that will become
X-ray binaries will do so quite early in the life of the cluster and will have
ceased to be luminous in the current epoch.  Thus Ivanova et al. predict that
X-ray binaries observed in Galactic globular clusters today are likely to be
products of exchange, tidal capture, or collisions.  Although exchange
interactions produce most of the dynamically formed neutron star binaries,
actual X-ray binaries will be produced in equal numbers by all three channels.
This is due to each channel having a different probability of producing a
mass-transferring system.  Finally Ivanova et al. predict $\sim$7.5 UCXBs in
the entire Galactic globular cluster system and $\sim$ 180 qLMXB systems.
They also make specific predictions that certain clusters should have at least
one LMXB and all of these predictions are met.

In addition to X-ray binaries, Ivanova et al. 2008 \cite{ivanova08} also
consider the population of MSPs and double neutron star binaries.  Overall
they have difficulty reconciling the number of X-ray binaries and MSPs in
their simulations with the observed ratios.  If they assume that all
mass-gaining events lead to MSPs then the number produced is too high unless
the detection probability for MSPs in clusters is only $\sim$ 10\%.  They
speculate that either not all accretion leads to spin up, that the kicks for
evolution-induced collapse  (EIC) supernovae are larger than they model, or
their model for mass segregation does not work properly for these systems and
more should sink to the core and be disrupted.  They find that by adjusting
their assumptions about spin-up during various kinds of accretion they can
bring the number of MSPs down to a level that is consistent with observations
of 47~Tuc and Terzan~5.  They also note that the fraction of single MSPs to
MSPs in binaries tends to increase with increasing cluster density.  This
implies that destructive interactions become important for MSP binaries at
high density.

They find that the production of double neutron star binaries is very
inefficient.  There are only three primordial neutron stars in the $\sim 2
\times 10^{7}$ M$_{\odot}$ of stars that make up their entire simulation set
while only 14 double neutron stars are formed by dynamical interactions.
Given their chosen IMF and cluster ages, this implies that it takes $\sim$ 500
single NSs to produce one merging double neutron star binary in 11 Gyrs.
These results are very density dependent with the vast majority of the DNSs
forming in the highest density simulations.  They estimate that merging DNSs
in globular clusters could provide at most 10-30\% of short gamma ray bursts
but note that this agrees fairly well with the predictions of Grindlay et
al. (2006) \cite{GrindlayEtAl06}.

This result agrees with results of Downing et al. (2010) \cite{DowningEtAl10}
who use a similar-sized Monte Carlo dataset, primarily to study black
hole-black hole binaries, and find no DNSs in any of their simulations.  That
DNSs are not produced in similar numbers to X-ray binaries in globular
clusters may be a combination of the lower number of potential progenitors,
kicks for both progenitors, and the likelihood of merging or producing a WD
during interactions.

Ivanova et al. also report a significant number of LISA sources in their
simulations.  They predict that there should be an average of 10 LISA sources
per globular cluster at any given instant and on average 180 LISA sources are
produced per globular cluster per Gyr.  Scaling this to the Milky Way, they
find that there may be up to 1500 LISA binaries, both DWD and white
dwarf-neutron star, in the Galactic globular cluster system at the current
time.  They caution, however, that they may have overestimated the white
dwarf-neutron star binary population since it is not fully consistent with
UCXB observations.

Banerjee and Ghosh \cite{BanerjeeGhosh08a, BanerjeeGhosh08b} have
employed a novel method to investigate the compact binary population of
globular clusters.  They used the collisional Boltzmann equation as given by
Spitzer \cite{spitzer87} to calculate the evolution of the number distribution
of binaries in a uniform, non-evolving background of stars.  They did not
employ the Fokker-Planck approximation but rather considered mean encounter
rates with a randomly fluctuating component.  This is called a Wiener process
and is a mathematical description of Brownian motion.  The Wiener process
means that  the encounter rates, normally ordinary differential equations in
the Boltzmann equation, now become stochastic differential equations and must
be treated with a mathematical tool known as It\=o{} calculus.  Both Wiener
processes and It\=o{} calculus are described in Banerjee and Ghosh 2008b
\cite{BanerjeeGhosh08b}.

Banerjee and Ghosh specifically consider a bimodal population of 0.6
M$_{\odot}$ main sequence stars interacting with 1.4 M$_{\odot}$ neutron
stars.  They consider binary formation by tidal capture and exchange,
destruction by 3-body disruption of the binary or by exchange of the compact
component for a main sequence star, and hardening by magnetic breaking and
gravitational radiation.  They are able to re-produce the build-up of
short-period binaries found by other authors but, more interestingly they are
able to compare the mean rates predicted by the Boltzmann equation to the
effects of stochasticity in the Wiener process.  Thus they are able to
indicate if it is possible to predict the number of X-ray sources expected in
a given globular cluster or if stochastic effects will dominate.
Figure~\ref{BanerjeeGhosh8} shows both the mean number of X-ray binaries and
the stochastic fluctuation expected for various cluster parameters as well as
observed numbers from several Galactic globular clusters.  Although the number
of X-ray binaries in the models and observations follow the same trend, the
stochastic fluctuations are large and indicate it is probably impossible to
predict exact numbers of compact binaries in any given globular cluster. Thus,
although the method of Banerjee and Gosh is in some ways less detailed than
that of Ivanova et al., containing much more limited mass functions, a less
accurate treatment of individual encounters, and much more limited stellar
evolution, it is still valuable since it constrains the effects of
stochasticity in a controllable way.

% ==========
\epubtkImage{BanerjeeGhosh8.png}{
  \begin{figure}[htbp]
    \def\epsfsize#1#2{1.2#1}
    \centerline{\epsfbox{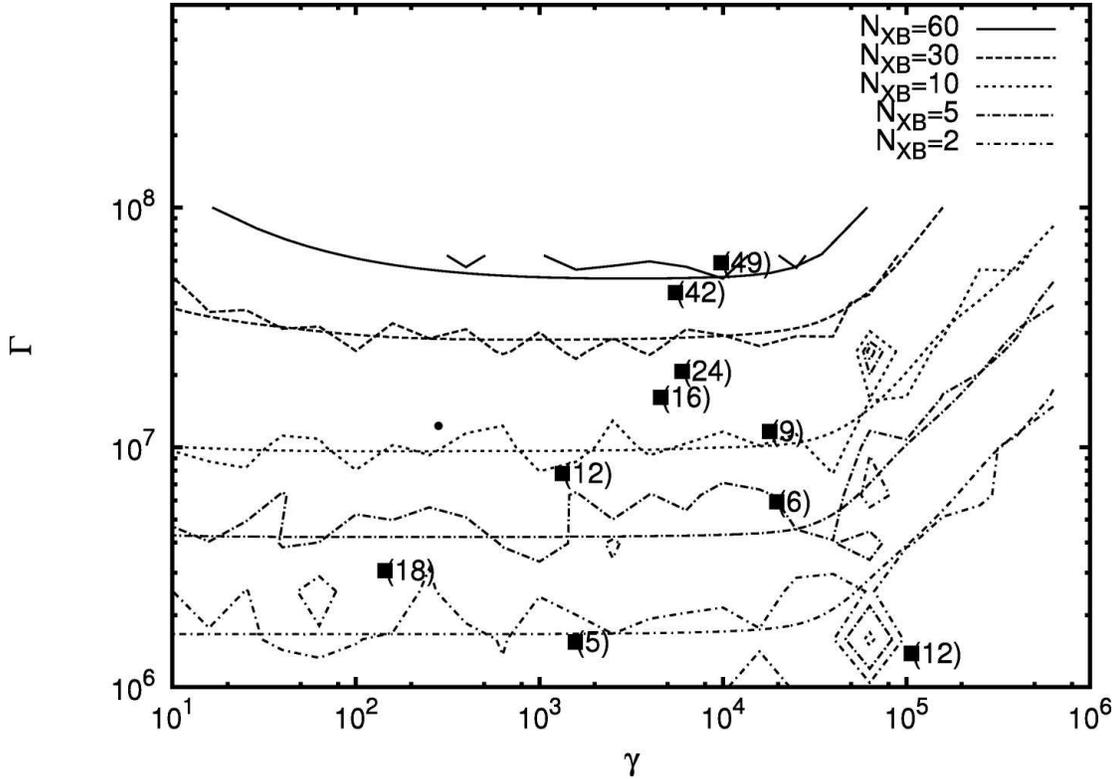}}
    \caption{Contours of constant number of X-ray binaries as a function of
      the ``Verbunt parameters''.  Here $\Gamma = \rho^{2}r_{c}^{3}/v_{c}$,
      $\gamma = \rho/v_{c}$, $\rho$ is the average stellar density, $r_{c}$ is
      the core radius and $v_{c}$ the core velocity dispersion.  $\Gamma$
      describes the two-body stellar encounter rate and $\gamma$ describes the
      rate of encounters between a single binary with the stellar background.
      The squares represent the location of various Galactic globular clusters
      in the $\gamma-\Gamma$ plane and the number in parentheses is the number
      of X-ray binaries observed in the cluster. Plot taken from Banerjee and
      Ghosh 2008b~\cite{BanerjeeGhosh08b}.}
    \label{BanerjeeGhosh8}
  \end{figure}}
% ==========

Recently Bagchi and Ray \cite{BagchiRay09a,BagchiRay09b} have used scattering
experiments to re-investigate the eccentricities of MSP binaries in light of
improved observations.  They noted that observationally there are three,
statistically distinct, eccentricity groupings in MSP binaries, those with $e
\sim 0$, those with $2 \times 10^{-6} < e \le 0.01$ and those with $0.01 \le
e$.  They then performed scattering experiments, utilizing the STARLAB
package, to attempt to determine the origin of the different groupings.  They
found that most of the higher eccentricity binaries could be explained by
exchanges or mergers during binary-single or binary-binary interactions.  The
intermediate eccentricities could be explained by fly-by encounters while the
lowest eccentricities were binaries that had been circularized without
experiencing fly-bys.  They also found one binary that they could only explain
by invoking resonant interactions within a hierarchical
triple~\cite{Kozai62}.

Of a more esoteric nature is the work of Trenti et al. 2008
\cite{TrentiEtAl08} considering the number triple systems that might contain
at least one pulsar.  They studied triple formation using direct N-body
simulations with between 512 and 19~661 bodies, both equal masses and with a
mass function, and various initial binary fractions.  They find that triple
systems are about two orders of magnitude less common than binary systems
within globular cluster cores but, given the number of MSP binaries currently
know in globular clusters, this implies that an MSP triple should be detected
soon.  They note, however, that triples are very difficult to observationally
confirm if the scales of the inner and outer systems are well-separated.  It
is also clear that if a triple with a single MSP will be so rare, a triple
system with more than one MSP is even more unlikely.

There are several overall conclusions that can be drawn about the population
of neutron star binaries in globular clusters.  Dynamical interactions can
certainly enhance the number of X-ray binaries in globular clusters but the
current consensus is that the enhancement will by factors of a few, not
factors of hundreds.  Recent simulations show that detailed descriptions of
dynamics, binary stellar evolution and stellar collisions are all necessary in
order for the results to be reliable.  The same interactions that produce
X-ray binaries are also effective at producing MSPs, although many of the
resulting binaries will be subsequently disrupted and the MSPs will continue
their lives as single objects rather than binaries.  It also seems that
neutron star-white dwarf binaries may be fairly common in globular clusters.
By contrast detailed simulations show that DNS binaries are \emph{not} common
in globular clusters since few exist primordially and they are not produced
efficiently by stellar dynamics.  The individual results of these studies are
highly dependent on the detailed treatment of various phases of binary stellar
evolution as well as details of the kick distribution.  It is worth noting
that many authors have successfully predicted the number of X-ray binaries in
globular clusters since the early 1980s.  On one hand this is shows that the
X-ray binary enhancement in globular clusters can be easily understood.  On
the other hand, it probably means that X-ray binaries are a rather generic
outcome of globular cluster dynamics.  Therefore observed numbers of X-ray
binaries probably cannot tell us much about which processes are important in
specific globular clusters.

\subsubsection{Binaries with Black Hole Primaries}
\label{sec:BHbinaries}
Black holes are non-luminous and can only be detected in the electromagnetic
spectrum if they are accreting matter, normally from a binary companion.
Black hole-black hole binaries (DBHs) can only be detected in gravitational
radiation.  So far there are no confirmed detections of black holes in
globular clusters.

Black holes are the most massive compact stellar remnants, up to $35-80 \Msun$
\cite{Fryer99, BelczynskiEtAl10}, and can, on average, be several times more
massive than the average mass of stars in even a fairly young globular
cluster.  Thus black holes are strongly affected by mass segregation, may form
Spitzer-unstable subsystems in the dense core of the cluster and will be
preferentially exchanged into binaries during interactions.  As such, the
black hole binary population in globular clusters will be particularly
strongly affected by dynamics.

DBHs are of great interest for gravitational wave detection since they are
quite massive compared to neutron star mergers and may be detectable by the
next generation of ground based detectors at Gigaparsec distances.  Strong
dynamical interactions should produce many hard DBHs in globular clusters.  It
is not clear, however, that globular clusters will a strong source of DBH
mergers since the binaries could be ejected from the cluster cores before they
reach the gravitational wave regime.  Sigurdsson and Hernquist (1993)
\cite{sigurdsson93} presented the first detailed study of black hole-black
hole binaries in globular clusters.  They used purely analytic estimates to
show that black holes would indeed mass segregate and form a Spitzer unstable
subsystem in the cluster core.  In this subsystem black holes interact with
each other to form binary and triple systems. Interactions between these
systems and the remaining black holes remove both from the cluster very
efficiently and Sigurdsson and Hernquist predict that there should be only
zero to four black holes left in the cores of old galactic globular clusters,
with perhaps a few more in the cluster halos.  Remarkably, this basic picture
of mass segregation, dynamical interaction, binary formation and ejection has
not changed much since Sigurdsson and Hernquist's first publication.

Studies of black hole binaries remained dormant until the direct N-body
simulations of Portegies Zwart and McMillan (2000) \cite{portegieszwart00b}.
Using small simulations of between 2048 and 4096 particles with a bimodal mass
distribution, 0.5-1\% of bodies ten times more massive than the rest, they
essentially confirmed the results of Sigurdsson and Hernquist.  They found
that the black holes mass segregate rapidly, form binaries through three-body
interactions, and are then ejected from the clusters.  Overall they found only
$\sim$8\% of black holes were retained by their parent clusters while
$\sim$61\% were ejected as singles and $\sim$31\% were ejected as binaries.
The found, however, that many of the ejected, dynamically-formed black
hole-black hole binaries were tight enough to merge in the galactic field
outside the boundaries of the clusters within a Hubble time.  Indeed,
extrapolating their results to globular clusters they estimated up to one
advanced LIGO detection of a dynamically-formed DBH binary
merger per day.  Thus Portegies Zwart and McMillan concluded that globular
cluster would be an important source of DBH binaries.  It
must be noted, however, that these simulations did not include stellar
evolution or natal kicks for black holes.

During the mid-2000s the focus shifted towards the possibility that, as
proposed by Miller and Hamilton (2002) \cite{miller02}, IMBHs could be formed
by repeated mergers of stellar-mass black holes in dense cluster cores.
G\"u{}ltekin et al. (2004) \cite{Gultekin04} performed numerical three-body
scattering experiments in a static cluster background, using the output of the
last scattering experiment as the input of the next, to follow the
merger-induced growth of black holes of various seed masses, initially in
binaries with $10 \Msun$ companions.  They used the static background to
calculate an encounter timescale between interactions and, if the quadrapole 
gravitational wave radiation timescale was less than the encounter timescale,
a merger occurred.  They found that it was not possible to create a merged
black hole with a mass $\gtrsim 240 \Msun$ for any reasonable 
seed mass.  Either the proto-IMBH was ejected or the cluster core was totally
depleted of black holes before the proto-IMBH could grow to a greater mass
(see Figure~\ref{fig:NBHsEjected}).  They confirmed, however, that many of the
ejected binaries would pass thorough the LISA band and then merge in the
VIRGO/LIGO band within a Hubble time.  They also found that many of the
binaries would retain significant eccentricity in the LISA band.

% =========
\epubtkImage{NBHsEjected.png}{
  \begin{figure}[htbp]
    \def\epsfsize#1#2{0.2#1}
    \centerline{\epsfbox{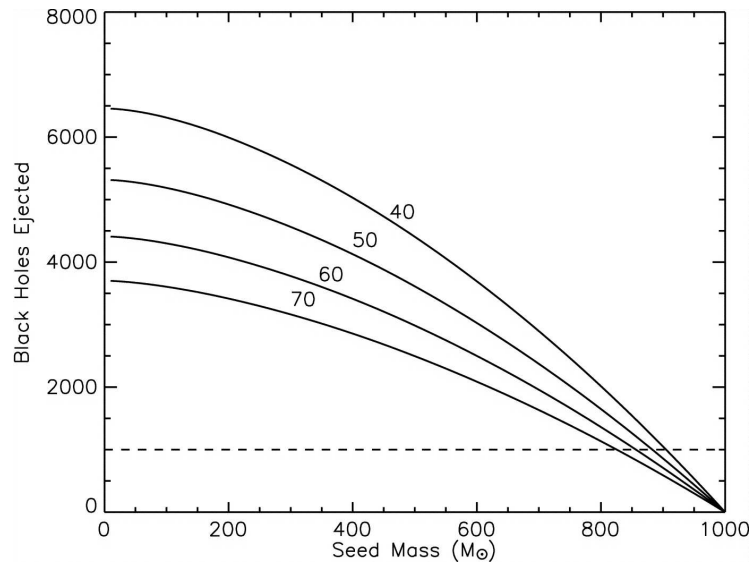}}
    \caption{Total number of black holes ejected in building up to a $1000
    \Msun$ IMBH as a function of seed mass.  The four numbered curves are made
    assuming different cluster escape velocities.  The dotted line is $10^{3}$
    black holes, the assumed population for a young cluster.  As can be seen,
    a very large seed mass is necessary before IMBH formation by merging black
    holes becomes a possibility.  Figure taken from \cite{Gultekin04}}
    \label{fig:NBHsEjected}
  \end{figure}}
% ==========

O'Leary et al. (2006) \cite{Oleary06} considered the same problem using a
slightly different approximation.  They assumed all black holes were
completely mass-segregated and formed a Spitzer-unstable subsystem interacting
only with itself.  The they followed the evolution of this subsystem, embedded
in a globular cluster potential, for a range of initial conditions using a
Monte Carlo selection of encounters and numerical integration of binary-single
and binary-binary interactions.  In some of their simulations they included a
prescription for gravitational wave recoil after a DBH merger.  Without recoil
they found the could build a merged black hole of a few hundred solar masses,
similar to G\"u{}ltekin et al. although slightly more massive.  With recoil,
however, the found that few, if any, merged products remained within the
cluster and essentially no black hole growth occurred at all.  Thus O'Leary et
al. also conclude that IMBH formation through merging stellar mass black holes
is impossible.

O'Leary et al. find, however, a large number of DBH binaries
are formed through various interactions, many of which will merge within a
Hubble time.  They quantified the detection rate for advanced LIGO using
simplified cosmological assumptions and assuming optimal orientation.  The
predict 10s of black hole merger detections per year, the exact number
depending strongly on the initial conditions of the cluster model, and find
that between 50\% and 70\% of these will take place in binaries that have been
ejected from the clusters.  Again, the clusters produce many black hole-black
hole binaries but do not retain them.

Both G\"u{}ltekin et al. and O'Leary et al. included only self-interacting
black hole systems in their simulations.  By contrast Sadowski et al. 2008
\cite{SadowskiEtAl08} modeled entire globular clusters using the same Monte
Carlo method as Ivanova et al. \cite{IvanovaEtAl06} and assumed that the black
holes remained in thermal equilibrium with the rest of the cluster.  This is,
in many ways, the opposite dynamical assumption to that made by O'Leary et
al. since it assumes that the black holes are minimally mass segregated.  It
maximizes the number of interactions between black holes and other cluster
stars while minimizing the number of interactions between black holes and
other black holes.  The main result of this is that there are fewer black
hole-black hole binaries formed but those that do form are far less likely to
experience a further encounter with another black hole.  Interactions with
lower-mass cluster members are less energetic and thus less likely to disrupt
or eject DBHs.  In practice the lower destruction and ejection rate wins out
and Sadowski et al. predict a very large number of DBH binaries in globular
clusters.  Indeed they estimate an advanced LIGO detection rate of 25-3000 per
year, 90\% of which will occur within the boundaries of the clusters.  When
compared with the results of O'Leary et al., this proves that details of the
treatment of global star cluster dynamics are very important.

Two recent sets of studies have tried to address the problem of more accurate
global dynamics.  The Monte Carlo simulations of Downing et al. 2010 and 2011
\cite{DowningEtAl10, DowningEtAl11} and the direct N-body simulations of
Banerjee et al. 2010 \cite{BanerjeeEtAl10}.  Downing et al. use the Monte
Carlo code of Giersz \cite{giersz98, Giersz06} to perform 160 fully
self-consistent simulations of globular clusters with 500 000 particles, a
realistic mass function, stellar evolution, 10-50\% primordial binaries, and
various combinations of initial concentration and metallicity.  The results of
the simulations tend to support the approximation of O'Leary et al., namely
the black holes mass segregate very quickly and form a dense system in the
core of the cluster.  Downing et al. find that many DBH binaries are formed,
particularly by three-body interactions, but many of these binaries are either
destroyed in subsequent interactions or are ejected from the cluster.  In
dense cluster DBH production is strongly time-dependent.  Many DBHs are formed
early but this number drops substantially after a few Gyrs as the clusters are
depleted of black holes.  By contrast the clusters with a low initial
concentration do not suffer as strong a depletion in their black hole
population and produce DBHs slowly but constantly throughout their lives.  The
number of black hole binaries produced is highly stochastic with large
simulation-to-simulation variations, even between models with the same initial
conditions, as shown in Table~\ref{tab:NBHBHgcs}.  Depending on the initial
conditions Downing et al. predict between 1-100 DBH merger
detections per year for advanced LIGO with an average of $\sim$15 per year.
This is consistent with, although slightly higher than, O'Leary et al.  The
detection rate is degenerate between some simulations with different initial
conditions and may be stochastic as well.  This means that the black
hole-black hole merger detection rate alone will probably not carry much
astrophysical information.  Downing et al. also predict the possibility of
eccentric DBH binaries in the LISA band.

% ==========
\begin{table}
  \centering 
  \begin{tabular}[c]{l r r r r}
    \hline
    \hline
    Simulation & $t = 3 t_{rh}$ & $t = 9 t_{rh}$ & $t = 25 t_{rh}$ & $t =
    14$ Gyr \\
    \hline
    10sol21  &   $1 \pm  1$ &        -     &        -     &   $1 \pm   1$ \\
    10sol37  &   $1 \pm  1$ &  $14 \pm 11$ &        -     &  $14 \pm  11$ \\
    10sol75  &   $0 \pm  0$ &   $8 \pm  6$ &  $49 \pm 19$ &  $52 \pm  19$ \\
    10sol180 &   $0 \pm  0$ &  $12 \pm  6$ &  $54 \pm 21$ & $123 \pm  27$ \\
    50sol21  &   $1 \pm  1$ &        -     &        -     &   $3 \pm   2$ \\
    50sol37  &   $3 \pm  2$ &  $36 \pm 10$ &        -     &  $50 \pm  11$ \\
    50sol75  &   $1 \pm  1$ &  $26 \pm  8$ & $115 \pm 24$ & $147 \pm  28$ \\
    50sol180 &   $0 \pm  1$ &  $11 \pm  5$ & $111 \pm 23$ & $354 \pm  33$ \\
    10low21  &  $22 \pm 10$ &        -     &        -     &  $27 \pm  10$ \\
    10low37  &  $23 \pm  4$ &  $44 \pm  6$ &        -     &  $44 \pm   6$ \\
    10low75  &  $18 \pm 10$ &  $32 \pm 13$ &  $54 \pm  6$ &  $54 \pm  16$ \\
    10low180 &  $26 \pm  8$ &  $51 \pm  8$ &  $79 \pm 20$ & $112 \pm  24$ \\
    50low21  & $104 \pm 16$ &        -     &        -     & $127 \pm  16$ \\
    50low37  &  $93 \pm 22$ & $175 \pm 26$ &        -     & $184 \pm  29$ \\
    50low75  &  $64 \pm 10$ & $155 \pm 22$ & $173 \pm 22$ & $202 \pm  38$ \\
    50low180 & $103 \pm 19$ & $205 \pm 35$ & $294 \pm 50$ & $453 \pm 109$ \\
    \hline
    \hline
  \end{tabular}
  \caption[BH-BH binaries]{The cumulative number of BH-BH binaries after $3$,
    $9$, and $25$ $t_{rh}$, and also after $1 T_{H}$.  The first column
    gives the initial conditions.  10 or 50 in the first place indicates
    10\% or 50\% primordial binaries.  sol or low indicates a metallicity of
    $z = 0.02$ or $z = 0.001$.  The number in the third place indicates
    $r_{t}/r_{h} =$ 21, 37, 75 or 180.  Each column is averaged over ten
    independent realizations and includes the $1\sigma$ uncertainty.  A dash
    in a column indicates that the cluster did not reach that number of
    half-mass relaxation times within one Hubble time.  Taken from Downing
    et al. 2010~\cite{DowningEtAl10}}
  \label{tab:NBHBHgcs}
\end{table}
% ==========

Banerjee et al. use NBODY6 to perform direct N-body simulations of young to
intermediate age massive clusters with between 5000 and 100 000 stars with
stellar evolution and a realistic IMF.  The also perform a series of 3000 to
4000 body simulations surrounded by a reflective boundary to model the core of
a globular cluster.  They find that the black holes mass-segregate quickly,
on timescales of 50-100 Myr, almost independently of the mass of the cluster.
Once mass-segregated the black holes interact strongly, making many black
hole-black hole binaries during three-body encounters.  Like most previous
authors, Banerjee et al. find that the black holes and black hole binaries are
efficiently disrupted or ejected and that their clusters are depleted of
black holes on a 4.5 Gyr timescale.  Thus they predict that current galactic
globular clusters contain no black holes.  They also estimate That
intermediate age massive clusters will produce $\sim$31 black hole binary
merger detections per year for advanced LIGO.  This is of the same order as
found by Downing et al. and slightly higher than the rate found by O'Leary et
al. while it is about an order of magnitude lower than that found in the
previous N-body simulations of Portegies Zwart and McMillan.  They caution,
however, that this rate is not for the old globular clusters that were the
focus of the previous studies and thus care must be taken in the comparison.

There have been recent efforts to include Post-Newtonian (PN) dynamics into
direct N-body simulations.  This has been done both for stellar mass black
holes \cite{KupiEtAl06} and for black holes around a central massive object
\cite{Aarseth07}.  The algorithm seems to work reasonably well but it is not
fully clear when a PN treatment is necessary.  For central massive objects
where there are many stars in the influence radius the two-body and cluster
space and time scales are not well-separated so relativistic dynamics may be
important for the global cluster evolution.  By contrast for stellar-mass
objects the two body and cluster scales are well-separated so it is probably
not important to include PN dynamics as part of the global simulation.

In 2010 Ivanova et al. \cite{IvanovaEtAl10} considered the possibility of
black hole-white dwarf X-ray binaries as UCXBs in globular clusters.  This
study employed scattering experiments but a rather more simplified treatment
of cluster dynamics and the interaction between binaries and the cluster.
Ivanova et al. determine that stellar collisions, exchange interactions and
three-body binary formation can all produce black hole-white dwarf X-ray
binaries, however in most cases the binary is too wide to become an X-ray
source within a Hubble time.  Thus they predict that most black hole-white
dwarf X-ray binaries must have experienced dynamical hardening, either through
fly-by encounters or as part of a hierarchical triple where the inner black
hole-white dwarf binary has its separation reduced by the Kozai mechanism
\cite{Kozai62}.  Each formation mechanism, coupled with dynamical hardening,
can explain the observed number of ULXBs in globular clusters but all require
that $\geq 1$\% of black holes remain in the cluster upon formation.  This
places limits on the possible natal kick distribution of black holes.

Unlike some of the other species mentioned, the basic behaviour of black
hole-black hole binaries in globular clusters is well-understood.  The black
holes mass segregate, interact, form binaries and are swiftly ejected.  The
focus of future investigations will be on event rates and details of mass
distributions.  It seems certain, however, that globular clusters will be a
major source of DBH binary mergers.  When compared with the results for field
DBH binaries of Belczynski et al. 2007 \cite{BelczynskiEtAl07} all of the
studies cited indicate that dynamically formed binaries will dominate the DBH
binary merger detection rate for advanced LIGO by factors of several.  Some
more recent results \cite{BelczynskiEtAl10b} indicate that the number of field
black hole binaries may be larger than previously estimated but dynamically
formed binaries are still expected to be a major source of detections.
Including younger clusters before they have been depleted in black holes may
be important for estimating event rates and will probably only cause them to
increase.  Should the next generation of ground-based detectors perform
according to expectations, detection of a DBH binary inspiral is almost
certain.

%%%%%%%%%%%%%%%%%%%%%%%%%%%%%%%%%%%%%%%%%%%%%%%%%%%%%%%%%%%%%%%%%%%%%%%%%%%%%%%
%%%%%%%%%%%%%%%%%%%%%%%%%%%%%%%%%%%%%%%%%%%%%%%%%%%%%%%%%%%%%%%%%%%%%%%%%%%%%%%

\subsection{Semi-empirical methods}
\label{sec:semi-empirical_methods}
The small number of observed relativistic binaries can be used to infer the
population of dark progenitor systems~\cite{benacquista01c}. For example, the
low-mass X-ray binary systems are bright enough that we see essentially all of
those that are in the galactic globular cluster system. If we assume that the
ultracompact ones originate from detached WD--NS systems, then we can estimate
the number of progenitor systems by looking at the time spent by the system in
both phases. Let $N_\mathrm{X}$ be the number of ultracompact LMXBs and
$T_\mathrm{X}$ be their typical lifetime. Also, let $N_\mathrm{det}$ be the
number of detached WD--NS systems that will evolve to become LMXBs, and
$T_\mathrm{det}$ be the time spent during the inspiral due to the emission of
gravitational radiation until the companion white dwarf fills its Roche
lobe. If the process is stationary, we must have
\begin{equation}
  \frac{N_\mathrm{X}}{T_\mathrm{X}} = \frac{N_\mathrm{det}}{T_\mathrm{det}}.
  \label{population_ratio}
\end{equation}
The time spent in the inspiral phase can be found from integrating the period evolution equation
\begin{equation}
\frac{dP}{dt} = -k_0P^{-5/3},
\end{equation}
to obtain
\begin{equation}
  T_\mathrm{det} = \frac{3}{8k_0}\left(P_0^{8/3}-P_\mathrm{c}^{8/3}\right),
  \label{detached_time}
\end{equation}
where 
\begin{equation}
k_0 = \frac{96}{5}\left(2\pi\right)^{8/3}\frac{G^{5/3}}{c^5}\frac{M_1M_2}{\left(M_1+M_2\right)^{1/3}},
\end{equation}
$P_0$ is the period at which the progenitor emerges from the common
envelope, and $P_\mathrm{c}$ is the period at which RLOF from the white dwarf
to the neutron star begins. Thus, the number of detached progenitors can be
estimated from
\begin{equation}
  N_\mathrm{det} = \frac{N_\mathrm{X}}{T_\mathrm{X}}
  \frac{3}{8k_0}\left(P_0^{8/3}-P_\mathrm{c}^{8/3}\right).
  \label{detached_number}
\end{equation}

There are six known ultracompact LMXBs with orbital periods
small enough to require a degenerate white dwarf companion to the neutron
star. There are four other LMXBs with unknown orbital periods. Thus, $4 \leq
N_\mathrm{X} \leq 10$. The lifetime $T_\mathrm{X}$ is rather uncertain,
depending upon the nature of the mass transfer and the timing when the mass
transfer would cease. A standard treatment of mass transfer driven by
gravitational radiation alone gives an upper bound of $T_\mathrm{X} \sim 10^9
\mathrm{\ yr}$~\cite{rappaport87}, but other effects such as tidal heating or
irradiation may shorten this to $T_\mathrm{X} \sim 10^7 \mathrm{\
  yr}$~\cite{applegate94, rasio00a}. The value of $P_0$ depends critically
upon the evolution of the neutron star--main-sequence binary, and is very
uncertain. Both $k_0$ and $P_\mathrm{c}$ depend upon the masses of the white
dwarf secondary and the neutron star primary. For a rough estimate, we take
the mass of the secondary to be a typical He white dwarf of mass
$0.4\,M_{\odot}$ and the mass of the primary to be $1.4\,M_{\odot}$. Rather
than estimate the typical period of emergence from the common envelope, we
arbitrarily choose $P_0 = 2000 \mathrm{\ s}$. We can be certain that all
progenitors have emerged from the common envelope by the time the orbital
period is this low. The value of $P_\mathrm{c}$ can be determined by using
the Roche lobe equation of Eggleton~\cite{eggleton83}:
\begin{equation}
R_{\rm L} = \frac{0.49q^{2/3}a}{0.6q^{2/3}+\ln{\left(1+q^{1/3}\right)}},
\end{equation}
and the radius of the white dwarf as
determined by Lynden-Bell and O'Dwyer~\cite{lyndenbell01}. Adopting the
optimistic values of $N_\mathrm{X} = 10$ and $T_\mathrm{X} = 10^7 \mathrm{\
  yr}$, and evaluating Equation~(\ref{detached_time}) gives $T_\mathrm{det}
\sim 10^7 \mathrm{\ yr}$. Thus, we find $N_\mathrm{det} \sim 1 \mbox{\,--\,}
10$, which is within an order of magnitude of the numbers found through
dynamical simulations and encounter rate
estimations.

Current production of ultracompact WD--NS binaries is more likely to arise
through collisions of neutron stars with lower mass red giant stars near the
current turn-off mass. The result of such a collision is a common envelope
that will quickly eject the envelope of the red giant and leave behind the
core in an eccentric orbit. The result of the eccentric orbit is to hasten the
inspiral of the degenerate core into the neutron star due to gravitational
radiation~\cite{peters63}. Consequently, $T_\mathrm{det}$ can be significantly
shorter~\cite{ivanova05b}. Adopting a value of $T_\mathrm{det} \sim 10^6$
gives $N_\mathrm{det} < 100$.

Continuing in the spirit of small number statistics, we note that there is one
known radio pulsar in a globular cluster NS--NS binary (B2127+11C) and about
50 known radio pulsars in the globular cluster system as a whole (although
this number may continue to grow)~\cite{lorimer01}. We may estimate that
NS--NS binaries make up roughly $1/50$ of the total number of neutron stars in
the globular cluster system. A lower limit on the number of neutron stars
comes from estimates of the total number of active radio pulsars in clusters,
giving $N_\mathrm{NS^2} \sim 10^5$~\cite{kulkarni90}. Thus, we can estimate
the total number of NS--NS binaries to be $\sim$~2000. Not all of these will
be in compact orbits, but we can again estimate the number of systems in
compact orbits by assuming that the systems gradually decay through
gravitational radiation and thus
\begin{equation}
  \frac{N_\mathrm{compact}}{N_\mathrm{NS^2}} =
  \frac{T_\mathrm{compact}}{T_\mathrm{coalesce}},
  \label{NS_binary_ratio}
\end{equation}
where $N_\mathrm{compact}$ is the number of systems in compact orbits
($P_\mathrm{orb} < 2000 \mathrm{\ s}$), $T_\mathrm{compact}$ is the time spent
as a compact system, and $T_\mathrm{coalesce}$ is the typical time for a
globular cluster NS--NS binary to coalesce due to gravitational radiation
inspiral. Adopting the coalescence time of B2127+11C as typical,
$T_\mathrm{coalesce} = 2 \times 10^8 \mathrm{\ yr}$~\cite{prince91}, and
integrating Equation~(\ref{detached_time}) for two $1.4\,M_{\odot}$ neutron
stars, we find $N_\mathrm{compact} \sim 25$. Again this value compares
favorably with the values found from encounter rate estimations.

%%%%%%%%%%%%%%%%%%%%%%%%%%%%%%%%%%%%%%%%%%%%%%%%%%%%%%%%%%%%%%%%%%%%%%%%%%%%%%%
%%%%%%%%%%%%%%%%%%%%%%%%%%%%%%%%%%%%%%%%%%%%%%%%%%%%%%%%%%%%%%%%%%%%%%%%%%%%%%%
%%%%%%%%%%%%%%%%%%%%%%%%%%%%%%%%%%%%%%%%%%%%%%%%%%%%%%%%%%%%%%%%%%%%%%%%%%%%%%%

\newpage

%%%%%%%%%%%%%%%%%%%%%%%%%%%%%%%%%%%%%%%%%%%%%%%%%%%%%%%%%%%%%%%%%%%%%%%%%%%%%%%
%%%%%%%%%%%%%%%%%%%%%%%%%%%%%%%%%%%%%%%%%%%%%%%%%%%%%%%%%%%%%%%%%%%%%%%%%%%%%%%
%%%%%%%%%%%%%%%%%%%%%%%%%%%%%%%%%%%%%%%%%%%%%%%%%%%%%%%%%%%%%%%%%%%%%%%%%%%%%%%

\section{Prospects of Gravitational Radiation}
\label{sec:grav_radiation}
In the coming decades, several gravitational wave observatories will begin
detecting signals from relativistic binaries. It is likely that several of
these binaries will have their origins within globular clusters. In this
section, we review the detectors and their expected sources from globular
clusters.

Ground-based interferometers such as the Laser Interferometer
Gravitational-wave Observatory (LIGO)~\cite{ligoweb} in the United States,
Virgo~\cite{virgoweb} in Europe, or the Large-scale Cryogenic Gravitational
wave Telescope (LCGT)~\cite{lcgtweb} in Japan, will be sensitive to
gravitational radiation in the frequency range from a few Hz to a few kHz. In
this frequency range, these detectors will be sensitive to the inspiral and
merger of binaries containing neutron stars or black holes. These systems are
somewhat rare in the Galactic globular cluster system, and the probability of
a coalescence occurring within our Galaxy while the detectors are operating is
vanishingly small. The enhanced design of LIGO was operational until the end
of 2010, but did not detect any known gravitational wave signal. The enhanced
design of Virgo will continue operating until 2012, but with the absence of
any other interferometers operating at similar sensitivities it is unlikely
that it will detect any known gravitational wave signals before shutting
down. Both Virgo and LIGO will be upgrading to advanced designs which will
provide significant increases in sensitivity. Like advanced LIGO (aLIGO) and
advanced Virgo (AdV), LCGT is a next-generation detector. It is unique in that
it will be in an underground facility with and it will operate
cryogenically. All of these detectors should be sensitive to double neutron
star mergers at distances of a few hundred Mpc, and double black hole mergers
at distances of about a Gpc. At these distances, the number of extragalactic
globular clusters is large enough that coalescences of neutron star or black
hole binaries may be commonplace. Unfortunately, the angular resolution of the
network of next-generation detectors will be much too coarse to identify any
detected signal with a globular cluster. However, the gravitational waveform
of the coalescence of these objects depends upon the orientation of the spins
of the components relative to the orbital angular momentum. Although it is not
entirely clear if field binaries will tend to have their spins aligned, there
is no reason for this to be the case with binaries that have been formed
dynamically through exchange of compact objects. Estimates of the event rates
of compact object binary coalescence during the advanced detector era tend to
weight the binary neutron star coalescences toward field binaries and the
binary black hole coalescences toward cluster binaries. Successful parameter
estimation for the properties of coalescing binaries in advanced detectors
will require accurate modeling of the waveform, including spin effects for
both black holes and neutron stars, and tidal effects for neutron stars.

Binary systems containing white dwarfs will be brought into contact at orbital
periods of a few hundred seconds. At contact, the system will either coalesce
on a dynamical timescale if the mass transfer is unstable, or it will begin to
spiral out if the mass transfer is stable. In any case, these systems will
never produce gravitational radiation in the frequency band of ground-based
interferometers. Space-based interferometers are capable of achieving
sufficient sensitivity in the millihertz band. Thus, the sources for
space-based interferometers will include white dwarf, neutron star, and black
hole binaries. At these orbital periods, the systems are not very relativistic
and the gravitational waveforms can safely be described by the quadrapole
formula of Peters \& Mathews~\cite{peters63,peters64}. An angle averaged
estimate of the typical strain amplitude is~\cite{lisa96}
\begin{equation}
  h_0 = 1.5 \times 10^{-21} \left(\frac{f_\mathrm{gw}}{10^{-3}
  \mathrm{\ Hz}}\right)^{2/3} \left(\frac{1 \mathrm{\ kpc}}{r}\right)
  \left(\frac{{\cal M}}{M_{\odot}}\right)^{5/3}\!\!\!\!\!\!\!.
\end{equation}
At a typical globular cluster distance of $r \sim 10 \mathrm{\ kpc}$ and
typical chirp mass of ${\cal M} \sim 0.5\,M_{\odot}$, a relativistic WD--WD or
WD--NS binary with $P_\mathrm{orb} = 400 \mathrm{\ s}$ will have a
gravitational wave amplitude of $10^{-22}$. This value is within the range of
proposed space-based gravitational wave observatories such as
LISA~\cite{lisa96}.

Many globular clusters lie off the plane of the galaxy and are relatively
isolated systems with known positions. The angular resolution of LISA improves
with signal strength. By focusing the search for gravitational radiation using
known positions of suspected sources, it is possible to increase the
signal-to-noise ratio for the detected signal. Thus, the angular resolution of
LISA for globular cluster sources can be on the order of the angular size of
the globular cluster itself at $f_\mathrm{gw} > 1 \mathrm{\ mHz}$. Consequently,
the orbital period distribution of a globular cluster's population of
relativistic binaries can be determined through observations in gravitational
radiation. We will discuss the prospects for observing each class of
relativistic binaries covered in this review.

WD--WD binaries that are formed from a common envelope phase will be briefly
visible while the recently revealed hot core of the secondary cools. These
objects are most likely the ``non-flickerers'' of Cool et al.~\cite{cool98}
and Edmonds et al.~\cite{edmonds99}. WD--WD binaries formed through exchange
interactions may very well harbor white dwarfs which are too cool to be
observed.  In either case, hardening through dynamical interactions will
become less likely as the orbit shrinks and the effective cross section of the
binary becomes too small. These objects will then be effectively invisible in
electromagnetic radiation until they are brought into contact and RLOF can
begin. During this invisible phase, the orbital period is ground down through
the emission of gravitational radiation until the orbital period is a few
hundred seconds~\cite{benacquista99}. With a frequency of 1 to 10~mHz,
gravitational radiation from such a binary will be in the band of
LISA~\cite{lisa96}.

WD--NS binaries that are expected to be progenitors of the millisecond pulsars
must pass through a phase of gravitational radiation after the degenerate core
of the donor star emerges from the common envelope phase and before the
spin-up phase begins with the onset of mass transfer from the white dwarf to
the neutron star. The orbital period at the onset of RLOF will be on the order
of 1 to 2 minutes and the gravitational wave signal will be received at LISA
with a signal-to-noise of 50\,--\,100 at a frequency of around 20~mHz for a
globular cluster binary.

Binaries with significant eccentricity will have a spectrum of harmonics of
the orbital frequency, with the relative strength of the $n$th harmonic for
eccentricity $e$ given by~\cite{peters63}
\begin{eqnarray}
  g(n, e) & = & \frac{n^4}{32}\bigg\{\!\left[J_{n-2}(ne) -
  J_{n-1}(ne) + \frac{2}{n}J_n(ne) + J_{n+1}(ne) -
  J_{n+2}(ne)\right]^2
  \nonumber \\
  & & \qquad + (1-e^2)\left[J_{n-2}(ne) \!-\! 2J_n(ne) \!+\!
  J_{n+2}(ne)\right]^2 \!+\! \frac{4}{3n^2}\left[J_n(ne)\right]^2
  \!\bigg\},
\end{eqnarray}%
where $J_n$ is the Bessel function. The higher harmonics of sufficiently
eccentric binaries ($e > 0.7$) can be detected by LISA even though the
fundamental orbital frequency is well below its sensitivity band of
1\,--\,100~mHz~\cite{benacquista02}. Dynamical interactions may produce an
eccentric population of 30\,--\,140 white dwarf binaries that would be present
in the LISA data after a 5~year
observation~\cite{willems07}.\epubtkUpdateA{Added last sentence and reference
  to Willems~(2007).}

Although the globular cluster population of NS--NS binaries is expected to be
quite small ($\sim$~10), they may have high eccentricities. The binary pulsar
B2127+11C is an example of a NS--NS binary in a globular cluster. In terms of
the unknown angle of inclination $i$, the companion mass to the pulsar is
$M_2\sin{i} \sim 1\,M_{\odot}$ and its eccentricity is $e =
0.68$~\cite{lorimer01}. These binaries may also be detectable by LISA. If the
globular cluster systems of other galaxies follow similar evolution as the
Milky Way population, these binaries may be potential sources for LIGO as
gravitational radiation grinds them down to coalescence. With their high
eccentricities and large chirp mass, black hole binaries will also be good
potential sources for gravitational radiation from the galactic globular
cluster system~\cite{benacquista01b, benacquista02}.

The relatively close proximity of the galactic globular cluster system and the
separations between individual globular clusters allows for the identification
of gravitational radiation sources with their individual host
clusters. Although the expected angular resolution of LISA is not small enough
to allow for the identification of individual sources, knowledge of the
positions of the clusters will allow for focused searches of the relativistic
binary populations of the majority of the galactic globular clusters. Armed
with a knowledge of the orbital periods of any detected binaries, concentrated
searches in electromagnetic radiation can be successful in identifying
relativistic binaries that may have otherwise been missed.

There may be a few double black hole systems within the Galactic globular
cluster system that are either within the LISA band, or are sufficiently
eccentric that their peak gravitational wave power is in a high harmonic that
is within the LISA band. In either case, such systems would be easily
detectable by a space-based gravitational wave detector due to their large
chirp masses. However, a greater number of double black hole systems are
expected to have been ejected from their birth globular clusters and these
would now be inhabitants of the Galactic halo. If these systems have been
formed dynamically, they are likely to be composed of two relatively massive
black holes, and they may have chirp masses that are high enough for them to
be detected at extragalactic distances.

\newpage

%%%%%%%%%%%%%%%%%%%%%%%%%%%%%%%%%%%%%%%%%%%%%%%%%%%%%%%%%%%%%%%%%%%%%%%%%%%%%%%
%%%%%%%%%%%%%%%%%%%%%%%%%%%%%%%%%%%%%%%%%%%%%%%%%%%%%%%%%%%%%%%%%%%%%%%%%%%%%%%
%%%%%%%%%%%%%%%%%%%%%%%%%%%%%%%%%%%%%%%%%%%%%%%%%%%%%%%%%%%%%%%%%%%%%%%%%%%%%%%

\section{Summary}
\label{section:summary}
Relativistic binaries are tracers of the rich dynamical evolution of globular
clusters.  The properties of this binary population are the result of an
interplay between the gravitational dynamics of large $N$-body systems, the
dynamics of mass transfer, the details of stellar evolution, and the effect of
the gravitational field of the galaxy.  The gravitational dynamics of globular
clusters can enhance the population of short period binaries of main-sequence
stars as well as inject compact objects such as white dwarfs and neutron stars
into stellar binary systems.  Once they are in such systems, the details of
stellar evolution and mass transfer in close binary systems govern the likely
end products of the dynamical interaction between the two stars. Furthermore,
most models of the evolution of the core of a globular cluster rely on the
gradual hardening and ejection of binary systems to delay the onset of core
collapse. The hardening of binaries in the core of globular clusters will
produce relativistic binaries, but it will also eventually eject these systems
as they gain larger and larger recoil velocities in each subsequent
encounter. The threshold for ejection from a globular cluster depends both
upon the gravitational potential of the cluster itself and the gravitational
potential of its environment generated by the Milky Way.  As the globular
cluster orbits the Milky Way, its local environment changes. Consequently, if
other dynamical processes (such as gravothermal oscillations) do not dominate,
the globular cluster's population of relativistic binaries may also reflect
the past orbital history of the globular cluster.

Over the last decade, observational techniques and technology have improved to
the extent that significant discoveries are being made regularly. At this
point, the bottleneck in observations of binary millisecond pulsars, low-mass
X-ray binaries, and cataclysmic variables is time, not technology. As these
observational techniques are brought to bear on more clusters, more
discoveries are bound to be made. In the next decade, the possibility of using
gravitational wave astronomy to detect relativistic binaries brings the
exciting possibility of identifying the populations of electromagnetically
invisible objects such as detached white dwarf and neutron star binaries and
black hole binaries in globular clusters. These observations can only help to
improve the understanding of the complex and interesting  evolution of these
objects and their host globular clusters.

\newpage

%%%%%%%%%%%%%%%%%%%%%%%%%%%%%%%%%%%%%%%%%%%%%%%%%%%%%%%%%%%%%%%%%%%%%%%%%%%%%%%
%%%%%%%%%%%%%%%%%%%%%%%%%%%%%%%%%%%%%%%%%%%%%%%%%%%%%%%%%%%%%%%%%%%%%%%%%%%%%%%
%%%%%%%%%%%%%%%%%%%%%%%%%%%%%%%%%%%%%%%%%%%%%%%%%%%%%%%%%%%%%%%%%%%%%%%%%%%%%%%

\section{Acknowledgements}
\label{section:acknowledgements}
MJB acknowledges the kind hospitality of the Aspen Center for Physics, and the
anonymous referees whose extensive knowledge of the subject has helped fill in
the gaps in his understanding. Extensive use was made of both the arXiv
pre-print server and the ADS system. Previous versions of this work were
supported by NASA EPSCoR grant NCCW-0058, Montana EPSCoR grant NCC5-240, NASA
Cooperative Agreement NCC5-579, and NASA APRA grant NNG04GD52G. This current
version has been supported by the Center for Gravitational Wave Astronomy,
which is supported by NASA Grant NNX09AV06A and NSF award \#0734800.

JMBD is supported by the European Gravitational Wave Observatory through VESF
grant EGO--DIR-50-2010.  He also wishes to acknowledge financial support
provided by the University of Texas at Brownsville for a visit during which
work on the most recent edition of this review was initiated.

\newpage

%%%%%%%%%%%%%%%%%%%%%%%%%%%%%%%%%%%%%%%%%%%%%%%%%%%%%%%%%%%%%%%%%%%%%%%%%%%%%%%
%%%%%%%%%%%%%%%%%%%%%%%%%%%%%%%%%%%%%%%%%%%%%%%%%%%%%%%%%%%%%%%%%%%%%%%%%%%%%%%
%%%%%%%%%%%%%%%%%%%%%%%%%%%%%%%%%%%%%%%%%%%%%%%%%%%%%%%%%%%%%%%%%%%%%%%%%%%%%%%

\bibliography{refs}

\end{document}